\documentclass[
twocolumn,
amsmath,amssymb,amsfonts,
longbibliography,
aps
,pra
,superscriptaddress
,nofootinbib
,notitlepage]{revtex4-1}
\usepackage{graphicx}
\usepackage{dcolumn}
\usepackage{bm}
\usepackage{braket}
\usepackage[dvipsnames]{xcolor}

\usepackage[normalem]{ulem}

\usepackage[colorlinks=False,linkcolor=blue,urlcolor=black,citecolor=blue,bookmarksopen=true]{hyperref}
\PassOptionsToPackage{hyphens}{url}\usepackage{hyperref}

\newcommand{\cL}{\hat{\mathcal{L}}}

\newcommand{\tl}{\text{tl}}
\newcommand{\ctl}{\text{c-tl}}
\newcommand{\ac}{\text{a-c}}
\newcommand{\mean}[1]{\langle #1 \rangle}
\newcommand{\hc}{\hat{c}}
\newcommand{\hcD}{\hat{c}^{\dagger}}
\newcommand{\hb}[1]{\hat{b}_{#1}}
\newcommand{\hbD}[1]{\hat{b}_{#1}^{\dagger}}
\newcommand{\tc}{\textrm{c}}
\newcommand{\aD}{a^{\dagger}}
\newcommand{\Tr}{\textrm{Tr}}

\newcommand{\update}[1]{{#1}} 
\newcommand{\CK}[1]{{#1}}

\begin{document}
\title{Proposal for a continuous wave laser with linewidth well below the standard quantum limit}

\author{Chenxu Liu}
\email[email: ]{chenxu\_liu@pitt.edu}
\altaffiliation[Now at: ]{Department of Physics, Virginia Tech, Blacksburg, VA, 24061, USA}
\affiliation{Department of Physics and Astronomy, University of Pittsburgh, Pittsburgh, PA, 15260, USA}
\affiliation{Pittsburgh Quantum Institute, University of Pittsburgh, Pittsburgh, PA, 15260, USA}

\author{Maria Mucci}
\affiliation{Department of Physics and Astronomy, University of Pittsburgh, Pittsburgh, PA, 15260, USA}
\affiliation{Pittsburgh Quantum Institute, University of Pittsburgh, Pittsburgh, PA, 15260, USA}

\author{Xi Cao}
\affiliation{Department of Physics and Astronomy, University of Pittsburgh, Pittsburgh, PA, 15260, USA}
\affiliation{Pittsburgh Quantum Institute, University of Pittsburgh, Pittsburgh, PA, 15260, USA}

\author{M. V. Gurudev Dutt}
\affiliation{Department of Physics and Astronomy, University of Pittsburgh, Pittsburgh, PA, 15260, USA}
\affiliation{Pittsburgh Quantum Institute, University of Pittsburgh, Pittsburgh, PA, 15260, USA}

\author{Michael Hatridge}
\affiliation{Department of Physics and Astronomy, University of Pittsburgh, Pittsburgh, PA, 15260, USA}
\affiliation{Pittsburgh Quantum Institute, University of Pittsburgh, Pittsburgh, PA, 15260, USA}

\author{David Pekker}
\email[email: ]{pekkerd@pitt.edu}
\affiliation{Department of Physics and Astronomy, University of Pittsburgh, Pittsburgh, PA, 15260, USA}
\affiliation{Pittsburgh Quantum Institute, University of Pittsburgh, Pittsburgh, PA, 15260, USA}

\date{\today}

\maketitle

\section{Abstract}
Due to their high coherence, a laser is a ubiquitous tool in science. We show that by engineering the coupling between the gain medium and the laser cavity as well as the laser cavity and the output port, it is possible to eliminate most of the noise due to photons entering as well as leaving the laser cavity. 
Hence, it is possible to reduce the laser linewidth by a factor equal to the number of photons in the laser cavity below the standard quantum limit.  
We design and theoretically analyze a superconducting  circuit that uses Josephson junctions, capacitors and inductors to implement a microwave laser, including the low-noise couplers that allow the design to surpass the standard quantum limit. Our proposal relies on the elements of superconducting quantum information, and thus is an example of how quantum engineering techniques can inspire us to re-imagine the limits of conventional quantum systems.

\begin{figure*}
    \centering
    \includegraphics[width=0.8 \textwidth]{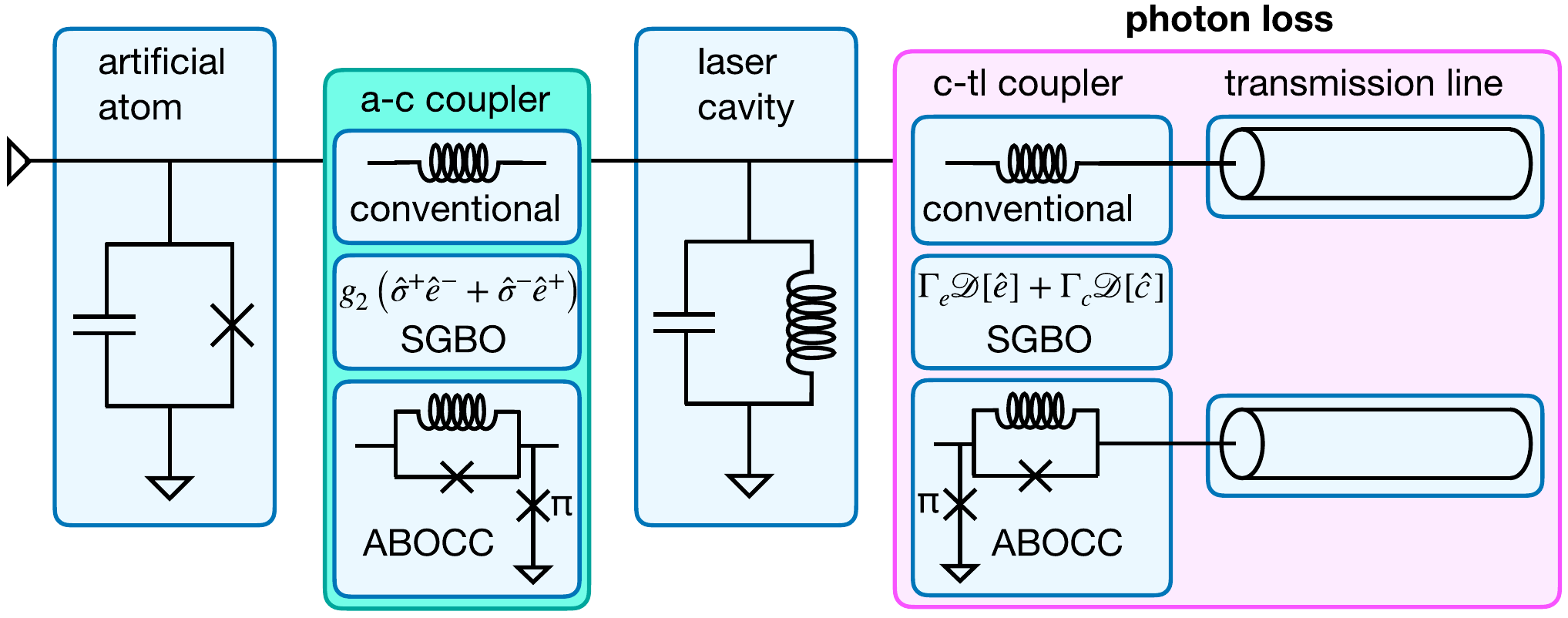}
    \caption{\update{Schematic of the superconducting Josephson laser. The laser is} composed of: an artificial two-level atom (transmon qubit) that is incoherently pumped from the ground to the excited state at the rate $\Gamma_p$, an atom-cavity (a-c) coupling circuit, a laser cavity made of an LC resonator, a cavity-transmission line (c-tl) coupling circuit, and an output transmission line. The coupling circuits come in three flavors: (1) conventional: linear inductors, (2) Susskind-Glogower bare operator (SGBO) couplers, and (3) Approximate Bare Operator Coupling Circuits (ABOCC). As the SGBO scheme is a purely theoretical construct, it is represented by the a-c coupling Hamiltonian and the photon loss operator. }
    \label{fig:model}
\end{figure*}

\section{Introduction}

\update{The key property of a laser, which is crucial to its applications in quantum optics, quantum information, and metrology, is its high coherence or narrow linewidth.}
The main components of a laser, depicted in Fig.~\ref{fig:model}, are (1)~one or more atoms with an inverted population (also called the gain medium), (2)~an atom-cavity coupler, (3)~a lasing cavity, (4)~a cavity-output coupler and an output channel (transmission line). 
The standard quantum limit (SQL) for the phase coherence time was first introduced by \update{Schawlow and Townes}~\cite{Schawlow1958}, who showed that the minimum possible laser linewidth is determined by the linewidth of the laser cavity divided by twice the number of photons in the cavity. \update{This} raises the question whether it is possible to surpass this limit?
Previous work on laser theory~\cite{Wiseman1999}, quantum-cascade lasers~\cite{Bartalini2010}, superradiant lasers~\cite{Bohnet2012}, and number-squeezed lasers~\cite{Yamamoto1986} has focused on the gain medium (1) and the atom-cavity coupler (2). Specifically, in Ref.~\cite{Wiseman1999}, Wiseman showed theoretically that by using Susskind-Glogower operators~\cite{Susskind1964} to couple the gain medium to the laser cavity, it is possible to eliminate pump noise, but not loss noise~\cite{Barnet1989}. This decreases the minimum laser linewidth, though only by a factor of two.

At optical frequencies, the inherent light-matter coupling is rather weak and consequently optical devices tend to be only weakly nonlinear. Over the past two decades, significant progress has been made on building strongly nonlinear optics. At optical frequencies, using small-size high-Q optical cavities coupling to atoms, strong coupling can be achieved in cavity QED system by reducing the optical mode volume~\cite{Kimble1998, Mabuchi2002, McKeever2003, Walther2006}. On the other hand, at microwave frequencies, circuit QED achieves extremely strong light-artificial atom interactions by utilizing the extreme nonlinearity and small size (compared to microwave-frequency photons) of Josephson junctions~\cite{Girvin2014}. Circuit QED devices include the various flavors of superconducting quantum computing platforms with components like fluxonium~\cite{Manucharyan2009} and transmon qubits~\cite{Koch2007, Schreier2008}, resonant cavities, microwave waveguides, and quantum limited parametric amplifiers~\cite{ClerkRMP2008,Bergeal2010}. There is also experimental precedent for building conventional lasers using superconducting circuits with linear couplers~\cite{Astafiev2007, Chen2014, Rolland2019}, as well as devices based on parametrically driven, weakly nonlinear oscillators ~\cite{Cassidy2017, Simon2018}. 

In a very recent work, Baker \update{et al.}~\cite{Wiseman2020arXiv} pointed out that it should be possible to reduce the linewidth of a laser by a factor of $\sim \langle n \rangle^2$ below the SQL, where $\langle n \rangle$ is the mean photon number inside the cavity. They called this new limit on laser linewidth the \CK{``Heisenberg limit''}. Further, Baker \update{et al.} proposed a microwave circuit in which photon gain and loss processes were engineered using a pair of \CK{``photon treadmills''} to add and remove photons from the microwave laser (maser) cavity. 
\update{Baker's photon treadmill proposal represents a substantial increase in complexity, as this maser requires multiple, pulsed light sources which make the maser's linewidth a direct trad-eoff against complex controls.}

\update{In the present work, we begin by showing that simple engineering of both the cavity-output coupling, in addition to the atom-cavity coupling (see Fig.~\ref{fig:model}),  can be used to suppresses the phase noise in the lasing cavity.
Although our simple engineering results in a decrease of the laser linewidth by a factor $\sim \langle n \rangle$ below the SQL, as compared with $\sim \langle n \rangle^2$ obtained by Baker \update{et al.}, it provides us with useful notions for how to construct lasers that operate well below the SQL using only static couplers that do not require coherent light pulses. Using these notions, we establish our main result: a blueprint for building a maser in which Josephson junctions and linear inductors are used to construct non-linear coupling circuits. These coupling circuits approximate the behavior of Susskind-Glogower operators for a range of cavity photon occupancies. We show that by using these circuits to couple the laser cavity to both the gain medium and the output port it is possible to suppress the linewidth of the resulting maser beyond the SQL by a factor of $\langle n \rangle^{-1.098}$.
Although we do not achieve the proposed `Heisenberg limit' of $\langle n \rangle^2$ reduction, our scheme requires only a single, continuous, incoherent pump and is thus a much simpler light source to control. We believe that this is an important advantage for the development of sub-SQL lasers at optical frequencies as well as for applications of these devices both at microwave and optical frequencies. }

\section{Result}
\subsection{Overview of the results}
The starting point for our exploration is a proposal due to Wiseman~\cite{Wiseman1999} for reducing laser spectral linewidth. The laser linewidth, coming from the phase noise of the laser light, has two equal contributions from both the atom-pump process and the cavity loss process~\cite{Barnet1989}. In~\cite{Wiseman1999} Wiseman proposed using \CK{`bare'} (so called because they lack a photon-number scaling pre-factor) raising and lowering operators
\begin{align}
\hat{e}=\sum_i |i\rangle \langle i+1| \quad \quad \hat{e}^\dagger=\sum_i |i+1\rangle \langle i|,
\label{eq:SGBO}
\end{align}
which were first introduced by Susskind and Glogower~\cite{Susskind1964}, to couple atoms of the gain medium to the resonant cavity of the laser. As these operators commute with the phase $\hat{\phi}$ of the optical field in the cavity (which can be verified by observing that $\hat{e}=e^{i\hat{\phi}}$), Wiseman's proposal eliminates pump noise and therefore reduces the minimum linewidth by a factor of two \update{(labeled $D_{\text{ST2}}$)} below the standard quantum limit, also known as the Schawlow-Townes (ST) limit \update{(labeled  $D_{\text{ST}}$)}~\cite{Scully1997, Wiseman1999},
\update{i.e., 
\begin{equation}
    D_{\text{ST2}} = \frac{1}{2} D_{\text{ST}}, \quad D_{\text{ST}} = \frac{\Gamma_c}{4 \mean{n}},
    \label{eq:ST_limit}
\end{equation}
where $\Gamma_c$ is the cavity linewidth and $\mean{n}$ is the mean photon number inside the cavity.
}

\update{Our first result is a simple argument that establishes the result that was presented by Baker \update{et al.}~\cite{Wiseman2020arXiv}, that it is possible to surpass the SQL on laser linewidth by manipulating both that the atom-cavity and the cavity-output channel couplings. We show that the application of Wiseman's scheme to both couplings can be used to eliminate both the pump and the cavity loss noise. However}, we must also add some conventional loss to the laser in order to stabilize it, the result is a linewidth that is $\langle n \rangle$ times narrower than the ST limit. Having a mathematical scheme for building an ultranarrow linewidth laser, we still need an experimentally viable method for building bare operators. 

Our second\update{, and main,} result is an Approximate Bare Operator Coupling Circuit (ABOCC), composed of Josephson junctions and inductors, that approximates the desired coupling Hamiltonian over a range of cavity photon occupancies. For laser applications, our ABOCC is an attractive alternative to previous proposals to build bare operators that relied on adiabatic rapid passage~\cite{Wiseman1999, Oi2013, Govia2014, Rosenblum2016, Um2016, Radtke2017} as it doesn't require additional drives. We argue that an ABOCC laser, in which couplers are ABOCCs, is a practical laser design that achieves the ultranarrow linewidth promised by SGBO laser. In the remainder of this paper, we first calculate the behavior of the purely theoretical SGBO laser compared to an ideal conventional laser, and then describe the physically realizable ABOCC in detail and describe the potential performance of a laser based on pair of ABOCCs.

\begin{figure}[t]
    \centering
    \includegraphics[width=0.8 \columnwidth]{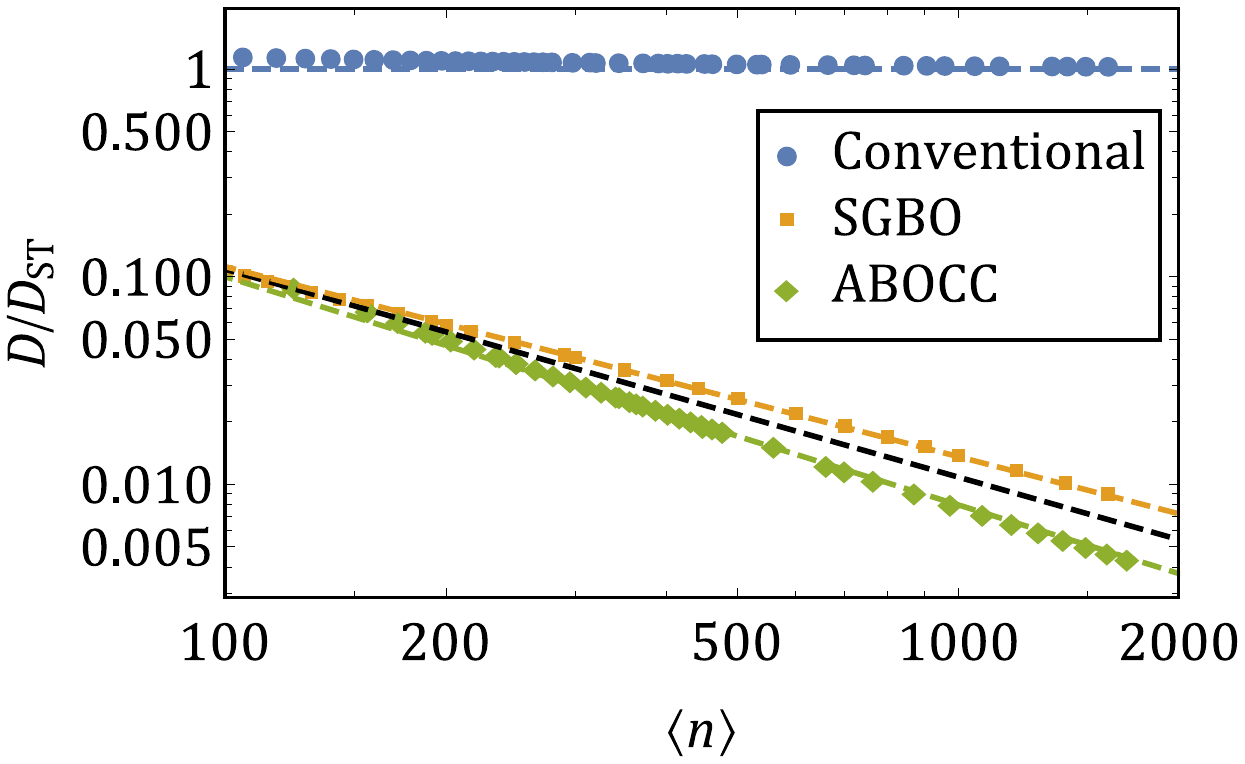}
    \caption{Laser linewidth as a function of the average number of photons $\langle n \rangle$ in the laser cavity for conventional, Susskind-Glogower Bare Operator (SGBO), and Approximately Bare Operator Coupling Circuit (ABOCC) lasers. The laser linewidth is in units of the generalized Schawlow-Townes linewidth. Engineering of the atom-cavity coupling and the photon loss allows both the SGBO and the ABOCC lasers to achieve a linewidth significantly narrower than the best conventional laser. The dots represent the numerically calculated linewidth ratio. The dashed blue line represents $D/D_\text{ST}=1$, the expected linewidth for the conventional laser in the $\mean{n}\rightarrow \infty$ limit. The orange and green dashed lines are linear fits on the log-log scale of the linewidth ratio \update{versus the mean photon number $\mean{n}$ of the SGBO and ABOCC laser, respectively.} The SGBO laser $D/D_{\text{ST}} \sim \mean{n}^{-0.914}$ (orange) and ABOCC laser $D/D_{\text{ST}} \sim \mean{n}^{-1.098}$ (green). The black dashed line is a guide to the eye with $D/D_{\text{ST}} \sim \mean{n}^{-1}$.
    }
    \label{fig:linewidth}
\end{figure}

\begin{figure*}[t]
    \centering
    \includegraphics[width=0.9\textwidth]{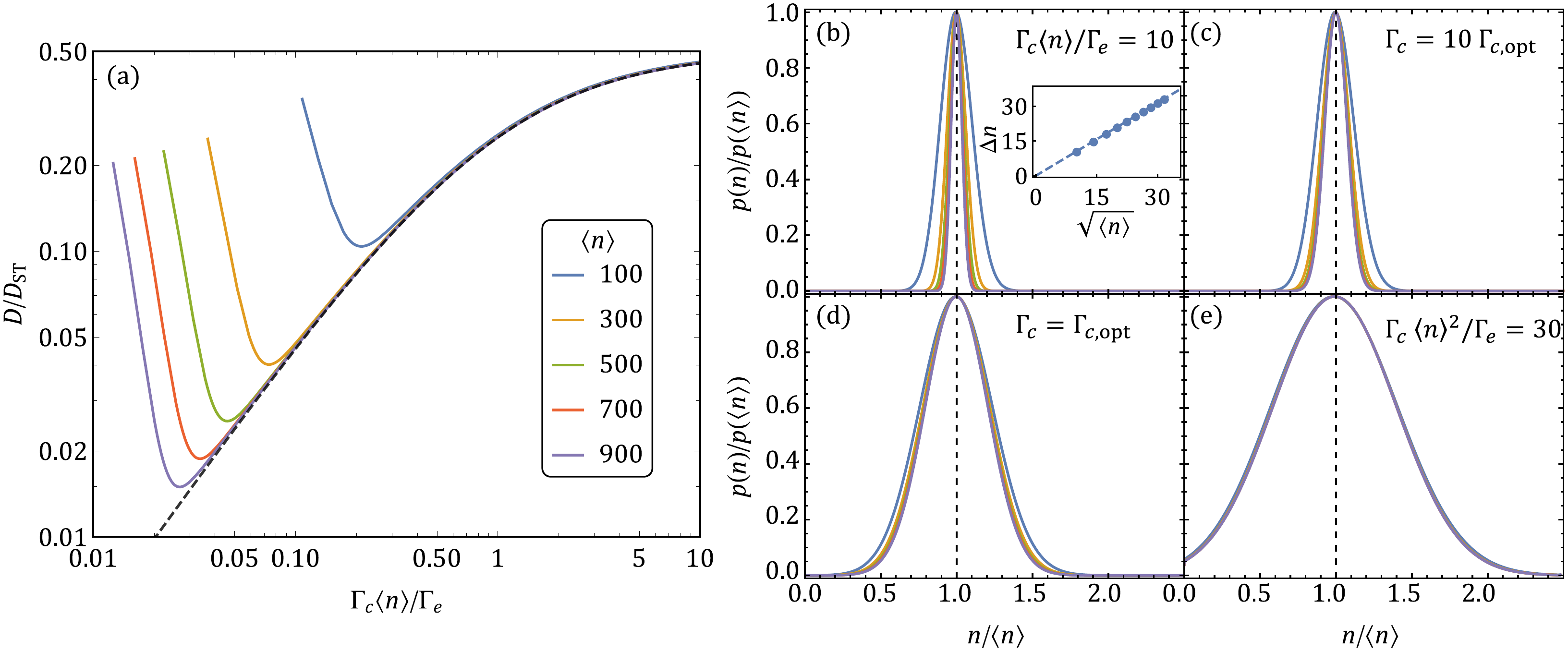}
    \caption{\update{The linewidth and the photon distribution of the Susskind-Glogower Bare Operator (SGBO) laser.} (a) Minimum laser linewidth can be achieved by tuning the ratio of power emitted by conventional ($\langle n\rangle \Gamma_\text{c}$) and bare operator ($\Gamma_\text{e}$) loss for different cavity occupancies. The dotted line represents Eq.~\eqref{eq:EWlinewidth}. Photon number distributions are shown in (b-e) for four cases: conventional loss is dominant (b); intermediate conventional loss (c); minimum linewidth (d); and   conventional loss is very weak (e). \update{The lines in all subfigures are for the SGBO laser with $\langle n \rangle = 100$ (blue), $300$ (orange), $500$ (green), $700$ (salmon) and $900$ (purple).}
    }
    \label{fig:EWlaser}
\end{figure*}

In order to compare the linewidth $D$ of different laser designs to the Schawlow-Townes limit $D_\text{ST}$ we need to generalize the Schawlow-Townes formula for the cases in which the cavity-transmission line coupler is not linear. We do so by replacing the cavity linewidth $\Gamma_\text{c}$ by the ratio of the laser luminosity to the energy of the photons in the cavity $\Gamma_\text{c}\rightarrow P_\text{out}/(\hbar \omega_\text{c} \langle n \rangle)$ \update{in Eq.~\eqref{eq:ST_limit}} thus obtaining the formula 
\begin{align}
D_{\text{ST}}=\frac{P_\text{out}}{4 \hbar \omega_\text{c} \langle n \rangle^2}. \label{eq:DGST}
\end{align}
For conventional lasers, Eq.~\eqref{eq:DGST} is identical to the standard Schawlow-Townes linewidth formula. To ensure that each type of laser is performing at its optimal,  we fix the mean photon number in the cavity and minimize the ratio $D/D_\text{ST}$ by tuning the laser parameters. For example, for the case of the conventional laser, we tune the atom-cavity coupling strength ratio $g/\Gamma_\text{c}$ and the atom incoherent pump rate ratio $\Gamma_p/\Gamma_\text{c}$. 

In Fig.~\ref{fig:linewidth} we plot the optimum laser linewidth, relative to the ST limit, as a function of the average number of photons in the laser cavity for three types of lasers: the conventional laser, the SGBO laser, and the ABOCC laser. All data in this figure were obtained numerically using the spectral method to analyze the master equation (see methods). We observe that for the conventional laser, the ratio $D/D_\text{ST}$ approaches unity as $n$ becomes large. At the same time, we observe that the laser linewidth for the SGBO laser as well as the ABOCC laser is significantly narrower and goes as $D \sim D_\text{ST}/\langle n \rangle$.

\begin{figure*}[t]
    \centering
    \includegraphics[width=\textwidth]{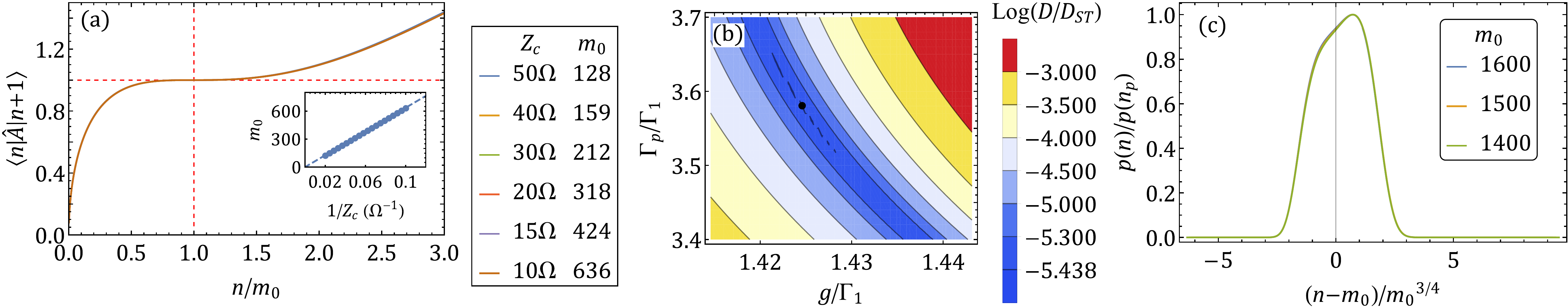}
    \caption{\update{The Approximately Bare Operator Coupling Circuit (ABOCC) operator and the linewidth and photon distribution of the ABOCC laser.} (a) The matrix element $\langle n \vert \hat{A}_{\text{c},m_0} \vert n+1 \rangle$ as a function of the photon number in the cavity shows a plateau on which the matrix element is independent of the photon number thus approximating the bare operator. The matrix elements, that we computed for cavity impedance \update{$Z_c=50~\Omega$ (blue), $40~\Omega$ (orange), $30~\Omega$ (green), $20~\Omega$ (salmon) $15~\Omega$ (purple) and $10~\Omega$ (brown)} which controls $m_0$ as shown in the inset (the impedance of the qubit was set to $Z_a=47.71~\Omega$), appear almost indistinguishable after re-scaling the photon number by $m_0$. 
    (b) Tuning the ABOCC laser by varying the pump power $\Gamma_\text{p}$ and the atom-cavity coupling strength $g$ that is controlled by $I_\text{J:a-c}$. (c) The photon number distributions of the optimum linewidth point \update{at $m_0 = 1600$ (blue), $1500$ (orange) and $1400$ (green)}. All three lines appear to collapse after appropriate re-scaling of both axis.
    }
    \label{fig:ABOCClaser}
\end{figure*}

\subsection{Suppressing the loss noise}

We start by extending Wiseman's strategy for decreasing the laser linewidth to make the SGBO laser.
Following Wiseman, we replace the linear inductive coupling between the atom and the cavity by the bare operator coupling 
\begin{align}
    H^{(\text{SGBO})}_\text{a-c}=g_2\left(\hat{\sigma}^+\hat{e}^- + \hat{\sigma}^-\hat{e}^+ \right),
\end{align}
where $e$ and $e^\dagger$ are defined in Eq.~\eqref{eq:SGBO}. We extend Wiseman's scheme by setting the cavity loss super-operator to be 
\begin{align}
    \cL^{(\text{SGBO})}_\text{c-tl} = \Gamma_\text{e} {\cal D}[\hat{e}] + \Gamma_\text{c} {\cal D}[\hat{a}],
\end{align}
where $\Gamma_\text{e}$ controls the rate of loss by the bare operators while $\Gamma_\text{c}$ controls the rate of the conventional loss mechanism. This extension can be thought as a form of bath engineering. A small amount of conventional loss is essential for stabilizing the laser as, without it, neither the rate at which photons are pumped into the cavity, nor the rate at which photons leave the cavity depends on the number of photons in the cavity, and hence the laser becomes unstable.

We can achieve this tremendous reduction in linewidth as the bare operator couplings allow photons to enter and leave the laser cavity without inducing phase noise (and hence they do not directly contribute to the linewidth). On the other hand, the photon number operator is conjugate to the phase operator, and therefore in the presence of only bare operator couplings the distribution of photon numbers in the cavity becomes infinitely broad. Adding conventional loss makes the photon number distribution have finite width, thereby stabilizing the laser at the cost of introducing phase noise. While both the conventional and the bare operator loss are contributing to the laser luminosity, only the conventional loss is contributing to the laser linewidth, and therefore the ratio of the SGBO laser linewidth $D_\text{SGBO}$ to the generalized Schawlow-Townes limit \update{can be approximated by}
\begin{equation}
\frac{D_\text{SGBO}}{D_\text{ST}}\update{\sim} \frac{\langle n \rangle\Gamma_\text{c}}{2\left(\Gamma_\text{e}+\langle n \rangle \Gamma_\text{c}\right)}. \label{eq:EWlinewidth}
\end{equation}
\update{
Here, the factor of 2 in the denominator accounts for the elimination of pump noise, similar to the the laser proposed in Ref.~\cite{Wiseman1999}.
}
In Fig.~\ref{fig:EWlaser}a we plot the linewidth ratio $D_\text{SGBO}/D_\text{ST}$ as a function of $\Gamma_\text{c} \langle n \rangle/\Gamma_\text{e}$, the ratio between power emitted by conventional and bare operator loss while keeping the mean photon number fixed (see methods). We observe that as we decrease $\Gamma_\text{c} \langle n \rangle/\Gamma_\text{e}$, the ratio $D_\text{SGBO}/D_\text{ST}$ first follows Eq.~\eqref{eq:EWlinewidth}, then saturates at a point that depends on the number of photons in the cavity, and then begins increasing again. The origin of saturation and increase can be understood by looking at the distribution of photon numbers in the laser cavity.

When $\Gamma_\text{c} \langle n \rangle/\Gamma_\text{e}$ is large, conventional loss dominates and the distribution of photon numbers in the cavity has a width $\sim \sqrt{\langle n \rangle}$ (Fig.~\ref{fig:EWlaser}b). As $\Gamma_\text{c} \langle n \rangle/\Gamma_\text{e}$ decreases, the distribution of photon numbers in the cavity broadens (Fig.~\ref{fig:EWlaser}c). This continues until the distribution width $\Delta n$ becomes roughly half of $\langle n \rangle$, at which point the linewidth saturates and the photon distribution becomes universal (Fig.~\ref{fig:EWlaser}d). As $\Gamma_\text{c} \langle n \rangle/\Gamma_\text{e}$ is decreased even further, the photon number distribution becomes even broader (Fig.~\ref{fig:EWlaser}e) and the probability to have no photons in the cavity becomes appreciable. The state with no photons in the cavity does not have a well defined phase. Consequently, the occupation of this state dominates the broadening of the laser linewidth for small $\Gamma_\text{c} \langle n \rangle/\Gamma_\text{e}$.

\subsection{Engineering the ABOCC coupling circuits}
We now take on the challenge of engineering the ABOCC laser using circuit QED devices. In the ABOCC laser system, we use a transmon qubit as a pumping atom (Fig.~\ref{fig:model}). We use bath engineering to achieve an incoherent drive on the transmon qubit to achieve population inversion (see \update{Supplementary Note~1}). The transmon qubit is modeled by a two-level system. The microwave cavity is modeled as a LC resonator, which couples to the transmon qubit and the output transmission line through two ABOCCs to mimic the SG operators.

We start with the atom-cavity coupling. The key property of the $\hat{e}$ operator is that the matrix element $|\langle n-1 | \hat{e} | n \rangle| = 1$ is independent of $n$, while for the standard photon annihilation operator $\hat{c}$, $|\langle n-1 | \hat{c} | n \rangle| = \sqrt{n}$. We have come up with the coupling circuit, depicted in Fig.~\ref{fig:model}, composed of an RF-SQUID (a Josephson junction shunted by a linear inductor) with an additional $\pi$ junction to ground (the $\pi$ junctions could be made from, for example, a second RF-SQUID that is flux-biased). The linear inductor in the RF-SQUID provides the linear coupling between the atom and the laser cavity (the cavity and the transmission line). The Josephson junction provides the nonlinear coupling, whose strength is controlled by the Josephson critical current of the junction. By tuning the critical current and the linear inductance, the Boson amplification factor ($\sqrt{n}$ factor) can be largely suppressed within some photon number range.

The ABOCC coupling the atom to the cavity is described by the Hamiltonian
\begin{align}
    H^{(\text{ABOCC})}_\text{a-c}&=\frac{\phi_0^2 \hat{\delta}^2}{2 L_\text{a-c}} - E_\text{J:a-c} \cos \hat{\delta} +  E_\text{J:c} \cos \hat{\varphi}_\text{c}, 
    \label{eq:HABOCC_classical}
\end{align}
where $\hat{\delta}=\hat{\varphi}_\text{a} -\hat{\varphi}_\text{c}$ and $\hat{\varphi}_\text{a}$ and $\hat{\varphi}_\text{c}$ are the superconducting phase operators of the transmon and the cavity; $\phi_0=\Phi_0/2\pi$; $E_\text{J:a-c}=\Phi_0 I_\text{J:a-c}$ and $L_\text{a-c}$ are the Josephson energy and the linear inductance of the RF-SQUID part of the ABOCC  and $E_\text{J:c}$ is the Josephson energy of the $\pi$ junction. After quantizing the microwave cavity field, applying Baker-Campbell-Hausdorff (BCH) formula and rotating wave approximation, the atom-cavity coupling induced by ABOCC is,
\begin{align}
    H^{(\text{ABOCC})}_\text{a-c}= g \left(\hat{\sigma}^+ \hat{A}_{\text{c},m_0} + \hat{\sigma}^- \hat{A}^\dagger_{\text{c},m_0}\right), \label{eq:HABOCC}
\end{align}
where $\hat{\sigma}$ are the Pauli operators for the atomic degree of freedom, the atom-cavity coupling strength $g$ is controlled by the the critical current ($I_{\text{J:a-c}}$) of the Josephson junction in the ABOCC and the normalization parameter $\mathcal{N}$ of the ABOCC cavity operator $\hat{A}_{\text{c},m_0}$ and $\hat{A}_{\text{c},m_0}$ is an effective cavity photon annihilation operator,
\begin{align}
\label{eq:ABOCC_Ac}
    \hat{A}_{\text{c},m_0} &= \frac{1}{ \mathcal{N} } \left(\frac{2\phi_0^2 \tilde{\varphi}_\text{a} \tilde{\varphi}_\text{c} }{L_\text{a-c} E_{\text{J:a-c}} } c
    \right. \\
     & \left.    -\sin \left( \tilde{\varphi}_\text{a} \right) e^{-\frac{\tilde{\varphi}_\text{c}^2}{2}}
    \sum_{n=0}^\infty \frac{(-1)^n \tilde{\varphi}_\text{c}^{2n+1} (c^\dagger)^n c^{n+1} }{n! (n+1)!}\right).  \nonumber 
\end{align}
and $\tilde{\varphi}_\text{a,c}=\frac{1}{\phi_0}\sqrt{\frac{\hbar Z_\text{a,c}}{2}}$ and $\mathcal{N}$ is a normalization factor that ensures that $\langle m_0 \vert A_{\text{c},m_0} \vert m_0 + 1 \rangle=1$.

The operator $\hat{A}_{\text{c},m_0}$ is designed to work when there are $m_0$ photons in the laser cavity, where $m_0$ depends on the transmon and cavity impedances (see inset of Fig.~\ref{fig:ABOCClaser}a and methods). In Fig.~\ref{fig:ABOCClaser}a we plot the matrix element $\langle n | \hat{A}_{\text{c},m_0} | n+1 \rangle$ as a function of $n/m_0$. We observe that the matrix element has a plateau, centered on $n=m_0$, around which it is independent of $n$. The plateau is obtained by combining the sinusoidal current phase relation of the Josephson junction with the linear current phase of the inductor in the coupling circuit (see methods). On this plateau, $\hat{A}_{\text{c},m_0}$ behaves approximately like the bare operator $\hat{e}$. 
\update{In Fig.~\ref{fig:ABOCClaser}a the matrix element traces with different $m_0$’s are indistinguishable. This indicates that the range of $n$'s over which $\hat{A}_{\text{c},m_0}$ behaves like $\hat{e}$ also scales with $m_0$. }

We introduce another ABOCC to engineer the nonlinear coupling between the laser cavity and the transmission line (bath). After applying BCH fomular and RWA, we further assume the transmission line field can be treated as a Markovian bath and trace out the bath degrees of freedom (see method). After tracing out the bath, the effective nonlinear loss for the cavity field can be expressed in the Lindblad form as,
\begin{align}
    \cL^\text{ABOCC}_\text{c-tl}&=\Gamma_1 {\cal D}[\hat{B}_{\text{c},m_0}], \label{eq:ABOCCloss}
\end{align} 
where the rate constant
\begin{align}
     \Gamma_1&=\mathcal{N}^2 \frac{E^2_{\textrm{J:c-tl}}}{\hbar^2}\left( \frac{\hbar Z_\text{tl}}{2 \phi_0^2}\right)\frac{1}{\omega_\text{c}}
     \label{eq:ABOCClossRate}
\end{align}
the nonlinear cavity operator
\begin{align}
    \hat{B}_{\text{c},m_0}&= \frac{1}{\mathcal{N}} \left( \mathcal{C}_{\textrm{tl}} e^{- \tilde{\varphi}_\text{c}^2/2}  \sum_{n = 0}^{\infty} 
    \frac{(-1)^n  \tilde{\varphi}_\text{c}^{2n+1}}{n! \cdot (n+1)!} \left( c^{\dagger} \right)^n c^{n+1} \right. \nonumber \\
    & \qquad \quad + \left. \frac{2 \phi_0^2 \tilde{\varphi}_c}{L_{\text{c-tl}} E_{\textrm{J:c-tl}}} c \right),
    \label{eq:ABOCClossOp}
\end{align}
and $Z_{\tl}$ is the transmission line characteristic impedance, $\omega_c$ is the frequency of the cavity, $\mathcal{C}_{\textrm{tl}}=\exp\left( - \frac{1}{\phi_0^2} \frac{\hbar Z_\tl}{4 \pi} \right)  \left[ 1 + \frac{\theta}{\omega_c} + \frac{\theta^2}{2 \omega_c^2} + O\left(\frac{\theta^3}{\omega_c^3}\right)\right]$ and $\theta \ll \omega_\text{c}$ is the bandwidth of the transmission line, $\mathcal{N}$ is a normalization constant (see \update{Supplementary Note~4}).

The ABOCC laser, in which we use ABOCCs for both atom-cavity and cavity-transmission line couplings, is stable and does not require conventional loss like the SGBO laser. To tune up the ABOCC laser we first choose the desired number of photons in the cavity and then set the impedances of the inverted transmon qubit, the cavity and the transmission line so that both ABOCCs have the desired value of $m_0$.  Next, we vary the transmon qubit's inverting incoherent pump strength $\Gamma_\text{p}$ (see~\cite{Ma2019, Mucci2020} 
and online supplementary notes for how to implement this drive) and the atom-cavity coupling strength $g$ while fixing the cavity-transmission line coupling strength in order to minimize the ratio $D_\text{ABOCC}/D_\text{ST}$ (see Fig.~\ref{fig:ABOCClaser}b).

The general performance characteristics of the ABOCC laser are very similar to those of the SGBO laser. Both have a linewidth which is a factor of $1/\langle n \rangle$ narrower than the generalized Schawlow-Townes limit (see Fig.~\ref{fig:linewidth}). At the optimal operating point, both have a photon distribution in the cavity with a width that scales with $\langle n \rangle^{3/4}$ as opposed to $\langle n \rangle^{1/2}$ for conventional lasers (see Fig.~\ref{fig:EWlaser}d, \ref{fig:ABOCClaser}c and \update{Supplementary Note~6}). There is, however, a difference in the shape of the distributions. In the case of the SGBO laser the photon distribution width is limited by occupation of the empty state. On the other hand in the ABOCC laser it is limited by the width of the plateau on which the operator $\hat{A}_{\text{c},m_0}$ behaves like the bare operator (see Fig.~\ref{fig:ABOCClaser}a). 

\section{Discussion}
\update{We pause to comment on what is a laser and what are the crucial ingredients for building a sub-SQL laser. We start by asking what are the key properties a laser: is it a lasing threshold? is it stimulated emission? or is it turning broad-band pump light into narrow linewidth output light? Both the device in Ref.~\cite{Wiseman2020arXiv} and our device can be thought of as single-atom masers, in which just one atom as opposed to a collection of atoms is used to pump the resonant cavity. There is considerable literature on the properties of conventional single-atom lasers and masers and whether these devices should be considered to be true lasers though they do not have a lasing threshold~\cite{Meschede1985, Filipowicz1986, Lugiato1987, Mu1992, Gunnar1994, Wiseman1997, McKeever2003, Boozer2004}. 
However, we concentrate on single-atom lasers as a more technically tractable approach to the problem of constructing a sub-SQL device. Instead, the crucial point is abandoning both stimulated emission and conventional photon loss from the resonant cavity, which are the key features of both conventional and single-atom laser, in favor of quantum-engineered atom-cavity and cavity-output couplings. While the sub-SQL masers proposed here and in Ref.~\cite{Wiseman2020arXiv} do not have a lasing threshold nor employ stimulated emission, they do share the crucial property that they can take noisy input light and turn it into ultranarrow output light as required by Refs.~\cite{Wiseman1997, Wiseman1999}.}

Our results raise the question:  given our demonstrated ability to greatly exceed the standard quantum limit,  what is the \update{ultimate} quantum limit on the linewidth of a laser? Baker et al. argue that the ST limit divided by $\langle n \rangle^2$, which the authors call the Heisenberg limit, is the ultimate limit on laser linewidth~\cite{Wiseman2019,Wiseman2020arXiv}. Our laser circuit already far exceeds the ST limit, operating at the geometric mean of the ST and the Heisenberg limit. However, in passive nonlinear optical/microwave systems, there appears to still be some room for future improvements. Specifically, in contrast to Ref.~\cite{Wiseman2020arXiv}, where the photon number noise scales as $\langle n \rangle$, our proposed ABOCC laser shows $\langle n \rangle^{3/4}$ scaling. We speculate that by further optimizing the coupling circuits it should be possible to realize photon number noise that scales with $\langle n \rangle$ and laser linewidth that approaches the Heisenberg limit.

In summary, we have shown that by engineering the photon loss operator it is possible to build a laser that is $\langle n \rangle$ times narrower than the standard quantum limit, where $\langle n \rangle$ is the number of photons in the laser cavity. We have also developed a realistic roadmap for constructing this type of laser using standard circuit QED components: capacitors, inductors and Josephson junctions. These are exactly the same components that are used in a wide variety of superconducting quantum information devices. \update{The device we propose could be an ultra-coherent, cryogenic light source for microwave quantum information experiments.

Further, t}he photon field in the cavity of the proposed laser is highly squeezed (see Supplementary Note~6) with the photon number distribution width scaling with the photon number (as opposed to the square-root of the photon number that is observed in conventional lasers). The proposed device can be thought of as approaching the Heisenberg limit on phase estimation. 
\update{We envision that the proposed device could be modified to provide designer quantum light which is an important resource for continuous variable/linear optical quantum computing~\cite{Gottesman2001, Menicucci2014, Marshall2016}, readout of quantum states in superconducting quantum computers~\cite{Didier2015, Eddins2018}, quantum metrology~\cite{caves1981, Bondurant1984, Ma2017, XJZuo2020}, and quantum communication~\cite{Yuen1978, Slusher1990, Gottesman2003, Vahlbruch2008}. 
In addition to being an interesting source of quantum light for quantum information experiments, the proposed device also shows how even well-understood quantum optical objects such as lasers can be re-imagined with the techniques of quantum information and the tools of superconducting circuits.
}

\section*{Methods}
\subsection{Eigenspectrum method to solve laser linewidth}
To describe the lasers we use the master equation
\begin{align}
    \dot{\rho}=\cL_\text{a}[\rho] - \frac{i}{\hbar} [H_{\text{a-c}},\rho] - \frac{i}{\hbar} [H_{\text{c}},\rho] + \cL_\text{c-tl}[\rho].
    \label{eq:master1}
\end{align}
The Hamiltonian $H_{\text{a-c}}$ describes that atom-cavity coupling and the super-operator $\cL_\text{c-tl}[\rho]$ describes loss of cavity photons to the transmission line. The super-operator $ \cL_\text{a}[\rho]= - \frac{i \omega_a}{2} [\sigma^z,\rho] + \Gamma_p {\cal D}[\sigma^+] \rho$ describes the artificial atom, where $\sigma^z$ is the atom Pauli matrix and ${\cal D}[\sigma^+] \rho = \sigma^+ \rho \sigma^- -\frac{1}{2} ( \sigma^- \sigma^+ \rho + \rho  \sigma^- \sigma^+ )$ describes the action of the incoherent pump. The incoherent drive on the transmon qubit is achieved through bath engineering. We couple a SNAIL qubit, which has third order nonlinearity to achieve two-photon pumping process, to the transmon qubit. We drive the two-photon process that pump both SNAIL and transmon qubits from their ground states to the first excited state. The SNAIL qubit also couples to a lossy cavity, which makes the SNAIL qubit to have a fast population decay back to the ground state. Therefore, the transmon qubit experiences an effective incoherent drive from its ground state to the excited state (see \update{Supplementary Note~1}). The Hamiltonian $H_{\text{c}}=\omega_c c^\dagger c$ describes the cavity energy levels, where $c$ is the photon annihilation operators in the cavity and we set $\omega_c=\omega_a$. 

To obtain the linewidth of the laser we numerically find the eigenspectrum of the time evolution super-operator defined by the right hand side of Eq.~\eqref{eq:master1}. 
The master equation Eq.~\eqref{eq:master1} can be thought of as an eigenvalue problem of the super-operator. The spectrum of eigenvalues has one zero eigenvalue $\lambda_0=0$, which corresponds to the steady state solution of the laser system, and a number of negative eigenvalues which correspond to the decaying modes. The laser linewidth can be calculated by the Fourier spectrum of the two-time correlation function of the light field, which consists of a linear combination of decaying exponentials with the decay time set by these negative eigenvalues. Almost all of the weight is carried by the eigenstate with the largest nonzero eigenvalue of the super-operator (see \update{Supplementary Note~3} for details). We use this largest non-zero eigenvalue of the super-operator to estimate the laser linewidth.

\subsection{Optimizing SGBO laser}
To optimize the SGBO laser (Fig.~\ref{fig:EWlaser}a): (1) We work in the strong atom-cavity coupling regime by fixing $g_2 = 1000 \sqrt{\Gamma_P \Gamma_\text{e}}$. In this regime we estimate the mean photon number to be $\langle n \rangle \sim (\Gamma_\text{p} - 2\Gamma_\text{e})/(2 \Gamma_\text{c})$. (2) We tune $\Gamma_\text{p}/\Gamma_\text{e}$ and $\Gamma_\text{c}/\Gamma_\text{e}$ to optimize the laser linewidth while fixing the mean number of photons in the cavity.

\subsection{Deriving ABOCC nonlinear cavity operators}
The ABOCC coupling the atom to the cavity is described by the Hamiltonian in Eq.~\eqref{eq:HABOCC_classical}. 
Note that the nonlinear atom-cavity coupling is given by $\cos \hat{\delta}$ term, which can be expanded as,
\begin{align}
    \cos \hat{\delta} &= \frac{1}{2} e^{-\frac{\tilde{\varphi}_c^2}{2}} \left(e^{-i \tilde{\varphi}_\text{a} \hat{\sigma}_x} e^{i \tilde{\varphi}_c c^{\dagger}} e^{i \tilde{\varphi}_c c}  
    \right.\\ \nonumber 
    & \qquad \qquad \qquad  \left. 
    + e^{i \tilde{\varphi}_\text{a} \hat{\sigma}_x} e^{-i \tilde{\varphi}_c c^{\dagger}} e^{-i \tilde{\varphi}_c c} \right)
\end{align}
where we assume the atom is a 2-level system, whose operators can be expressed using Pauli operators, i.e., $\hat{\varphi}_\text{a} = \tilde{\varphi}_{\text{a}} \hat{\sigma}_x$, and  $\tilde{\varphi}_\text{a,c}=\frac{1}{\phi_0}\sqrt{\frac{\hbar Z_\text{a,c}}{2}}$. This term induces a nonlinear self-energy of the cavity field $\cos(\tilde{\varphi}_{\text{a}}) \cos{\hat{\varphi}_c}$ can be cancelled by tuning the critical current of the $\pi$-junction in ABOCC. The rest of the terms with the cavity operators can be expanded using the Baker–Campbell–Hausdorff (BCH) formula, and applying the rotating wave approximation we obtain Eq.~\eqref{eq:HABOCC}. 

The nonlinear coupling between the cavity and transmission line through ABOCC is analyzed using similar method. At first, the free field of the transmission line is quantized (see \update{Supplementary Note~2} for details)~\cite{ClerkRMP2008}. The ABOCC coupling Hamiltonian is in the similar form of Eq.~\eqref{eq:HABOCC_classical}, except the phase difference operator
\begin{align}
    \hat{\delta}_{\ctl} = \tilde{\varphi}_\tc \left( \hc + \hcD \right) + i \sum_{k} \tilde{\varphi}_\tl (k) \left( \hb{k} - \hbD{k} \right).
\end{align}
where $\tilde{\varphi}_\tl(k) = \frac{1}{\phi_0} \sqrt{\frac{\hbar Z_\tl}{2}} \sqrt{\frac{v_p}{l \omega_k}}$ is the transmission line quantization constant, $v_p$, $Z_{\tl}$ and $l$ is the wave speed, the characteristic impedance and total length of the transmission line, respectively.

Because of the nonlinear nature of the cavity-transmission line coupling, there are multi-photon exchange processes, which can be seen from the expanding the $\cos (\hat{\delta}_{\ctl})$ terms in the coupling Hamiltonian. We re-order the expanded terms by the number of transmission line (bath) operators, which is also the number of photon exchange between the cavity and the transmission line. We assume the transmission line is a vacuum bath and apply RWA and Markov approximation to trace out the transmission line degrees of freedom, all the terms in different orders in the coupling Hamiltonian give decay channels, which are represented by different Lindblad operators. The first order terms in the Hamiltonian gives the leading order among these decay channels (see \update{Supplementary Note~4}). We focus on the leading order and ignore the higher order terms. The first order term give the Lindblad operator given in Eq.~\eqref{eq:ABOCCloss} with loss constant and nonlinear cavity operator in Eqs.~\eqref{eq:ABOCClossRate} and~\eqref{eq:ABOCClossOp}.

\section*{Data and code availability}
The code written in this study have been deposited in the Zenodo under DOI: \url{10.5281/zenodo.5016168}~\cite{code_DOI}. The data used for generating all the figures in both the main manuscript and the supplementary notes are generated by the codes. 

\section*{Acknowledgments}
We thank Andrew Daley, John Jeffers, and Howard Wiseman for helpful comments. C. Liu acknowledges support from a Pittsburgh Quantum Institute graduate student fellowship. Research was also supported by the Army Research Office under Grants Number W911NF-18-1-0144 and W911NF-15-1-0397, and by M. Hatridge's NSF CAREER grant (PHY-1847025). G.D. was partially supported by NSF EFRI ACQUIRE 1741656. The views and conclusions contained in this document are those of the authors and should not be interpreted as representing the official policies, either expressed or implied, of the Army Research Office or the U.S. Government. The U. S. Government is authorized to reproduce and distribute reprints for Government purposes notwithstanding any copyright notation herein. 

\section*{Author Contributions}
C.L. derived the master equation, wrote and ran the numerical codes, analyzed the results and produced the figures. M.V.G.D. pointed out Wiseman's work from 1999 that inspired this project. Following this inspiration, all co-authors contributed to the discussions that led to the development of the physical picture. M.M., X.C., and M.H. invented the two-qubit artificial atom and provided experimentally realizable parameters that were used by C.L. in maser calculations. D.P. and C.L. invented the ABOCC circuit and wrote the manuscript with input from all the co-authors. D.P. supervised the project. 

\section*{Competing Interests}
The authors declare no competing interests.

\def\bibsection{
\section*{References}
} 

\bibliography{laser_ref}

\newpage

\onecolumngrid

\renewcommand{\figurename}{Supplementary Figure}
\renewcommand{\tablename}{Supplementary Table}
\renewcommand{\thesection}{Supplementary Note \arabic{section}}

\begin{center}
\textbf{
\large{Proposal for a continuous wave laser with linewidth well below the standard quantum limit: Supplementary Information}
}
\end{center}

This supplement is composed of 6 supplementary notes.
\begin{itemize}
    \item Supplementary note 1: In this note we describe how to engineer an artificial 3-level atom using two qubits.
    \item Supplementary note 2: In this note we derive the superoperator that describes the photon loss to the transmission line.
    \item Supplementary note 3: In this note we describe the relation between the spectrum of the time evolution superoperator and the laser linewidth.
    \item Supplementary note 4: In this note we construct the photon loss operator that is induced by the AOCC circuit.
    \item Supplementary note 5: In this note we analyze the robustness of the ABOCC laser to imperfections.
    \item Supplementary note 6: In this note we show that the eigenstates of the ABOCC lowering operator are highly squeezed. This is in contrast to the conventional lowering operator that has coherent eigenstates.
\end{itemize}

\section{Engineering an artificial atom with population inversion} \label{sec:incoherent_pump}

In our main text, we used an incoherent drive to pump the transmon qubit from its ground state to the excited state. In this note we provide a specific design recipe for how to build such an incoherent pump.

\begin{figure}[h!]
    \centering
    \includegraphics[width = 0.9 \textwidth]{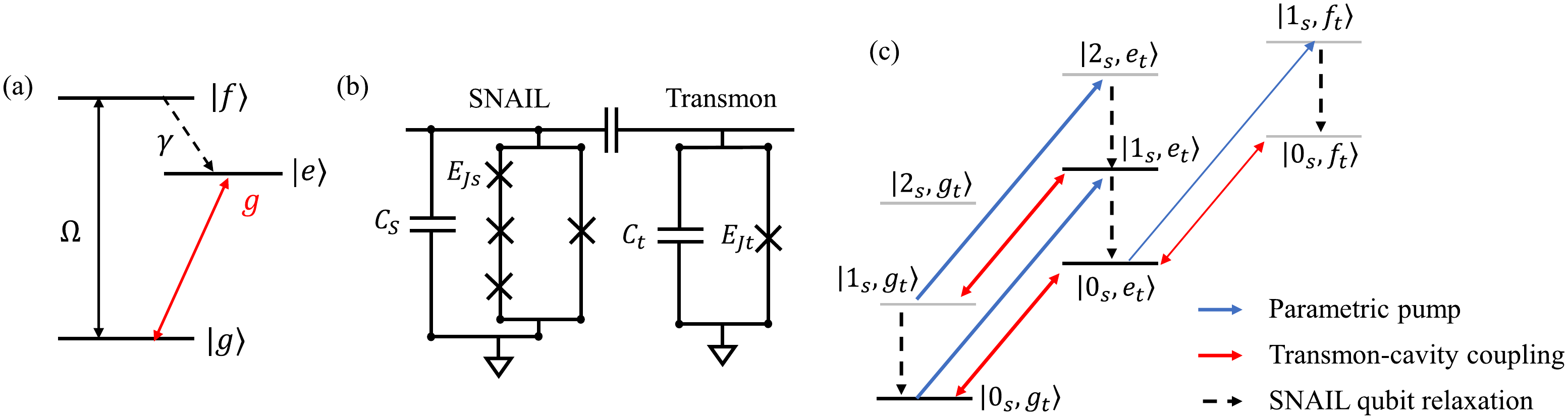}
    \caption{The abstract three-level model of the laser pump media, the circuit diagram and the level structure of the proposed laser pump media. (a) the three-level model of the laser pump media (atom) where only coherent drives are permitted. (b) the circuit diagram of the SNAIL qubit and transmon qubit composite system, in which the effective incoherent drive of the transmon qubit can be achieved. (c) the level structure of the SNAIL qubit and the transmon qubit. The gray thin lines are the levels that are weakly populated in the pumping scheme discussed in \ref{sec:incoherent_pump}. The blue lines are the parametric drive on the composite system, the red arrows show the transition of the composite system due to the coupling to the laser cavity. The dashed arrows show the relaxation process of the SNAIL qubit. The thick solid lines show the transition that are on resonance, while the thin solid lines show the ones that are not on resonance (because of the anharmonicity of the transmon qubit).}
    \label{fig:SNAIL_transmon_levels}
\end{figure}

Our goal is to build an effective three-level-atom (see Fig.~\ref{fig:SNAIL_transmon_levels}a), where the transition between the ground state $\ket{g}$ and the second excited state $\ket{f}$ is coherently driven while the second excited state experiences a fast decay to the first excited state $\ket{e}$. If the decay process is sufficiently fast, as the population of the atom is driven to the state $\ket{f}$, it quickly relaxes to $\ket{e}$ and we achieve population inversion on the two lasing levels $\ket{g}$ and $\ket{e}$. However, the single-photon transition between $\ket{g}$ and $\ket{f}$ for a transmon qubit is forbidden by selection rule. Therefore, we propose coupling a SNAIL qubit to the transmon qubit to form a composite system (see Fig.~\ref{fig:SNAIL_transmon_levels}b). The key feature of the SNAIL qubit is that it has third order nonlinearity that makes the $\ket{g} \rightarrow \ket{f}$ transition allowed. 

The level structure of the two qubit system is shown in Fig.~\ref{fig:SNAIL_transmon_levels}c. We use $\ket{g_t}$, $\ket{e_t}$ and $\ket{f_t}$ to represent the ground, first, and second excited states of the transmon qubit. For the SNAIL qubit we use $\ket{0_s}$, $\ket{1_s}$ and $\ket{2_s}$, etc., to represent the ground, first excited, second excited, etc. states. 

Coupling of the SNAIL and transmon qubits results in the hybridization of their states. Consequently, third order non-linearity of the SNAIL qubit can be used to drive the $\ket{0_s, g_t} \rightarrow \ket{1_s, e_t}$ transition as the state $\ket{1_s, e_t}$ is hybridized with the state $\ket{2_s, g_t}$. This type of transitions are labeled by blue arrows in Fig.~\ref{fig:SNAIL_transmon_levels}c. If the SNAIL qubit is also coupled to an output port, such that the relaxation of the SNAIL qubit [see Fig.~\ref{fig:SNAIL_transmon_levels}c, black dashed arrows] is fast compared to the pump process (and also the transmon-cavity coupling, see Fig.~\ref{fig:SNAIL_transmon_levels}c red arrows), then the two qubit system can form an effective three-level atom, in which $\vert 0_{s}, g_{t} \rangle$ plays the role of the ground state ($\vert g \rangle$ in Fig.~\ref{fig:SNAIL_transmon_levels}a), $\vert 1_{s}, e_{t} \rangle$ the role of the second excited state ($\vert f \rangle$ in Fig.~\ref{fig:SNAIL_transmon_levels}a), and $\vert 0_{s}, e_{t} \rangle$ the role of the first excited state ($\vert e \rangle$ in Fig.~\ref{fig:SNAIL_transmon_levels}(a)).

The Hamiltonian of a SNAIL qubit coupled to a transmon qubit is
\begin{subequations}
\label{eq:composite_hamiltonian}
\begin{align}
    & H = H_{\textrm{T}} + H_{\textrm{S}} + H_{\textrm{couple}} \\
    & H_{\textrm{T}} = \omega_t \hat{t}^{\dagger} \hat{t} + k_t \hat{t}^{\dagger} \hat{t}^{\dagger} \hat{t} \hat{t} \\
    & H_{\textrm{S}} = \omega_s \hat{s}^{\dagger} \hat{s} + g_3 \left(\hat{s}^{\dagger} \hat{s}^{\dagger} \hat{s} + \hat{s}^{\dagger} \hat{s} \hat{s} \right) \\
    & H_{\textrm{couple}} = g_2 \left( \hat{s}^{\dagger} \hat{t} + \hat{t}^{\dagger} \hat{s} \right)
\end{align}
\end{subequations}
where $\hat{s}$ ($\hat{t}$) is the photon annihilation operator for the SNAIL (transmon) qubit. We have also truncated the nonlinear parts of the Hamiltonians of both  qubits to the lowest non-trivial order and dropped the rapidly rotating terms like $\hat{s}^{\dagger}\hat{s}^{\dagger}\hat{s}^{\dagger}$ and $\hat{s} \hat{s} \hat{s}$ in SNAIL qubit Hamiltonian. Finally, we make the assumption that the SNAIL and transmon qubits are strongly detuned as compared with the strength of the linear coupling, i.e., $2 \Delta = \vert \omega_s - \omega_t \vert \gg g_2$, and hence the modes of the two qubits are only weakly hybridized. In the dressed basis, with respect to $g_2$ coupling, the third order nonlinearity of the bare SNAIL mode results in a third order nonlinear coupling between the dressed SNAIL and transmon modes $H_3 = \hat{s}'^{\dagger}\hat{s}'\hat{t}' + h.c.$, where $\hat{s}'$ and $\hat{t}'$ are the dressed SNAIL and transmon operators.  Applying classical drive to the bare SNAIL mode 
\begin{equation}
    H_{\textrm{d}} = \Omega_{d} \exp \left[i \left( \omega_s + \omega_t \right) t \right] \hat{s} + h.c.,
    \label{eq:composite_coh_drive}
\end{equation}
induces the two-photon pump process (see blue arrows in Fig.~\ref{fig:SNAIL_transmon_levels}) via $H_3$. 

\begin{table}[h]
    \centering
    \begin{tabular}{l|c|c}
    \hline
    Parameter & Variable & Value \\
    \hline
       SNAIL qubit frequency & $\omega_s/(2\pi)$  &  $7.2$~GHz \\ 
       SNAIL qubit 3rd order nonlinearity & $g/(2\pi)$ & $50$~MHz \\
       transmon qubit frequency & $\omega_t/(2\pi)$ & $6.7$~GHz \\
       transmon qubit 4th order nonlinearity & $k/(2\pi)$ &$-0.3$~GHz\\
       SNAIL-transmon coupling & $g_2/(2\pi)$ & $50$~MHz\\
       SNAIL coherent pump rate & $\Omega_d/(2\pi)$ & $2.0$~GHz\\ 
       SNAIL loss rate & $\gamma/(2\pi)$ & $5$~MHz\\ 
    \hline
    \end{tabular}
    \caption{Parameters used for numerical integration of Eq.~\ref{eq:SNAIL_transmon_master_eq} when generating figure Fig.~\ref{fig:population_inversion}}.
    \label{tab:parameters}
\end{table}

\begin{figure}[ht!]
    \centering
    \includegraphics[width = 0.9 \textwidth]{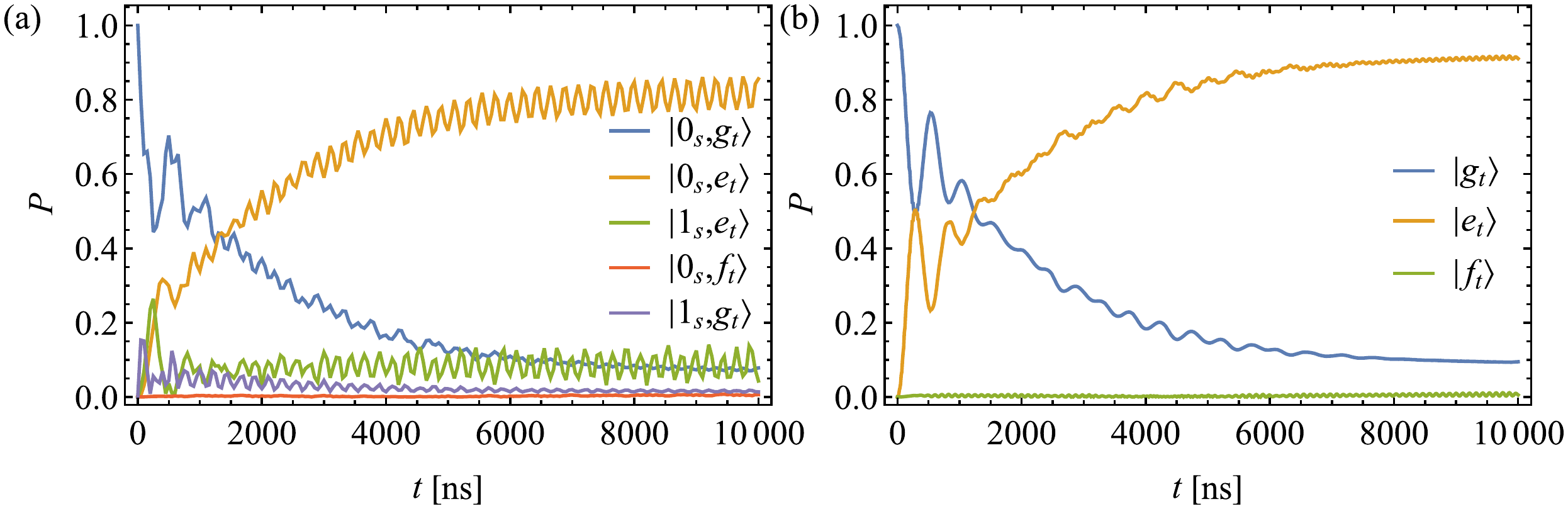}
    \caption{Population inversion of the composite quantum system made up of a SNAIL and a transmon qubit under the influence of a coherent drive. (a) Population of the lowest five levels of the composite system (as labeled) plotted as a function of time. (b) Population of the transmon levels, ground state (blue), excited state (orange) and second excited state (green), after tracing over the SNAIL qubit degrees of freedom. See Table~\ref{tab:parameters} for values of parameters used to construct this figure.
}
    \label{fig:population_inversion}
\end{figure}

The composite quantum system can be described by the master equation 
\begin{subequations}
\label{eq:SNAIL_transmon_master_eq}
    \begin{align}
        \partial_t \rho_{St} & = -i [H + H_d, \rho_{St}] + \gamma \mathcal{D}[\hat{s}] \rho_{St} \\
        \label{eq:SNAIL_dissipator}
        \mathcal{D}[\hat{s}]\rho_{St} & = -\frac{1}{2} \left( \hat{s}^{\dagger} \hat{s} \rho_{St} + \rho_{St} \hat{s}^{\dagger} \hat{s} - 2 \hat{s} \rho_{St} \hat{s}^{\dagger} \right) 
    \end{align}
\end{subequations}
where $\rho_{St}$ is the density operator for the composite system of the coupled SNAIL and transmon qubits, the system Hamiltonian is given by Eq.~\eqref{eq:composite_hamiltonian}, and the classical drive Hamiltonian by Eq.~\eqref{eq:composite_coh_drive} and Eq.~\eqref{eq:SNAIL_dissipator} describes the dissipation of the SNAIL mode.

To demonstrate that the proposed two-qubit system functions as a three level atom, we numerically integrate the the master equation, Eq.~\eqref{eq:SNAIL_transmon_master_eq}. As the higher levels of the composite systems are weakly populated, we truncated the Hilbert space of the SNAIL qubit to allow a maximum of $6$ photons, and the space of the transmon qubit to $3$ photons. The populations of the lowest five states of the composite systems are shown in Fig.~\ref{fig:population_inversion}a. We observe that the coherent drive applied to the SNAIL qubit, together with the photon loss from the SNAIL qubit, induce an effectively population transfer from $\vert 0_s, g_t \rangle$ to $\vert 0_s, e_t \rangle$. We observe population inversion after $\sim 3 \mu$s. At long times the system reaches a steady state with a significant population inversion (with roughly $90\%$ occupancy of the state $\vert 0_s, e_t \rangle$ and $10\%$ of the state $\vert 0_s, g_t \rangle$). The residual population of the state $\vert 0_s, g_t \rangle$ is caused by the decay of the $\vert 0_s, e_t \rangle$ via its hybridization with the state $\vert 1_s, g_t \rangle$. The ripples on the population curves are caused by the classical drive on the SNAIL qubit which causes the SNAIL mode to have a fast oscillating component at the frequency $(\omega_s + \omega_t)$. In Fig.~\ref{fig:population_inversion}b, we plot the population of the ground, first and second excited states of the transmon qubit after tracing over the SNAIL qubit degrees of freedom. From this plot we observe that the transmon qubit is effectively being pumped from the ground state $\vert g_t \rangle$ to the first excited state $\vert e_t \rangle$. From the two plots in Fig.~\ref{fig:population_inversion}, we observe that the higher excited states of the composite systems (e.g., $\vert 0_s, f_t \rangle$ and $\vert 1_s, g_t \rangle$) have very little population, especially the second excited state of the transmon qubit $\vert f_t \rangle$, which justifies the truncation of the composite system Hilbert space in our numerical calculation.

\section{Describing the photon loss from the laser cavity to the transmission line induced by a linear inductive coupler}
\label{sec:photonLossLinear}
In this note we derive the effective photon loss operator for the laser cavity induced by a linear inductive coupling between the laser cavity (LC resonator) and the transmission line. We extend this description to the ABOCC in Sec.~\ref{sec:ABOCCloss}.

The quantization of the LC resonator and the transmission line is discussed in Ref.~\cite{ClerkRMP2008, Girvin2014}. The canonical position and momentum of the LC resonator are the node superconducting phase $\varphi_c$ and charge $Q_c$. Using these coordinates, the quadratic Hamiltonian of the LC resonator can be quantized, similar to the Harmonic oscillator, via
\begin{subequations}
    \begin{align}
    \label{eq:quantized_LC_Q}
        \hat{Q}_c = & - i \sqrt{\frac{\hbar}{2 Z_c}} \left( \hc - \hcD \right) \\
    \label{eq:quantized_LC_Phi}
        \hat{\Phi}_c = & \sqrt{\frac{\hbar Z_c}{2}} \left( \hc + \hcD \right)
    \end{align}
\end{subequations}
where $Z_c$ is the characteristic impedance of the LC resonator, $Z_c = \sqrt{L_c / C_c}$, and the Hamiltonian of the LC resonator, in second quantized form, is $H_c = \hbar \omega_c \hat{a}^{\dagger} \hat{a}$, where the frequency of the LC resonator is $\omega_c = 1/\sqrt{L_c C_c}$. The voltage on the LC resonator is $V_c = \dot{\Phi}_c$ and the current flow in the LC resonator is $I_c = \dot{Q}_c$. We can express the voltage and current operators using the raising and lowering operators via
\begin{subequations}
    \begin{align}
    \label{eq:quantized_LC_V}
        \hat{V}_c = & -\frac{i}{\hbar} [\hat{\Phi}_c, H_c] = -i \omega \sqrt{\frac{\hbar Z_c}{2}} \left(\hc - \hcD \right), \\
    \label{eq:quantized_LC_I}
        \hat{I}_c = & -\frac{i}{\hbar} [\hat{Q}_c, H_c] = \omega \sqrt{\frac{\hbar}{2 Z_c}} \left( \hc + \hcD \right).
    \end{align}
\end{subequations}

Here we consider a single-mode transmission line that couples to the LC resonator via a linear inductor. The generalized flux along the transmission line is
\begin{equation}
    \Phi(x,t) = \int_{-\infty}^{t} d \tau V(x,\tau) = \frac{1}{\phi_0} \phi(x,t),
\end{equation}
and the charge density along the transmission line $q(x)$ can be quantized. After the quantization of the transmission line fields, $\phi(x)$ and $q(x)$ are 
\begin{subequations}
\begin{align}
\label{eq:tl_quantized_qx}
    \hat{q}(x) & = \sum_{k=-\infty}^{+\infty} \sqrt{\frac{\hbar \omega_k C}{2 l}} \left( \hb{k} e^{i k x} + \hbD{k} e^{-i k x}\right) \\
\label{eq:tl_quantized_phix}
    \hat{\phi}(x) & = -i \sum_{k=-\infty}^{+\infty} \sqrt{\frac{\hbar Z_{\tl} v_p}{2 l \omega_k}} \left( \hb{k} e^{i k x} - \hbD{k} e^{-i k x} \right)
\end{align}
\end{subequations}
where $v_p = 1/\sqrt{LC}$ is the wave speed along the transmission line and the dispersion relation of the mode with momentum $k$ is $\omega_k^2 = v_p^2 k^2$, $Z_{\tl} = \sqrt{L/C}$ is the characteristic impedance of the transmission line, $l$ is the total length of the transmission line.

We further assume that the linear inductive coupling element ($L_{\textrm{c-tl}}$) that couples the LC resonator and the transmission line, connects to the $x=0$ point of the transmission line. The coupling Hamiltonian is,
\begin{equation}
    H_{\ctl} = \frac{\phi_0^2}{2 L_{\ctl}}\left( \varphi_c - \varphi_{\tl} \right)^2 =  \frac{\phi_0^2}{2 L_{\ctl}} \left[ \tilde{\varphi}_c \left( c + c^{\dagger}\right) + \sum_{k} i \tilde{\varphi}_{\tl}(k) \left(\hat{b}_k - \hat{b}_k^{\dagger}\right) \right]^2,
\end{equation}
where the dimensionless parameters $\tilde{\varphi}_c$ and $\tilde{\varphi}_\tl (k)$ are defined as
\begin{equation}
    \tilde{\varphi}_c = \frac{1}{\phi_0} \sqrt{\frac{\hbar Z_c}{2}}, \qquad \tilde{\varphi}_\tl(k) = \frac{1}{\phi_0} \sqrt{\frac{\hbar Z_\tl}{2}} \sqrt{\frac{v_p}{l \omega_k}}.
    \label{eq:quantization_const}
\end{equation}

Here we only consider the LC resonator and the transmission line parts of the Josephson micromaser to understand the dissipation induced on LC resonator by the transmission line. The Hamiltonian of the system under rotating wave approximation is
\begin{equation}
    H = H_{\textrm{c}} + H_{\textrm{tl}} + H_{\textrm{c-tl}} 
    = \hbar \omega_c \hcD \hc + \hbar \sum_{k} \omega_k \hbD{k} \hb{k} + i \hbar \sum_{k} \kappa_{k} \left( \hcD \hb{k} - \hbD{k} \hc \right),
\end{equation}
where $\kappa_{k} = \frac{\phi_0^2}{\hbar L_{\textrm{c-tl}}} \tilde{\varphi}_c \tilde{\varphi}_{\textrm{tl}}(k)$. We further treat the transmission line as a vacuum bath, and apply the Born-Markov approximation to simplify the dynamics of the cavity field as in Ref.~\cite[Chap.~8]{Scully1997}. The dynamics of the LC resonator mode can be described by the master equation
\begin{equation}
    \partial_t \rho(t) = -\frac{i}{\hbar} [H_c, \rho(t)] - \frac{\gamma}{2} \left( \hcD \hc \rho(t) + \rho(t) \hcD \hc -2 \hc \rho(t) \hcD \right),
\end{equation}
where $\rho$ is the density operator for the LC resonator, and the decay rate 
\begin{equation}
    \gamma = \frac{Z_{c}Z_{\tl}}{2 L_{\textrm{c-tl}} \omega_c}.
\end{equation}

\section{Obtaining the laser line shape and linewidth using the spectrum of the time-evolution super-operator}
\begin{figure}
    \centering
    \includegraphics[width = \textwidth]{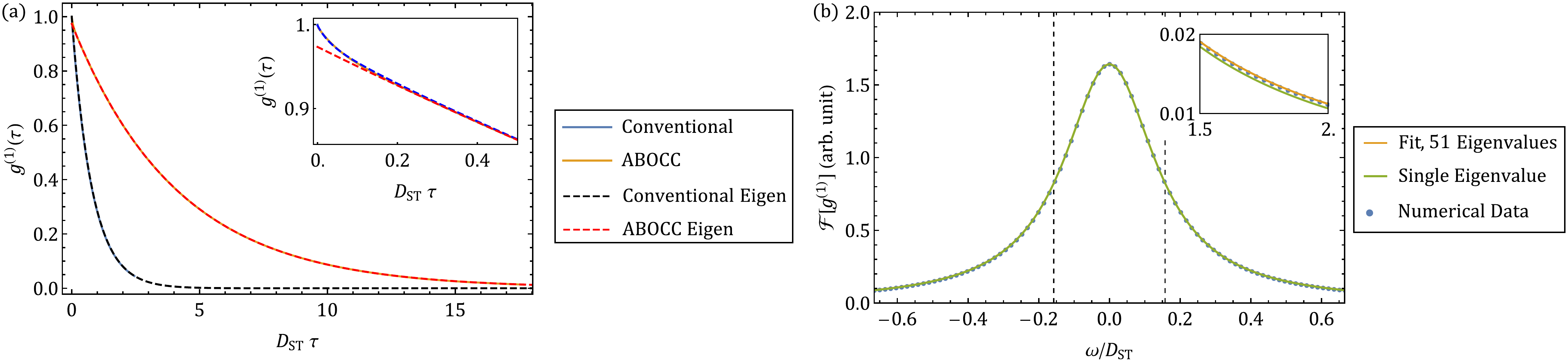}
    \caption{Comparison between the exact numerical time evolution versus the eigenspectrum method. In (a), we show the comparison between the numerical calculated two-time correlation function $g^{(1)}(\tau)$ versus the first eigenvalue of the laser systems. The numerical calculation is shown as solid lines (blue for conventional laser and orange for ABOCC laser), while the dashed lines are the exponential decay with the decay rate calculated by the eigen-spectrum of the super-operator. In the inset of (a), we zoom in on the ABOCC laser $g^{(1)}$ for short time period. The blue dashed line is the fitted decay of $g^{(1)}$ using the first 51 nonzero eigenvalues of the ABOCC laser system. In (b), we show the Fourier transform of the $g^{(1)}$ function. The fitted $g^{(1)}$ decay (orange line) matches with the numerical time evolution result (blue dots) well. The result using only the first nonzero eigenvalue is shown as the green line. Parameters: for conventional laser, $\Gamma_p = 30$, $\Gamma_c = 0.1$ and $g \sim 1.013$; for ABOCC laser: $m_0 = 43$ ($Z_c = 150~\Omega$), $\Gamma_1 = 1.0$, $\Gamma_p = 3.58$ and $g = 0.8747$.}
    \label{fig:compare_lw}
\end{figure}

The phase noise of the laser cavity field causes the phase of the laser light to fluctuate, which gives a finite linewidth to the laser. Before proceeding with a detailed analysis, we begin by summarizing the key points. The master equation Eq.~\eqref{eq:general_master_eq} can be thought of as an eigenvalue problem Eq.~\eqref{eq:eigenvalueProblem}. The spectrum of eigenvalues has one zero eigenvalue $\lambda_0=0$, which corresponds to the steady state solution of the laser, and a number of negative eigenvalues $\lambda_{i\neq0}<0$ which correspond to the decaying modes. As we show in this note, the spectral representation of the two-time correlation function $G(t+\tau,t)$ that describes the decay of coherence consists of a linear combination of decaying exponentials with the decay time set by these negative eigenvalues, see Eq.~\eqref{eq:Gspectral}. To obtain the laser line shape we go to Fourier space. In the Fourier representation, $G(\omega)$ consists of a linear combination of Lorentzians with width set by the negative eigenvalues. Moreover, almost all of the weight ($\sim 97.32\,\%$ for the ABOCC laser example in this note) is carried by the Lorentzian with the largest nonzero eigenvalue (that is narrowest of the Lorentzians). It is precisely this Lorentzian that forms the central peak of the laser line shape, while the remaining Lorentzians contribute to slightly broadening the ``pedestal'' at the base of the central peak, see Fig.~\ref{fig:compare_lw}.

For a conventional laser system phase noise results in the two-time correlation function decaying as~\cite{Scully1997}
\begin{equation}
    G(t+\tau,t) = \mean{\aD(t+\tau) a(t)} \sim \mean{n} \exp (-i \omega_c \vert \tau\vert - D \vert\tau\vert).
\end{equation}
The power spectrum of the laser is given by the Fourier transform of the two-time correlation function $G(t+\tau,t)$,
\begin{equation}
    S(\omega) = \frac{1}{\pi} \text{Re} \int \mean{\aD(t+\tau) a(t)} e^{i \omega \tau} d\tau = \frac{\mean{n}}{\pi} \frac{D}{(\omega - \omega_c)^2 + D^2},
\end{equation}
which is a Lorentzian with full-width at half minimum $2D$, where $D$ is the linewidth of the laser field.

For a system that couples to a Markovian bath, which can be described by the master equation
\begin{equation}
    \partial_t \rho(t) = \hat{\mathcal{L}} \rho(t)
    \label{eq:general_master_eq}
\end{equation}
where $\hat{\mathcal{L}}$ is the super-operator acting on the system density operator ($\rho$), the time-evolution of the density operator can be formally solved by
\begin{equation}
    \rho(t) =  V(t,t_0) \rho(t) = e^{\hat{\mathcal{L}} (t - t_0)} \rho(t_0)
    \label{eq:time_evolve_SO}
\end{equation}
where $V(t,t_0)$ is a time-evolution super-operator that acts on the system degrees of freedom and $\rho(t_0)$ is the initial state of the laser system~\cite{GardinerQNbook}. Also notice that the two-time correlator can be obtained via
\begin{equation}
    G(t+\tau,t) = \Tr_{S+B} \left[ U^{\dagger}(t+\tau) \aD U(t+\tau) U^{\dagger} (t) a U(t) R_0 \right] = \Tr_{S}\left\lbrace \aD  \Tr_{B} \left[U(\tau) a R(t) U^{\dagger}(\tau) \right] \right\rbrace,
\end{equation}
where $U$ is the time-evolution operator for the system and the bath, $R$ is the density operator for both the system and the bath and $\Tr_{S}$, $\Tr_{B}$, $\Tr_{S+B}$ are trace over system, bath, system and bath degrees of freedom, respectively. The term $\Tr_{B} \left[ U(\tau) a R(t) U^{\dagger}(\tau) \right]$ can be thought of as a time-evolution of the ``state'' $a R(t)$ by a time period $\tau$, which is $\lbrace V(\tau,0) [a \rho(t)] \rbrace $ according to Eq.~\eqref{eq:time_evolve_SO}. For the laser system, we are only interested in the laser linewidth when the system is stable, i.e., $\Tr_{S} \left( \aD(\tau) a \rho_s \right)$, where $\rho_{s}$ is the steady state of the system, then the two-time correlation function can be written as
\begin{equation}
    G(\tau) = \Tr_{S} \left[ \aD e^{\cL \tau} \left( a \rho_{s} \right)\right]
    \label{eq:steady_two_time_correlator}
\end{equation}

Similar to the time-evolution of a closed quantum system, where we study the eigenstates of the Hamiltonian operator to understand the system dynamics, we can also find the eigen-spectrum of super-operator $\cL$ (so called damping basis) to study the time-evolution of the open quantum systems whose dynamics is described by the master equation Eq.~\eqref{eq:general_master_eq}~\cite{Briegel1993, Ginzel1993, Briegel1994}. The right-eigenstates of the super-operator $\cL$ can be defined as
\begin{equation}
    \cL \hat{u}^{(i)} = \lambda^{(i)} \hat{u}^{(i)}
    \label{eq:eigenvalueProblem}
\end{equation}
where $\lambda_i$ is the corresponding eigenvalue. Notice that for any physical systems, the steady states are invariant under the time evolution, which means they are the nullspace of the super-operators. These states should be valid quantum states, i.e., they should have unit trace. Further, because the master equation should preserve the trace of the density operator, all the eigenstates with nonzero eigenvalues should have zero trace. 

Suppose $a\rho_s$ in Eq.~\eqref{eq:steady_two_time_correlator} can be expanded using the the right eigenstates as
\begin{equation}
    a \rho_s = c_0 \hat{u}^{(0)} + \sum_{i} c_i \hat{u}^{(i)}
\end{equation}
where $c$'s are the corresponding expansion coefficients and $c_0 \hat{u}_0$ corresponding to the steady state solution, the two-time correlation function can be calculated as,
\begin{align}
    G(\tau) & = c_0 \Tr_S \left[ \aD \hat{u}^{(0)}\right] + \sum_i c_i e^{\lambda^{(i)} \tau} \Tr_S \left[ \aD \hat{u}^{(i)} \right] \\
    & = \sum_i c_i e^{\lambda^{(i)} \tau} \Tr_S \left[ \aD \hat{u}^{(i)} \right] \label{eq:Gspectral}
\end{align}
where we take the fact that the density operator for steady state of a typical laser system is purely diagonal in the Fock basis, which kills the first term. Within all the right eigenstates of the super-operator $\cL$ which have significant contribution to the decay of the two-time correlation function ($c_i \Tr_S [\aD \hat{u}^{(i)}]$ term is large), the largest eigenvalue ($\cL$ is non-positive) contribute to the long-time performance, i.e., it controls the linewidth of the laser system.

To decompose the operator using the eigenstates of the super-operator $\cL$, notice that the super-operator is, in general, not Hermitian, the left eigenstates are needed to define the orthogonality relation. Following Ref.~\cite{Briegel1993}, the dual operator of the super-operator is defined as
\begin{equation}
    \Tr[A \cL(B)] = \Tr[\check{L}(A) B]
\end{equation}
for any operator $A$ and state $B$. Then the left-eigenstates of the super-operator $\cL$ is
\begin{equation}
    \check{v}^{(i)} \cL = \check{L} \check{v}^{(i)} = \lambda^{(i)} \check{v}^{(i)}
\end{equation}
with orthonormal relation $\Tr(\check{v}^{(i)} \hat{u}^{(j)}) = \delta_{i,j}$.

To numerically solve the eigenstates of the super-operator $\cL$, we notice in typical laser systems, the super-operator contains terms in the form of
\begin{equation}
    \cL \rho = \sum_i \cL^{(i)}\rho = \sum_i \hat{P}^{(i)} \rho \hat{Q}^{(i)}
\end{equation}
where $\hat{P}^{(i)}$ and $\hat{Q}^{(i)}$ operate on the system Hilbert space. After expanding the operators $\hat{P}^{(i)}$ and $\hat{Q}^{(i)}$ using a suitable basis of the system Hilbert space and acting on the system density operator, in the matrix form we get
\begin{equation}
     \rho^{(i)}_{j,k} = \left( \cL^{(i)}\rho \right)_{j,k} = \sum_{l,m} \hat{P}^{(i)}_{j,l} \rho_{l,m} \hat{Q}^{(i)}_{m,k} ,
\end{equation}
where $\rho^{(i)} \equiv \hat{P}^{(i)} \rho \hat{Q}^{(i)}$. We can redefine the density operator as a vector, and hence the super-operator becomes a matrix, whose eigenvectors can be easily solved. To explicitly show the matrix representation of the master equation, we re-group the indices as
\begin{equation}
    \rho^{(i)}_{j,k} = \left( \cL^{(i)}\rho \right)_{j,k} \rightarrow \rho^{(i)}_{jk} = \sum_{lm} \cL^{(i)}_{jk;lm} \rho_{lm} = \sum_{lm} \hat{P}^{(i)}_{j,l}\hat{Q}^{(i)}_{m,k} \rho_{lm} 
\end{equation}
and the super-operator becomes 
\begin{equation}
    \cL \rightarrow \cL_{jk;lm} = \sum_{i} \hat{P}^{(i)}_{j,l}\hat{Q}^{(i)}_{m,k}.
\end{equation}
The left and right eigenvectors of the matrix representation of the super-operator $\cL$ can be re-mapped back to the original index convention to get back the matrix representation in the system Hilbert space. The vector product of left and right eigenvectors are equivalent to the product definition of the operators, i.e., the trace of the two operator product.

When we consider the laser system, especially we consider the cavity field, the Hilbert space is infinite dimensional. We can truncate the Hilbert space of the cavity field, but, in practice, the dimension is still very large and hence it is not practical to find all the eigenvectors of the super-operator. However, we notice that the right eigenvector with the lowest nonzero $\vert \lambda^{(i)} \vert$ is very close to $a \rho_s$. The other eigenstates, with larger $\vert \lambda^{(i)} \vert$, decay faster 
than this eigenstate and do not contribute significantly to the linewidth of the laser.

In Fig.~\ref{fig:compare_lw}a, we compare the numerical calculated normalized two-time correlation function $g^{(\tau)} = G(\tau)/\mean{n}$ using direct time-evolution of the master equations versus the function $d_1 e^{\lambda^{(1)} \tau}$ where $d_1$ is a coefficient that we fit and $\lambda^{(1)}$ is the largest nonzero eigenvalue of the super-operator $\cL$. In both the conventional laser system and ABOCC laser system, the decay of the normalized two-time correlation function from the numerical time-evolution method (solid lines) match well with the super-operator eigenvalues (dashed lines). For the ABOCC laser system, we also notice that at the small time-scale, the decay of the two-time correlation function slightly differs from the single eigenvalue fit (see the inset of Fig.~\ref{fig:compare_lw}a).  This is caused by the finite overlapping to the other eigenstates which a responsible for the rapid initial decay. We fit the numerical time-evolved $g^{(1)}(\tau)$ function using first 20 eigenvalues (decay rates) of the super-operator of the ABOCC laser (blue dashed line), which gives a good fit to the  direct time-evolution curve. From the fitting, we extract the overlap constant for the first nonzero eigenvalue of the super-operator is $0.9732$, which means the main contribution of the long-time decay is given by the eigenstates with the largest nonzero eigenvalue. In Fig.~\ref{fig:compare_lw}b, we further examine the contribution to the linewidth in frequency domain. We Fourier transform the $g^{(1)}(\tau)$ data from both time-evolved method (blue dots) and the eigen-spectrum method (orange and green curves). We notice that the Fourier transform of fitted $g^{(1)}(\tau)$ using first $51$ eigenvalues of the super-operator (the orange line) gives good agreement with the time-evolve method (blue dots). If we just extract the lineshape from the first eigenvalue (Fig.~\ref{fig:compare_lw}b green curve), the lineshape is slightly different from the time-evolution result (see inset). The difference is due to the finite overlap with the eigenstates that have a larger decay constant. However, the faster dacaying states only contribute to the short-time dynamics (see Fig.~\ref{fig:compare_lw}a inset), which gives a slightly larger background on the Lorentzian lineshape and slightly increase the linewidth. However, the central peak of the spectrum is dominated by the first nonzero eigenvalue. In Fig.~\ref{fig:compare_lw}b, we observe that the Lorentzian based on the first nonzero eigenvalue (green curve) and the fitted time-evolution using the first $51$ eigenvalues, and the numerically calculated lineshape all match well.

Compared to direct time-evolution method of calculating the laser dynamics, especially the steady state photon distribution and the laser linewidth or two-time correlation function, the eigen-spectrum method is significantly more computationally efficient. To further improve computational efficiency, we truncate the density matrices to include only the main diagonal and the first few minor diagonals. We verify that this truncation does not affect the part of the eigen-spectrum that we are interested in by increasing the number of minor diagonals until convergence to the exact solution is reached for all values of $\langle n\rangle \leq 1000$ for the conventional and SGBO laser and $m_0\leq 1000$ for the ABOCC laser.

\section{Laser cavity loss with approximately bare operator coupling circuit (ABOCC)}
\label{sec:ABOCCloss}

\begin{figure}[h]
    \centering
    \includegraphics[width = 0.6 \textwidth]{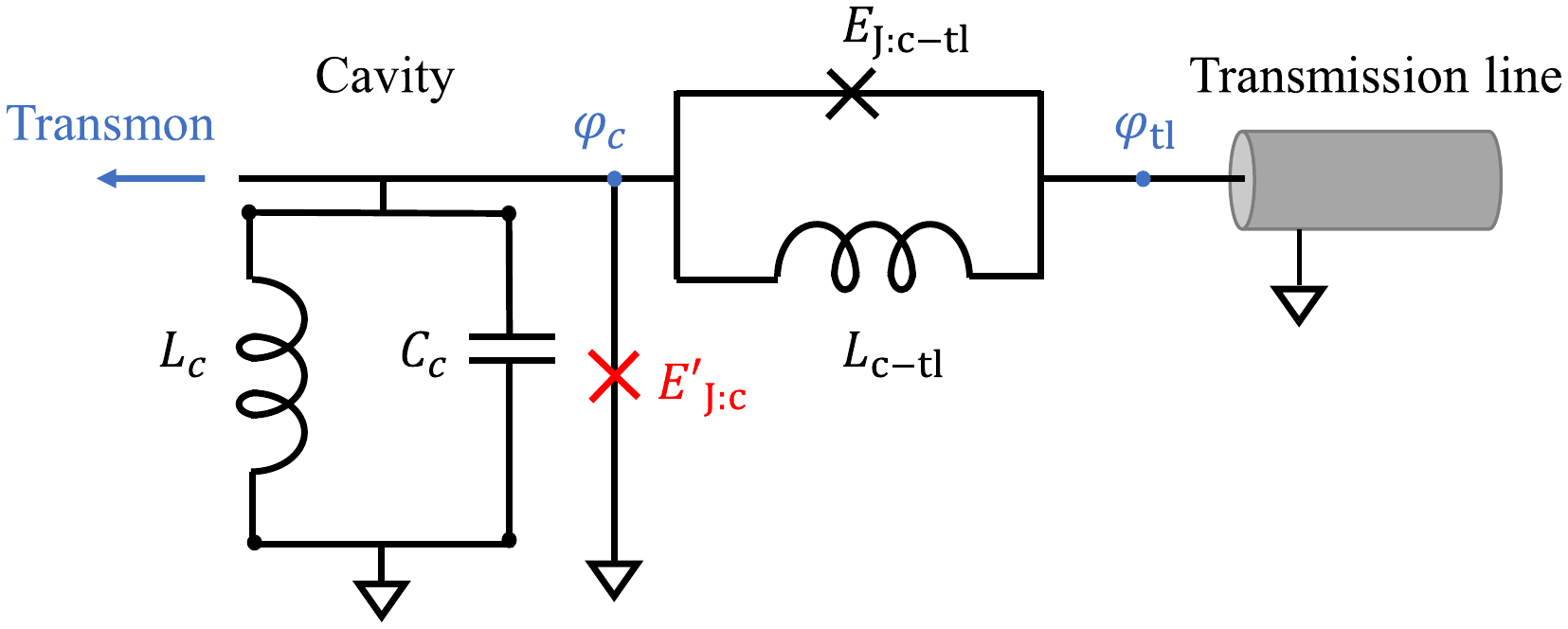}
    \caption{The circuit diagram of the ABOCC coupling circuit between the cavity adn the transmission line.}
    \label{fig:ABOCC_circuit}
\end{figure}

In our main text, the ABOCC coupling circuit between the cavity and the transmission line is shown in Fig.~1. The Hamiltonian for the ABOCC coupling is
\begin{equation}
    H = - E_{\textrm{J:c-tl}} \cos \left( \hat{\varphi}_\tc - \hat{\varphi}_\tl \right) + \frac{\phi_0^2}{2 L_\textrm{c-tl}} \left( \hat{\varphi}_\tc - \hat{\varphi}_\tl \right)^2 + E_{\textrm{J:c}}' \cos (\hat{\varphi}_\tc) 
    ,
    \label{eq:classical_couple_H_full}
\end{equation}
where $E_{\textrm{J:c-tl}}$ is the coupling Josephson junction energy, $L_{\textrm{c-tl}}$ is the coupling linear inductance, $E_{\textrm{J:c}}'$ is the $\pi$-junction Josephson energy,
and the node phases operators are  $\hat{\varphi}_\tc$ and $\hat{\varphi}_\tl$ as labeled in Fig.~\ref{fig:ABOCC_circuit}. The $\pi$-junctions are used to correct the dispersion given by the nonlinear coupling between the cavity and the transmission line. Here, we will ignore these two $\pi$ junctions at the beginning of the discussion. Further, we define a dimensionless parameter $r = \frac{\phi_0^2}{2 L_{\textrm{c-tl}}}/E_{\textrm{J:c-tl}}$. 
With the second quantization of the transmission line field and the LC resonator mode (see Appendix~\ref{sec:photonLossLinear}), the phase across the ABOCC circuit is
\begin{equation}
    \hat{\delta}_{\ctl} \equiv \hat{\varphi}_\tc - \hat{\varphi}_\tl = \tilde{\varphi}_\tc \left( \hc + \hcD \right) + i \sum_{k} \tilde{\varphi}_\tl (k) \left( \hb{k} - \hbD{k} \right).
\end{equation}
The coupling Hamiltonian (without the $\pi$ junctions) becomes
\begin{align}
    H/E_{\textrm{J:c-tl}} = - \cos \left( \hat{\delta}_{\ctl} \right) + r \hat{\delta}_{\ctl}^2  = \frac{1}{2} \left( e^{i \hat{\varphi}_\tc} e^{- i \hat{\varphi}_\tl} + e^{-i \hat{\varphi}_\tc}e^{ i \hat{\varphi}_\tl } \right) + \hat{\delta}_{\ctl}^2,
    \label{eq:realistic_coupling_h}
\end{align}
where we have used the fact that the resonator operators and the transmission line field operators commute. We use the Baker-Campbell-Hausdorff formula to transform the exponential of the operators
\begin{align}
     e^{\pm i \hat{\varphi}_\tc} = e^{-\tilde{\varphi}_\tc^2/2} e^{\pm i\tilde{\varphi}_\tc \hcD } e^{\pm i \tilde{\varphi}_\tc \hc}, 
     \qquad e^{\pm \hat{\varphi}_\tl (k)} = e^{-\tilde{\varphi}_\tl^2 (k)/2} e^{\mp \tilde{\varphi}_\tl \hbD{k}} e^{\pm\tilde{\varphi}_\tl \hb{k}}.
\label{eq:expansion_cos_real}
\end{align}
We will use this result in order to reach a normal ordering in which the LC resonator operators are to the right of the transmission line operators.

The expansion of Eq.~\eqref{eq:realistic_coupling_h} to third order in the transmission line field operators yields
\begin{equation}
    H/E_{\textrm{J:c-tl}} = h_{\tc} + h_{\tl} + h_{1} + h_{2} + h_{3} + ... 
    \label{eq:expansion_of_H_orders}
\end{equation}
where $h_{\textrm{c}}$ and $h_{\textrm{tl}}$ are the dimensionless Hamiltonian acting solely on the cavity field and the transmission line, respectively. The cavity-transmission line coupling is expanded in orders of the transmission line field operators and the first, second  and third order terms are labeled as $h_{1}$, $h_{2}$ and $h_{3}$. 
\begin{subequations}
\label{eq:self_coupling}
\begin{align}
h_{\tc} & = - \left( \prod_{k} e^{-\tilde{\varphi}_\tl^2 (k)/2}\right) \frac{1}{2} \left( e^{i \hat{\varphi}_\tc} + e^{-i \hat{\varphi}_\tc}\right) + r\hat{\varphi}_\tc^2 \equiv - \mathcal{C}_{\tl} \cos \left( \hat{\varphi}_\tc \right) + r\hat{\varphi}_\tc^2, \\
h_{\tl} & = -e^{-\tilde{\varphi}_\tc^2/2} \frac{1}{2}\left( e^{i \hat{\varphi}_\tl} + e^{-i \hat{\varphi}_\tl}\right) + r\hat{\varphi}_\tl^2 \equiv -\mathcal{C}_{\tc} \cos \left(\hat{\varphi}_\tl\right) + r\hat{\varphi}_\tl^2,
\end{align}
\end{subequations}
where, $\mathcal{C}_{\textrm{tl}}$ and $\mathcal{C}_c$ are two constants. Notice that $h_{\textrm{sys}}$ and $h_{\textrm{tl}}$, induced by the coupling circuit, contributes new nonlinearities to the cavity and the transmission line. These nonlinearities, especially the nonlinearity of the cavity field, will degrade the laser performance by shifting the cavity frequency as the number of photons in the cavity increases. To compensate for these dispersive effects on the cavity, we include a $\pi$ junction as shown in Fig.~\ref{fig:ABOCC_circuit}, in which the Josephson energies satisfy
\begin{equation}
E_{\textrm{J:c}}' = \mathcal{C}_{\tl} E_{\textrm{J:c-tl}}, \quad E_{\textrm{J:tl}}' = \mathcal{C}_{\textrm{c}} E_{\textrm{J:c-tl}}.
\end{equation}
This $\pi$ junctions cancel out the non-linear contributions of $h_{\textrm{tl}}$ thus reducing dephasing. In the following discussion we focus on the remaining terms, which are the cavity-transmission line coupling terms $h_{1}$, $h_{2}$ and $h_{3}$.

Expansion of Eq.~\eqref{eq:expansion_cos_real} yields
\begin{subequations}
\begin{align}
\label{eq:first_order_in_b}
    h_1 & = -i \mathcal{C}_{\tl} 
        \sum_{k} \tilde{\varphi}_{\tl}(k)  (b_{k}^{\dagger}-b_{k}) \sin \left(\hat{\varphi}_c \right)
    + 2 i r \hat{\varphi}_c \sum_{k} \tilde{\varphi}_{\tl}(k) \left( b_k - b_k^{\dagger} \right), \\
\label{eq:second_order_in_b}
    h_2 & = -\mathcal{C}_{\textrm{tl}} 
        \left\lbrace 
            \sum_{k} \frac{\tilde{\varphi}_{\tl}^2 (k)}{2} \left(  b_{k}^2 + \left(b_{k}^{\dagger}\right)^2 - b_{k}^{\dagger} b_{k}\right) 
         + {\sum_{k,q}}^\prime \tilde{\varphi}_{\tl}(k) \tilde{\varphi}_\tl (q) \left( b_{k} b_{q} + b_{k}^{\dagger} b_{q}^{\dagger} - 2 b_{k}^{\dagger} b_{q}\right)
    \right\rbrace
    \left[\cos \left( \hat{\varphi}_c\right) -1 \right], \\
\label{eq:third_order_in_b}
    h_3 & = - i \mathcal{C}_{\textrm{tl}} \left\lbrace \sum_{k} \frac{\tilde{\varphi}_{\tl}^3 (k)}{6}\left[\left(b_k^{\dagger}\right)^3 -3 \left(b_{k}^{\dagger}\right)^2 b_{k} + 3 b_{k}^{\dagger} b_{k}^{2} - b_{k}^3\right]\right.  + {\sum_{k,q}}^\prime \frac{\tilde{\varphi}_\tl^2 (k) \tilde{\varphi}_\tl (q)}{2} \left[\left( b_{k}^{\dagger}\right)^2 + b_{k}^{2} - 2 b_{k}^{\dagger} b_{k}\right]\left( b_{q}^{\dagger} - b_{q}\right) \\ \nonumber
    & \qquad + \left.
    {\sum_{k,q,p}}^\prime \tilde{\varphi}_\tl (k) \tilde{\varphi}_\tl (q) \tilde{\varphi}_\tl (p) \left( b_{k}^{\dagger} - b_{k}\right) \left( b_{q}^{\dagger} - b_{q}\right) \left( b_{p}^{\dagger} - b_{p}\right)
    \right\rbrace  \sin \left(\hat{\varphi}_c \right),
\end{align}
\end{subequations}
where the summation $\sum_{k}$ is from $-\infty$ to $+\infty$, the summations with prime ${\sum^\prime}_{k,q}$ and ${\sum^\prime}_{k,q,p}$ omit the terms in which any of the summation indices are equal.

We further assume that the coupling strength between the LC resonator and the transmission line, which is controlled by $E_{\textrm{J:c-tl}}$, is small compared to the LC resonator mode frequency and the transmission line dynamics, which allows us to apply the rotating-wave approximation (RWA). Applying the rotating wave approximation to $h_1$ of Eq.~\eqref{eq:first_order_in_b}, and restoring dimensions of energy, we obtain
\begin{equation}
\begin{aligned}
    H_1 = - i & E_{\textrm{J:t-cl}} \mathcal{C}_{\textrm{tl}} e^{- \tilde{\varphi}_c^2/2}  \sum_{k} \sum_{n = 0}^{\infty} (-1)^n  \frac{\tilde{\varphi}_{\tl}(k) \tilde{\varphi}_c^{2n+1}}{n! \cdot (n+1)!}
    \left[ 
        b_k^{\dagger}\left( c^{\dagger} \right)^n c^{n+1}
    - b_k \left(c^{\dagger}\right)^{n+1} c^n 
    \right] \\
    & + 2 i r E_{\textrm{J:c-tl}} \tilde{\varphi}_c \sum_{k} \tilde{\varphi}_{\tl}(k) \left( c^{\dagger} b_k - b_k^{\dagger} c \right).
\end{aligned}
\label{eq:first_order_RWA}
\end{equation}
Defining the nonlinear operator for the cavity field
\begin{equation}
    \hat{B}_1 =  \mathcal{C}_{\tl} e^{- \tilde{\varphi}_c^2/2}  \sum_{n = 0}^{\infty} (-1)^n  \frac{\tilde{\varphi}_c^{2n+1}}{n! \cdot (n+1)!} \left( c^{\dagger} \right)^n c^{n+1}
    + 2 r \tilde{\varphi}_c c,
    \label{eq:nonlinear_A1}
\end{equation}
and applying the Born-Markov approximation to trace over the transmission line degrees of freedom, the first order coupling contributes to a resonator dissipator term
\begin{equation}
    \Gamma_1 \mathcal{D}[\hat{B}_1] \rho = -\frac{\Gamma_1}{2} \left( \hat{B}_{1}^{\dagger} \hat{B}_{1} \rho + \rho \hat{B}_{1}^{\dagger} \hat{B}_{1} - 2 \hat{B}_{1} \rho \hat{B}_{1}^{\dagger} \right),
\end{equation}
where the decay rate is
\begin{equation}
\Gamma_1 = \frac{E^2_{\textrm{J}}}{\hbar^2}\left( \frac{\hbar Z_\tl}{2 \phi_0^2}\right)\frac{1}{\omega_c}.
\label{eq:first_order_rate}
\end{equation}

Before we move to the higher order nonlinear terms in the ABOCC Hamiltonian, we calculate the constant parameter $\mathcal{C}_{\textrm{TL}}$, which is given by Eq.~\eqref{eq:self_coupling}. In the quantum optics regime, we assume that the transmission line has a large length, where we can assume $l \rightarrow \infty$. In this limit, we can approximate the summation of $k$ by the integral of $k$ as
\begin{equation}
    \sum_{k} \rightarrow \frac{1}{2 \pi} \int dk. 
    \label{eq:sum_k_integral}
\end{equation}
Further, in quantum optics regime, especially for the system that Born-Markov approximation applies, the cavity frequency is the dominant frequency to the coupling bandwidth $\theta$ and the system-bath coupling strength. Here we explicitly assume that the cutoff frequencies for the system-bath coupling are $\omega_L = \omega_c - \theta/2$ and $\omega_H = \omega_c + \theta/2$, while the corresponding cutoff wave-vectors are $k_L$ and $k_H$~\footnote{Here there will be another factor of 2, which is because of the coupling to the left-propagating modes and the right-propagating modes.}. The constant $\mathcal{C}_{\textrm{TL}}$ can be calculated as
\begin{equation}
\begin{aligned}
    \mathcal{C}_{\textrm{tl}} & = \prod_k e^{- \frac{\tilde{\varphi}_\tl (k)}{2}} = \exp \left[ - \frac{1}{2} \sum_{k=k_L}^{k_H} \left( \frac{1}{\phi_0^2} \frac{\hbar Z_\tl}{2}\right) \frac{1}{k l}\right] \\
    & \rightarrow \exp \left[ - 2 \frac{1}{4 \pi} \left( \frac{1}{\phi_0^2} \frac{\hbar Z_\tl}{2}\right) \int_{k_L}^{k_H} \frac{1}{k} d k\right] \\
    & = \exp\left( - \frac{1}{\phi_0^2} \frac{\hbar Z_\tl}{4 \pi} \right) \frac{\omega_H}{\omega_L},
\end{aligned}
\label{eq:ctl_mid_1}
\end{equation}
where $\omega_H$ and $\omega_L$ are the high and low frequencies of the bandwidth.
Next, we adopt the assumption that $\theta/\omega_c \ll 1$, so the ratio $\omega_H/\omega_L$ can be expanded in the order of $\theta/\omega_c$, and $\mathcal{C}_{\textrm{tl}}$,
\begin{equation}
    \mathcal{C}_{\textrm{tl}} = \exp\left( - \frac{1}{\phi_0^2} \frac{\hbar Z_\tl}{4 \pi} \right) \left[ 1 + \frac{\theta}{\omega_c} + \frac{\theta^2}{2 \omega_c^2} + o\left(\frac{\theta^3}{\omega_c^3}\right)\right].
    \label{eq:ctl_const}
\end{equation}
If we choose the characteristic impedance of the transmission line as $Z_\tl = 50\ \Omega$. the lowest order approximation of the parameter $\mathcal{C}_{\textrm{tl}}$ is $0.9961$. 

The second order term in the expansion of the coupling Hamiltonian is given by Eq.~\eqref{eq:second_order_in_b}. Under the rotating wave approximation (and restoring dimensions), the second order term is
\begin{equation}
\begin{aligned}
    & H_2/(-E_{\textrm{J:c-tl}} \mathcal{C}_{\textrm{tl}} \mathcal{C}_c)  =
     -  \left( \sum_{k} \frac{\tilde{\varphi}_\tl^2(k)}{2} 
        \left( b_{k}^{\dagger}\right)^2 + {\sum_{k,q}}' \tilde{\varphi}_\tl(k) \tilde{\varphi}_\tl (q) b_{k}^{\dagger} b_{q}^{\dagger} \right) \sum_n \frac{(-1)^{n} \tilde{\varphi}_c^{2n+2}}{n! \cdot (n+2)!}\left(c^{\dagger}\right)^n c^{n+2} \\
    & \qquad \qquad \qquad - \left( \sum_{k} \frac{\tilde{\varphi}_\tl^2 (k)}{2}
         b_{k}^2 + {\sum_{k,q}}' \tilde{\varphi}_\tl(k) \tilde{\varphi}_\tl (q) b_{k} b_{q} \right) \sum_n \frac{(-1)^{n} \tilde{\varphi}_c^{2n+2}}{n! \cdot (n+2)!}\left(c^{\dagger}\right)^{n+2} c^{n} 
         \\
    & \qquad \qquad \qquad 
    - \left( \sum_{k} \tilde{\varphi}_\tl^2(k)
         b_{k}^{\dagger} b_{k} + {\sum_{k,q}}' 2 \tilde{\varphi}_\tl(k) \tilde{\varphi}_\tl (q) b_{k}^{\dagger} b_{q} \right) \sum_n \frac{(-1)^n \tilde{\varphi}_c^{2n}}{(n!)^2 }\left( c^{\dagger} c \right)^{n}.
\end{aligned}
\label{eq:second_order_RWA}%
\end{equation}
where the summation $\sum'$ does not contain the terms that have the same indices. 

Suppose the density operator of the cavity is given by $\rho(t)$ and the transmission line is assumed to be a vacuum bath. the master equation given by the nonlinear coupling Hamiltonian $H_2$ in Eq.~\eqref{eq:second_order_RWA} is
\begin{equation}
    \partial_t \rho(t) = -\frac{i}{\hbar} \textrm{Tr}_B [H_2(t), R(t_0)] - \frac{1}{\hbar^2} \textrm{Tr}_B \int_{t_0}^{t} \left[ H_2(t), [H_2 (\tau), R(\tau)]\right] d\tau
    \label{eq:master_second_order_0}
\end{equation}
where $R$ is the density operator for the system (cavity) and the bath (transmission line), which can be approximated by $R(t) \sim \rho(t) \otimes \rho_B$ and $\rho_B = \vert \textrm{vac} \rangle \langle \textrm{vac} \vert$ where $\vert \textrm{vac} \rangle$ is the vacuum state of the bath (transmission line). The partial trace of the bath DOFs is
\begin{equation}
    \textrm{Tr}_B \lbrace \cdot \cdot \rbrace = \sum_{\left\lbrace n_{k_1}, n_{k_2}, ...\right\rbrace} \langle n_{k_1}, n_{k_2}, ... \vert \cdot \cdot \vert n_{k_1}, n_{k_2}, ... \rangle 
\end{equation}
where $\vert n_{k_1}, n_{k_2}, ... \rangle = \vert n_{k_1} \rangle \vert n_{k_2} \rangle ...$, $\vert n_{k_i}\rangle$ is the $n_{k_i}$-photon Fock state of the mode $k = k_i$.

Note that because in $H_2$, all the terms are aligned in the normal order, the first term in Eq.~\eqref{eq:master_second_order_0} is zero. We will focus on the second term of Eq.~\eqref{eq:master_second_order_0}. After expansion of the commutation relation, the Eq.~\eqref{eq:master_second_order_0} is
\begin{equation}
\begin{aligned}
\dot{\rho} = -\frac{1}{\hbar^2} \textrm{Tr}_B \int_{t_0}^{t} d\tau & \left\lbrace H_2 (t) H_2 (\tau) \rho(\tau) \otimes \rho_B + \rho(\tau) \otimes \rho_B H_2(\tau) H_2(t)  
- H_2 (t) \rho(\tau) \otimes \rho_B H_2(\tau) \right. 
 \\
 &
\left.
- H_2 (\tau) \rho(\tau) \otimes \rho_B H_2 (t) 
\right\rbrace
\end{aligned}
\label{eq:master_second_order_1}
\end{equation}
Note that the Hamiltonian $H_2$ should be considered as the interaction picture Hamiltonian, where the transformation is $U = \exp (i H_0)$, where $H_0$ is
\begin{equation}
    H_0 = \hbar \omega_c c^{\dagger}c + \sum_{k} \hbar \omega_k \ b_k^{\dagger} b_k
\end{equation}
Next, we will work term by term in Eq.~\eqref{eq:master_second_order_1} to get master equation for the cavity field.

We start from the term 
\begin{equation}
    T_3 \equiv \frac{1}{\hbar^2} \int_{t_0}^{t} d\tau \textrm{Tr}_B \left\lbrace 
    H_2 (t) \rho(\tau) \otimes \rho_B H_2(\tau)
    \right\rbrace.
    \label{eq:second_order_t3}
\end{equation}
As the transmission line is assumed to be a vacuum bath, and the coupling Hamiltonian $H_2$ is in normal order, the partial trace will kill all the terms that contain lowering operators for the bath DOFs. The Hamiltonian terms that survive in $T_3$ partial trace are
\begin{subequations}
\begin{align}
\frac{H_{2,1,(\textrm{I})}(t) }{ (E_{\textrm{J:c-tl}} \mathcal{C}_{\textrm{tl}} \mathcal{C}_c )} & \equiv h_{2,1,(\textrm{I})}(t) = \sum_{k} h_{2,1,kk,(\textrm{I})} + {\sum_{k,q}}' h_{2,1,kq,(\textrm{I})} \\
h_{2,1,kk,(\textrm{I})}(t) & = \frac{\tilde{\varphi}_\tl^2(k)}{2} \left( b_{k}^{\dagger}\right)^2 \hat{B}_{2} e^{i 2 (\omega_k-\omega_c) t} \\
h_{2,1,kq,(\textrm{I})}(t) & =  \tilde{\varphi}_\tl (k) \tilde{\varphi}_\tl (q) b_{k}^{\dagger} b_{q}^{\dagger} \hat{B}_2  e^{i (\omega_k+\omega_q -2 \omega_c) t},
\end{align}
\end{subequations}
where the cavity nonlinear operator $\hat{B}_2$ is defined as
\begin{equation}
    \hat{B}_2 = \sum_n \frac{(-1)^{n} \tilde{\varphi}_c^{2n+2}}{n! \cdot (n+2)!}\left(c^{\dagger}\right)^n c^{n+2}
\end{equation}
and the Eq.~\eqref{eq:second_order_t3} is
\begin{equation}
\begin{aligned}
    \frac{T_3 \, \hbar^2}{(E_{\textrm{J:c-tl}} \mathcal{C}_{\textrm{tl}} \mathcal{C}_c )^2} & = \int_{t_0}^{t} d\tau \left\lbrace \sum_{k} \langle 2_k \vert h_{2,1,kk,(\textrm{I})}(t) \vert \textrm{vac} \rangle \rho \langle \textrm{vac} \vert h_{2,1,kk,(\textrm{I})}^{\dagger} (\tau) \vert 2_k \rangle \right. \\
    & \qquad \qquad + \left. 
    {\sum_{k,q}}' \langle 1_k, 1_q \vert h_{2,1,kq,(\textrm{I})}(t) \vert \textrm{vac} \rangle \rho \langle \textrm{vac} \vert h_{2,1,kq,(\textrm{I})}^{\dagger} (\tau) \vert 1_k,1_q \rangle
    \right\rbrace,
\end{aligned}
\label{eq:second_order_t3_2}
\end{equation}
where we apply the bath state orthogonality relations and remove all the zero terms.
Further, the term
\begin{subequations}
\begin{align}
    \langle 2_k \vert h_{2,1,kk,(\textrm{I})}(t) \vert \textrm{vac} \rangle & = \frac{\sqrt{2}\tilde{\varphi}_{\tl}^2(k)}{2} \hat{B}_2 e^{i 2 (\omega_k-\omega_c) t} \\
    \langle 1_k, 1_q \vert h_{2,1,kq,(\textrm{I})}(t) \vert \textrm{vac} \rangle & = \tilde{\varphi}_{\tl}(k) \tilde{\varphi}_{\tl}(q) \hat{B}_2 e^{i (\omega_k+\omega_q -2 \omega_c) t}
\end{align}
\end{subequations}
For the first term, which is involved in the time integral of the first line of Eq.~\eqref{eq:second_order_t3_2} (noted as $T_{3,1}$), after applying the Born-Markov approximation, and define $k_c = \omega_c / v_p$ and take Eq.~\eqref{eq:sum_k_integral}, the term $T_{3,1}$ is
\begin{subequations}
\begin{align}
    T_{3,1} & = (\mathcal{C}_{\textrm{tl}} \mathcal{C}_c )^2 \frac{E_{\textrm{J:c-tl}}^2}{\hbar^2} \frac{l}{2\pi v_p}\int_{t_0}^{t} d\tau \int_{\omega_L}^{\omega_H} d \omega_k \frac{\tilde{\varphi}_\tl^4(k)}{2} \hat{B}_{2}\rho(\tau)\hat{B}_{2}^{\dagger} e^{i 2 (\omega_k-\omega_c) t} \\
    & = (\mathcal{C}_{\textrm{tl}} \mathcal{C}_c )^2 \frac{1}{8} \left( \frac{\hbar Z_\tl}{2 \phi_0^2}\right) \frac{E_{\textrm{J:c-tl}}^2}{\hbar^2 \omega_c} \tilde{\varphi}_\tl^2 (k_c) \hat{B}_2 \rho(t) \hat{B}_{2}^{\dagger} \\
    & = \Gamma_{2,1} \hat{B}_2 \rho(t) \hat{B}_{2}^{\dagger}.
\end{align}
\label{eq:second_order_t31}%
\end{subequations}
Similar to the definition of the nonlienar operator of the cavity field in first order coupling, we can redefine the nonlinear operator for second order as
\begin{equation}
    \hat{B}_2^\prime = (\mathcal{C}_{\textrm{tl}} \mathcal{C}_c ) \hat{B}_2
    \label{eq:redefine_A2}
\end{equation}
and then the associated rate is
\begin{equation}
    \hat{\Gamma}_{2,1}^\prime = \frac{1}{ (\mathcal{C}_{\textrm{tl}} \mathcal{C}_c )^2} \Gamma_{2,1} = \frac{1}{8} \left( \frac{\hbar Z_\tl}{2 \phi_0^2}\right) \frac{E_{\textrm{J:c-tl}}^2}{\hbar^2 \omega_c} \tilde{\varphi}_\tl^2 (k_c)
\end{equation}
Compared with the rate associated with the first order coupling Hamiltonian, $\Gamma_1$ [see Eq. \eqref{eq:first_order_rate}],  the rate associated with this term is
\begin{equation}
    \Gamma_{2,1}^{\prime} /\Gamma_1 =  \frac{1}{8} \tilde{\varphi}_\tl^2 (k_c).
\end{equation}
For a realistic setup, where we assume the transmission is $1$~m long, the cavity frequency is $7.5$~GHz, and the speed of microwave along the transmission line is speed of light, and the characterestic impedance of the transmission line is $50\ \Omega$, the quantization parameter $\tilde{\varphi}_\tl (k_c) \sim 0.0124$. So this process is much slower than the first order coupling, which is controlled by the small parameter $\tilde{\varphi}_\tl^2 (k_c)$, which is equivalent to the small parameter $\frac{2\pi v_p}{\omega_c l}$ [see Eq.~\eqref{eq:small_para_nc}]. Especially, as the transmission line $l \rightarrow \infty$ (approaching theoretical limit), this term $\rightarrow 0$.

The second line of the Eq.~\eqref{eq:second_order_t3_2} (noted as $T_{3,2}$) is
\begin{equation}
    T_{3,2} = (\mathcal{C}_{\textrm{tl}} \mathcal{C}_c )^2 \frac{E_{\textrm{J:c-tl}}^2}{\hbar^2} \left( {\sum_{k,q}} - \sum_{k,q} \delta_{k,q} \right) \tilde{\varphi}_{\tl}^2(k) \tilde{\varphi}_{\tl}^2(q) \hat{A}_2 e^{i (\omega_k+\omega_q -2 \omega_c)(t-\tau)}
\end{equation}
Note the second summation term is similar to the calculation in Eq.~\eqref{eq:second_order_t31}, and it is $2 \Gamma_{2,1} \hat{B}_2 \rho(t) \hat{B}_{2}^{\dagger}$.

The first summation term
\begin{align}
T_{3,2} & +2 \Gamma_{2,1} \hat{B}_2 \rho(t) \hat{B}_{2}^{\dagger} = 
(\mathcal{C}_{\textrm{tl}} \mathcal{C}_c )^2 \frac{E_{\textrm{J:c-tl}}^2}{\hbar^2} \frac{l^2}{(2\pi v_p)^2} \int_{t_0}^{t} d\tau  \int_{\omega_L}^{\omega_H} d \omega_k \int_{\omega_L}^{\omega_H} d \omega_q  
\left\lbrace \tilde{\varphi}_\tl^2(k)  \tilde{\varphi}_\tl^2(q) \hat{B}_{2}\rho(\tau)\hat{B}_{2}^{\dagger} e^{i (\omega_k+\omega_q-2 \omega_c) t} \right\rbrace 
\label{eq:second_order_t32_1}
\end{align}
With Born-Markov approximation, we replace $\tilde{\varphi}_T{k}$ and $\tilde{\varphi}_T{q}$ by the central frequency mode $k_c = \omega_k / v_p$, and because of the fast oscillation term $e^{i (\omega_k+\omega_q-2 \omega_c) t}$, only the modes that satisfies $\omega_k + \omega_q = 2 \omega_c$ will have large contribution, we can approximate the integral of two modes frequencies by $\theta \int d\bar{\omega}$ where $\bar{\omega} = (\omega_k + \omega_q)/2$, and $\theta$ is the coupling bandwidth. Then the integral in Eq.~\eqref{eq:second_order_t32_1} is
\begin{subequations}
\begin{align}
    T_{3,2} & +2 \Gamma_{2,1} \hat{B}_2 \rho(t) \hat{B}_{2}^{\dagger} = 
(\mathcal{C}_{\textrm{tl}} \mathcal{C}_c )^2 \frac{E_{\textrm{J:c-tl}}^2}{\hbar^2} \frac{l^2 \theta}{(2\pi v_p)^2} \tilde{\varphi}_{\tl}^4 (k_c) \int_{t_0}^{t} d\tau \int d\bar{\omega}
\left\lbrace \hat{B}_{2}\rho(\tau)\hat{B}_{2}^{\dagger} e^{2 i (\bar{\omega}- \omega_c) t} \right\rbrace \\
& = (\mathcal{C}_{\textrm{tl}} \mathcal{C}_c )^2 \frac{1}{4\pi} \left( \frac{\hbar Z_\tl}{2 \phi_0^2}\right)^2 \frac{E_{\textrm{J:c-tl}}^2}{\hbar^2 \omega_c} \frac{\theta}{\omega_c} \hat{B}_2 \rho(t) \hat{B}_{2}^{\dagger} \\
& \equiv \Gamma_{2,2} \hat{B}_2 \rho(t) \hat{B}_{2}^{\dagger}
\end{align}
\label{eq:second_order_t32}
\end{subequations}
Note that this term is also slow compared to the first order dynamics. Similarly, to consistently compare with the first order rate in Eq.~\eqref{eq:first_order_rate}, we redefine the cavity operator $\hat{B}$ as Eq.~\eqref{eq:redefine_A2} and the rate associated rate $\Gamma_{2,2}^\prime$ as
\begin{equation}
    \Gamma_{2,2}^\prime = \frac{1}{(\mathcal{C}_{\textrm{tl}} \mathcal{C}_c )^2} \Gamma_{2,2} =\frac{1}{4\pi} \left( \frac{\hbar Z_\tl}{2 \phi_0^2}\right)^2 \frac{E_{\textrm{J:c-tl}}^2}{\hbar^2 \omega_c} \frac{\theta}{\omega_c}
    \label{eq:gamma_22p}
\end{equation}
and then the ratio for the rates
\begin{equation}
    \frac{\Gamma_{2,2}'}{\Gamma_1} = \frac{1}{4 \pi} \left( \frac{\hbar Z_\tl}{2 \phi_0^2}\right) \frac{\theta}{\omega_c},
\end{equation}
where if the transmission line impedance is $50\ \Omega$, the term $\left( \frac{\hbar Z_\tl}{2 \phi_0^2}\right) \sim 0.1560$ and in the quantum optics system assumption, $\theta / \omega_c \ll 1$. So this second order coupling dynamics is also slower than the first order coupling dynamics, and is controlled by the small parameter $\theta/\omega_c$.

Similarly, we can perform the same procedure for the other three terms and obtain the master equation induced by $H_2$
\begin{subequations}
\begin{align}
    \partial_t \rho(t) & = - \Gamma_{2} \mathcal{D}[\hat{B}_2] \rho(t) \\
    \mathcal{D}[\hat{B}_2] \rho(t) & = -\frac{1}{2} \left(\hat{B}_2^{\dagger} \hat{B}_2 \rho + \rho \hat{B}_2^{\dagger}\hat{B}_2 - 2 \hat{B}_2 \rho(t) \hat{B}_2^{\dagger} \right) \\ 
    \hat{B}_2 & = \mathcal{C}_{\textrm{tl}} \mathcal{C}_c \sum_n \frac{(-1)^{n} \tilde{\varphi}_c^{2n+2}}{n! \cdot (n+2)!}\left(c^{\dagger}\right)^n c^{n+2} \\
    \Gamma_{2} & = \left[ \frac{1}{4\pi} \left( \frac{\hbar Z_\tl}{2 \phi_0^2}\right)^2 \frac{E_{\textrm{J:c-tl}}^2}{\hbar^2 \omega_c} \right] \left[ \frac{\theta}{\omega_c} - \frac{\pi v_p}{2 l \omega_c} \right] 
\end{align}
\label{eq:second_order_result}%
\end{subequations}

Finally, I want to note that the above derivation is valid when the length of the transmission line is large. This is consistent with the Born-Markov approximation. We assume in a coupling bandwidth $\theta \ll \omega_c$, the number of modes in this bandwidth is still much greater than the system DOFs, so the transmission line must be considered to be long, in which the integer
\begin{equation}
   n_c \equiv \frac{k_c l}{2\pi} = \frac{\omega_c l}{2\pi v_p} \gg 1 ,
\label{eq:small_para_nc}
\end{equation} 
such that we can find an other integer $n_{\theta}$ which satisfies $\vert n_{\theta} - n_{c} \vert \gg 1$ and $n_{\theta}/n_c \ll 1$. In the regime where $l \rightarrow \infty$, the rate of the quantum process given by $H_2$ nonlinear system-bath coupling is given by $\Gamma_{2,2}'$ term [Eq.~\eqref{eq:gamma_22p}], and is controlled by small parameter $\theta /\omega_c$ which does not depend on the length of the transmission line.

Similar to the discussion for the second order term in the ABOCC coupling circuit, after we apply the rotating-wave approximation, the third order Hamiltonian is given by
\begin{subequations}
   \begin{align}
\label{eq:h3_term1}
     \frac{H_3}{E_{\textrm{J:c-tl}} \mathcal{C}_c \mathcal{C}_{\textrm{tl}}} = 
    & i \left\lbrace \left( \sum_{k} \frac{\tilde{\varphi}_\tl^3(k)}{6} ( b_{k}^{\dagger} )^3 + {\sum_{k,q}}' \frac{\tilde{\varphi}_\tl(k) \tilde{\varphi}_\tl(q)}{2} (b_{k}^{\dagger})^2 b_{q}^{\dagger}  
    + 
    {\sum_{k,q,p}}' \tilde{\varphi}_\tl(k) \tilde{\varphi}_\tl(q) \tilde{\varphi}_\tl(p) b_{k}^{\dagger} b_{q}^{\dagger} b_{p}^{\dagger} \right) \right.
    \\ \nonumber
    &  \quad \quad \times \left. \sum_n \frac{(-1)^n \tilde{\varphi}_c^{2n+3}}{n! \cdot (n+3)!}(c^{\dagger})^{n} c^{n+3} - h.c. \right\rbrace \\
\label{eq:h3_term2}
    -i & \left\lbrace \left[
        \sum_k \frac{\tilde{\varphi}_\tl^3(k)}{2} (b_{k}^{\dagger})^2 b_{k} + {\sum_{k,q}}' \tilde{\varphi}_\tl^2(k) \tilde{\varphi}_\tl(q) \left( b_{q}^{\dagger}b_{k}^{\dagger}b_{k} + \frac{1}{2} (b_{k}^{\dagger})^2 b_{q}\right) 
    {\sum_{k,q,p}}' 6 \tilde{\varphi}_{\tl}(k) \tilde{\varphi}_{\tl}(q) \tilde{\varphi}_{\tl}(p) b_{k}^{\dagger} b_{q}^{\dagger} b_{k}
    \right] 
    \right.
    \\ \nonumber
    & \qquad \left. \times
    \sum_{n} \frac{(-1)^n \tilde{\varphi}_c^{2n+1}}{n! \cdot (n+1)!} (c^\dagger)^{n} c^{n+1}
    - h.c. \right\rbrace
    \end{align} 
\end{subequations}
Following the same argument as in the discussion of the second order terms, the only Hamiltonian term that contributes to the system dynamics when the bath is in vacuum state is the first term in Eq.~\eqref{eq:h3_term1}. We can define a system nonlinear operator 
\begin{equation}
\hat{B}_3 = \mathcal{C}_c \mathcal{C}_{\textrm{tl}} \sum_n \frac{(-1)^n \tilde{\varphi}_c^{2n+3}}{n! \cdot (n+3)!}(c^{\dagger})^{n} c^{n+3}.
\end{equation}
In Eq.~\eqref{eq:h3_term1}, there are three terms, the first term, $\sum_{k} \frac{\tilde{\varphi}_\tl^3(k)}{6} (b_k^{\dagger})^3$ term, will give a Lindblad term in master equation of cavity field as $\mathcal{D}[\hat{B}_3] \rho(t)$ with rate $\Gamma_{3,1}$. This process is further suppressed by the small parameter $1/n_c$ [see Eq.~\eqref{eq:small_para_nc}] as
\begin{equation}
    \frac{\Gamma_{3,1}}{\Gamma_1} \propto n_c^{-2} = \left(\frac{2\pi v_p}{\omega_c l}\right)^2.
\end{equation}
The second term, $\sum_{k,q} (b_{k}^{\dagger})^2 b_{q}^{\dagger}$ term, will give a Lindblad term $\mathcal{D}[\hat{B}_3] \rho(t)$ with rate $\Gamma_{3,2}$, 
\begin{equation}
    \frac{\Gamma_{3,2}}{\Gamma_{1}} \propto \frac{1}{n_c} \frac{\theta}{\omega_c} = \left(\frac{2\pi v_p}{\omega_c l}\right) \frac{\theta}{\omega_c},
\end{equation}
where $\theta$ is the coupling bandwidth. The third term $\sum_{k,q,p} b_{k}^{\dagger} b_{q}^{\dagger} b_{p}^{\dagger}$ givens the same Lindblad term with rate $\Gamma_{3,3}$,
\begin{equation}
    \frac{\Gamma_{3,3}}{\Gamma_{1}} \propto \frac{\theta^2}{\omega_c^2}.
\end{equation}
In the limit where $l \rightarrow \infty$, the third term is dominant, but is still further suppressed by $\theta/\omega_c$, even compared with the second order dynamics. In the main text, we ignore the second and third order terms in the system-bath coupling Hamiltonian and only focus on the first order terms. 

Further, in the main text, we normalize the nonlinear cavity operator $\hat{B}_1$ by introducing a normalization constant $\mathcal{N}$, to make the flat region of function $\langle n \vert \hat{B}_{1} \vert n+1 \rangle$  to be unity. The normalized nonlinear cavity operator and associated decay rate (see Eq.~\eqref{eq:nonlinear_A1} and Eq.~(12) in main text) is
\begin{align}
    \hat{B}_1 =  \mathcal{C}_{\tl} e^{- \tilde{\varphi}_c^2/2}  \sum_{n = 0}^{\infty} (-1)^n  \frac{\tilde{\varphi}_c^{2n+1}}{n! \cdot (n+1)!} \left( c^{\dagger} \right)^n c^{n+1}
    + 2 r \tilde{\varphi}_c c, 
    \label{eq:ABOCC_loss_app}
\end{align}
where the dimensionless parameter 
\begin{align}
    r = \frac{\phi_0^2}{2 L_{\textrm{c-tl}}}/E_{\textrm{J:c-tl}}
\end{align}
and the corresponding decay rate (see Eq.~\eqref{eq:first_order_rate} and Eq.~(11) in Main text) is
\begin{align}
    \Gamma_1 = \mathcal{N}^2 \frac{E^2_{\textrm{J:c-tl}}}{\hbar^2}\left( \frac{\hbar Z_\tl}{2 \phi_0^2}\right)\frac{1}{\omega_c}.
\end{align}

\section{ABOCC laser robustness against imperfections}
\label{sec:robust}

When building the ABOCC laser system to achieve the narrow linewidth as we proposed in our paper, the fabrication and tuning errors may appear in the system. In this note, we consider several possible perturbations on the parameters in our ABOCC laser system, and explore how these perturbations can change the linewidth of our ABOCC laser from the perfect situation.

Notice that in real experiments, there are possible fabrication errors, e.g., the critical currents of the Josephson junctions, the linear inductance and the $\pi$-junctions inside the two ABOCCs may not match the perfect designed values; the transmon qubit frequency may mismatch the cavity frequency, and the transmon qubit impedance may also be shifted from the value we chose, which affect the parameters of our ABOCC laser. For example, the critical currents of the Josephson junctions inside the ABOCCs control the atom-cavity coupling strength $g$ and the cavity loss rate $\Gamma_1$, while combined with the linear inductance of the ABOCCs, they also control the shape of the ABOCC operators and change the slope of the first off-diagonal elements of the ABOCC operators. Since these experimental errors can have different effects on the laser parameters simultaneously and the authenticated simulation of the laser system on device level is beyond the scope of the current paper, we focus on the perturbation on the ABOCC laser parameters, and examine the robustness of the ultra-narrow linewidth which suppresses the standard limit against the perturbation on the laser parameters.

\subsection{Perturbation on the coupling strengths and pump strength}

In the fabrication of the ABOCC laser, the control on the critical current of the Josephson junctions inside the ABOCCs may not be accurate enough. As we mentioned above, the critical currents of the Josephson junctions control the coupling strength between the segments that the ABOCCs connect. In this subsection, we consider the perturbation on the atom-cavity coupling strength $g$ and the cavity loss rate $\Gamma_1$. Further, as we use $\Gamma_1$ as a frequency unit in our discussion, we effectively perturb the ratio $g/\Gamma_1$.

\begin{figure}[htbp]
    \centering
    \includegraphics[width = \textwidth]{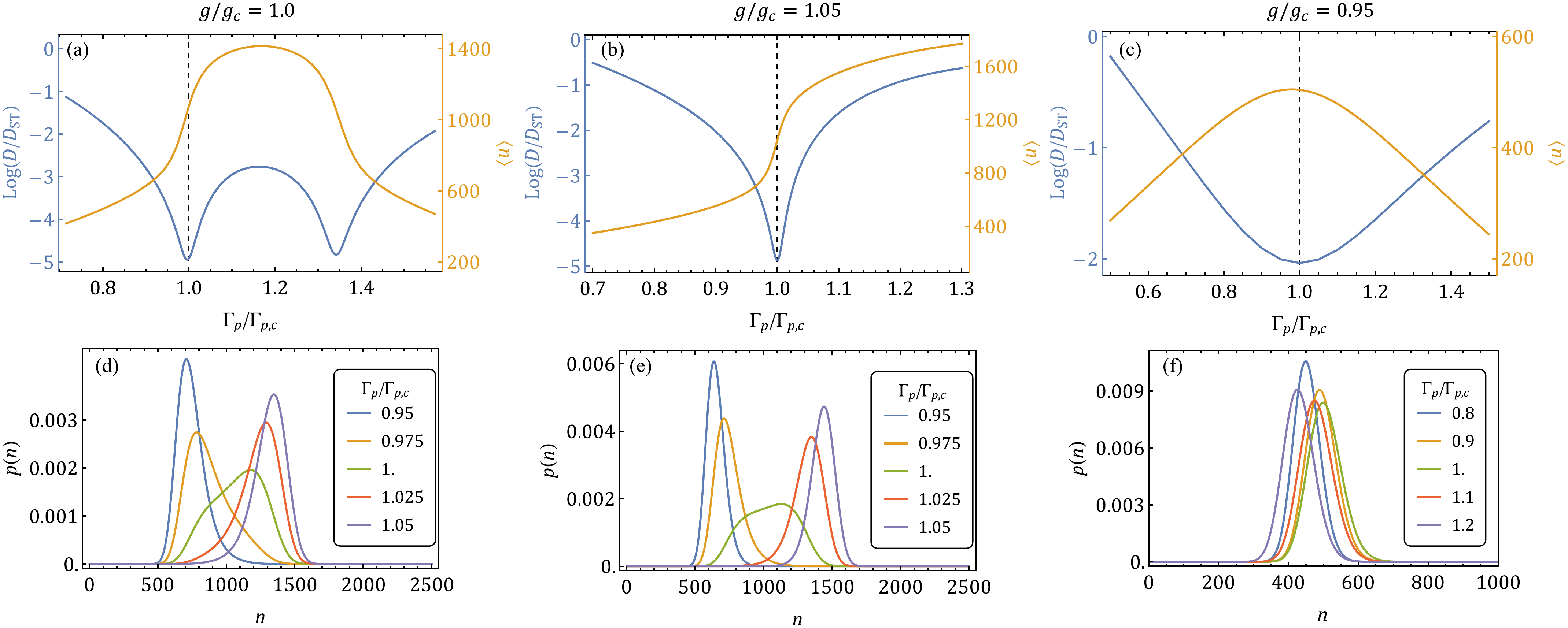}
    \caption{The linewidth and photon distribution of the Approximately Bare Operator Coupling Circuit (ABOCC) laser with atom-cavity coupling perturbation.
    The linewidth ratio $\text{log}(D/D_ST)$ (blue curves, left axis), mean photon number $\mean{n}$ (orange curves, right axis) for cavity-atom coupling strength $g/g_c = 1.0$, $1.05$ and $0.95$ are given in (a), (b) and (c). The photon distribution at different incoherent pumping strength $\Gamma_p$ for $g/g_c = 1.0$, $1.05$ and $0.95$ are plotted in (d), (e) and (f). The ABOCC operators have $m_0 = 1000$ in all three cases. $g_c$ is the optimum cavity-atom coupling strength for the narrowest laser linewidth. $\Gamma_{p,c}$ is the optimum incoherent pump strength for the given cavity-atom coupling.}
    \label{fig:perturb_g}
\end{figure}

From Fig.~4b in main text, we notice that there is a valley that the laser can still have ultra-narrow linewidth as we tune the parameter $g/\Gamma_1$. To reach this valley, for different coupling strength $g$, the incoherent pump strength $\Gamma_p$ needs to be fine tuned. In Fig.~\ref{fig:perturb_g}, we consider shifting the laser atom-cavity coupling $g$ from the optimum value $g_\text{c}$ by $\pm 0.05$. We show how the laser linewidth ($\text{log}(D/D_\text{ST})$, shown as blue lines in Fig.~\ref{fig:perturb_g}a-c) and mean photon number ($\mean{n}$, shown as orange lines in Fig.~\ref{fig:perturb_g}a-c) vary as the incoherent pump strength $\Gamma_p$ is tuned. Notice that for the optimum coupling strength $g/g_c = 1$, as we increase the pump strength to $\Gamma_p/\Gamma_{p,c} \sim 1.3$, the incoherent pump strength is strong compared to the coupling strength, which causes the self-quenching on the single-atom laser system, as pointed out in Ref.~\cite{Mu1992} for single-atom laser. In this regime, as we increase the pump strength, the fast incoherent drive destroys the coherence between the ground state and the excited state of the atom, which reduces the photon number inside the cavity. Similar performance can be found in the case where $g/g_c = 1.05$. Around the optimum pump strength $\Gamma_{p,c}$ for both $g/g_c = 1.0$ and $1.05$, the optimum linewidth is achieved by fine-tuning the incoherent pump strength $\Gamma_p$ to make the photon distribution living inside the flatten region of the ABOCC operator. This can be seen from the photon distributions shown in Fig.~\ref{fig:perturb_g}d and Fig.~\ref{fig:perturb_g}e green lines. As we decrease the atom-cavity coupling strength, the threshold for the incoherent pump induced self-quenching also decreases, which makes the achievable maximum mean photon number inside the cavity lower than the plateau in the ABOCC operators. Then the ABOCC laser cannot achieve the large suppression on the laser linewidth (see Fig.~\ref{fig:perturb_g}c and e).

\begin{figure}[h]
    \centering
    \includegraphics[width = 0.7 \textwidth]{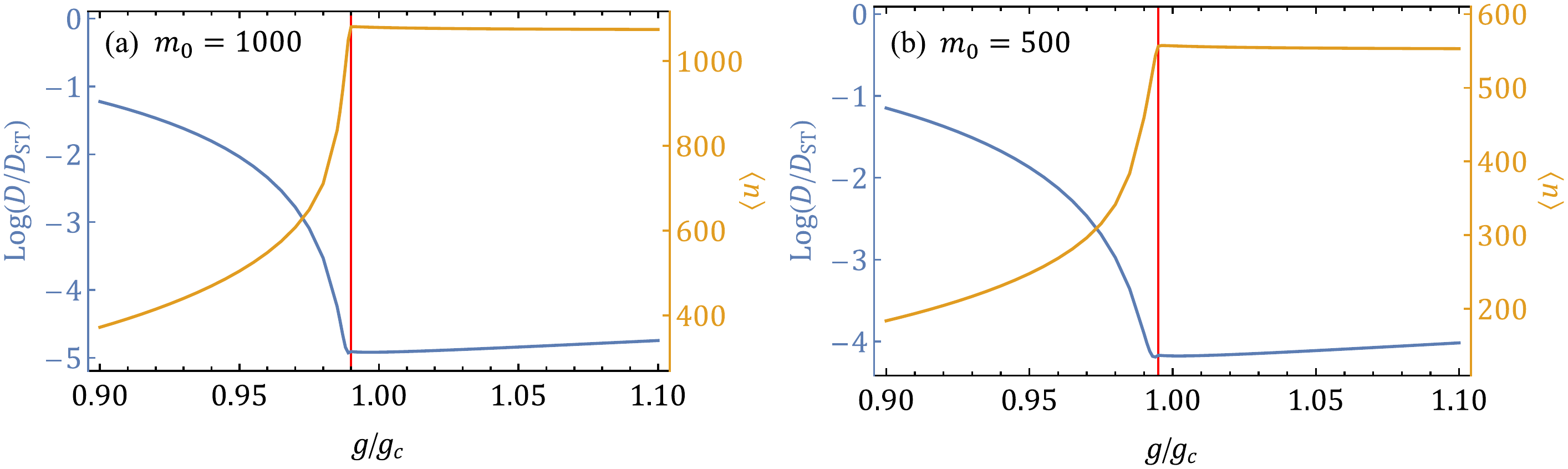}
    \caption{The linewidth ratio $D/D_{\text{ST}}$ (blue) and the mean photon number $\mean{n}$ (orange) as we perturb the cavity-atom coupling strength $g$ for Approximately Bare Operator Coupling Circuit (ABOCC) laser. The coupling strength $g$ is in the unit of optimum cavity-atom coupling strength $g_c$. The ABOCC lasers have plateau center photon number $m_0 = 1000$ (a) and $m_0 = 500$ (b). For each coupling strength $g$, the linewidth is optimized by tuning the incoherent pump strength $\Gamma_p$.}
    \label{fig:perturb_sweep_g}
\end{figure}

With a given cavity-atom coupling strength $g$, as long as the coupling strength is strong enough such that the maximum mean photon number can reach the designed $m_0$ value of the ABOCC operator, as shown in Fig.~\ref{fig:perturb_g}a and Fig.~\ref{fig:perturb_g}b, the optimum linewidth can achieved by tuning the pump strength. There are $\pm 5\%$ to $\pm 10\%$ error allowed to the pump strength ($\Gamma_p$) around the optimum pumping strength $\Gamma_{p,c}$, where we can still get a decent suppression ($ > 10$ times) beyond the standard limit. 

Because the incoherent pump strength is not a parameter determined by the device, we can tune the pump strength experimentally, we further explore in detail on the robustness of the ultra-narrow linewidth as we perturb the cavity-atom coupling strength $g$ but leave $\Gamma_p$ as a tunable parameter. In Fig.~\ref{fig:perturb_sweep_g}, we focus on two ABOCC lasers, whose ABOCC operators have central photon number for the flat plateau $m_0 = 1000$ (see Fig.~\ref{fig:perturb_sweep_g}a) and $m_0 = 500$ (see Fig.~\ref{fig:perturb_sweep_g}b). As we perturb the coupling strength $g$, we notice that the optimum cavity-atom coupling strength $g_c$ is very close to the edge of the strong self-quenching regime. For the ABOCC laser with $m_0 = 1000$, the coupling strength that unable to get cavity photon distributed on the ABOCC operator plateau is $g/g_c = 0.99$ while for $m_0 = 500$, the coupling strength is $g/g_c = 0.995$ (see the red vertical lines in Fig.~\ref{fig:perturb_sweep_g}). However, as we go beyond this point, there is a large parametric space for coupling strength $g$ to vary, which means the ABOCC laser can still get a decent suppression of the laser linewidth beyond the standard limit.

\subsection{Perturbation on the inductance ratio in ABOCC operators}

In the main text, we focus on the ABOCC laser system whose two ABOCC operators are identical and fine-tuned. Compare the ABOCC operator between the atom and the cavity ($\hat{A}_{c,m_0}$) given in Eq.~(9) in main text and the operator between the cavity and the transmission line $\hat{B}_{c,m_0}$ given in Eq.~\eqref{eq:ABOCC_loss_app} (and also Eq.~(12) in main text), to make identical ABOCC operators $\hat{A}$ and $\hat{B}$, we can fine-tune the impedance of the transmon qubit $Z_{a}$, the impedance of the cavity $Z_c$ and the transmission line impedance $Z_\tl$ and the inductance ratio $r_\ac \equiv L_\text{J:a-c} / L_\ac$ and $r_\ctl \equiv L_\text{J:c-tl} / L_\ctl$. In our proposal, we manually set
\begin{align}
    & r_{\ac} = r_\ctl, \\
    & \frac{\sin \left( \tilde{\varphi}_{a} \right)}{\tilde{\varphi}_{a}} = \mathcal{C}_{\tl}
    \label{eq:za_zc_relation}
\end{align}
to ensure the two ABOCC operators to be the same. We further set $Z_\tl = 50~\Omega$, which gives $Z_a = 47.71~\Omega$. The impedance ratios $r_\ac$ and $r_\ctl$, controls the slope of the flat region of the ABOCC operators, while the cavity impedance $Z_c$ controls the center of the flat plateau of the ABOCC operators, $m_0$ (see Fig.~4a in main text). To get a perfect ABOCC operator, $r_\ac = r_\ctl = 0.418$. 

\begin{figure}[h]
    \centering
    \includegraphics[width = \textwidth]{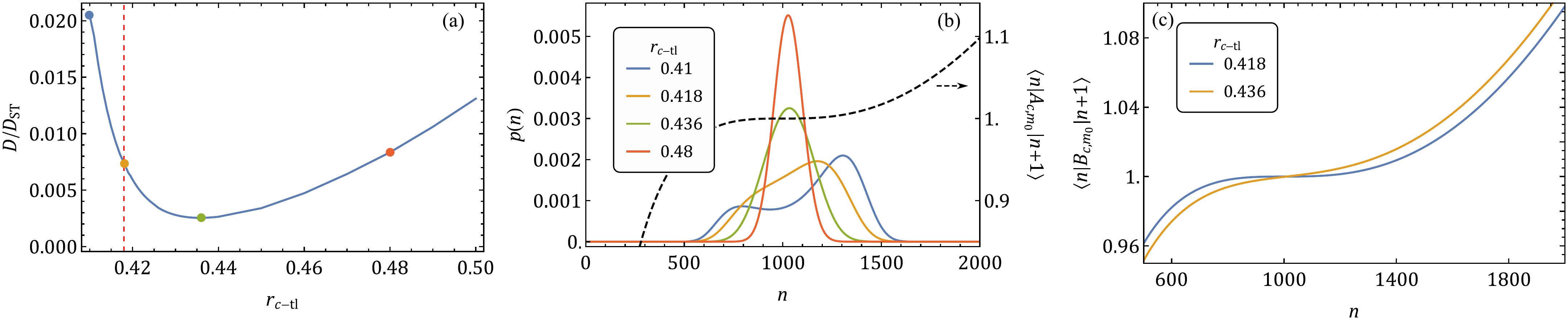}
    \caption{The laser linewidth, photon distribution and the Approximately Bare Operator Coupling Circuit (ABOCC) operator with perturbation on cavity-transmission line ABOCC operator inductance ratio. (a) ABOCC laser linewidth $D$ in the unit of standard limit $D_\text{ST}$ as we perturb the cavity-transmission line ABOCC operator inductance ratio $r_\ctl$. The red dashed vertical line is $r_{\ctl} = 0.418$, the ratio for the unperturbed ABOCC operator. The photon distribution (solid lines) for four $r_\ctl$ values [$r_\ctl = 0.41$ (blue), $0.418$ (orange), $0.436$ (green), $0.48$ (salmon)] are shown in (b). The corresponding $r_\ctl$ values are shown as the points with the same color in (a). The ABOCC operator for the cavity-atom coupling matrix elements $\langle n \vert \hat{A}_{c,m_0} \vert n+1 \rangle$ is shown as the black dashed line in (b). In (c), we compare the matrix elements $\langle n \vert \hat{B}_{c,m_0} \vert n+1 \rangle$ of ABOCC operators with $r_\ctl = 0.418$ (blue) and $r_\ctl = 0.436$ (orange). The larger $r_\ctl$ value gives larger slope around the plateau region.}
    \label{fig:perturb_r_ctl}
\end{figure}

As long as we ensure Eq.~\eqref{eq:za_zc_relation}, $m_0$ for two ABOCC operators are always equal. In this subsection, we at first discuss the robustness of the ultra-narrow ABOCC laser linewidth against on one of the ABOCC operator $r$ values. We then set the cavity-atom ABOCC operator to be the perfect (with $r = 0.418$), while we tune the loss ABOCC operator by tuning the $r_\ctl$. In Fig.~\ref{fig:perturb_r_ctl}a, for each $r_\ctl$, we optimize the laser linewidth by tuning the atom-cavity coupling strength $g$ and the incoherent pump strength $\Gamma_p$. We notice that the ultra-narrow laser linewidth does indeed require fine tuning of the ABOCC operator $r_\ctl$. However, it can still be tolerant to small perturbation. Further, as we increase the $r_\ctl$ from the optimum value ($0.418$, shown by the red vertical line), the linewidth ratio $D/D_\text{ST}$ decreases slightly beyond the linewidth achieved by two fine-tuned balanced ABOCC operators. Notice that the best ratio can be achieved at a slightly larger $r_\ctl$, i.e., the slope on the cavity loss ABOCC operator is slightly larger (see Fig.~\ref{fig:perturb_r_ctl}c orange line). The larger slope on cavity loss enhance the Boson amplification rate slightly around $m_0$, which shift the steady state of the laser photon distribution. The photon distribution is shift towards the center of the plateau of the perfect ABOCC operator (the ABOCC operator for the cavity-atom coupling). This effect slightly decreases the noise added by the ABOCC operator on pump side, which enhance the linewidth performance of the ABOCC laser. On the other hand, in terms of the fabrication error, the ABOCC laser can maintain a decent suppression on the linewidth beyond the ST limit with $\sim 10 \%$ ($0.42$ to $0.48$) on the $r_\ctl$ value.

\begin{figure}[h]
    \centering
    \includegraphics[width = \textwidth ]{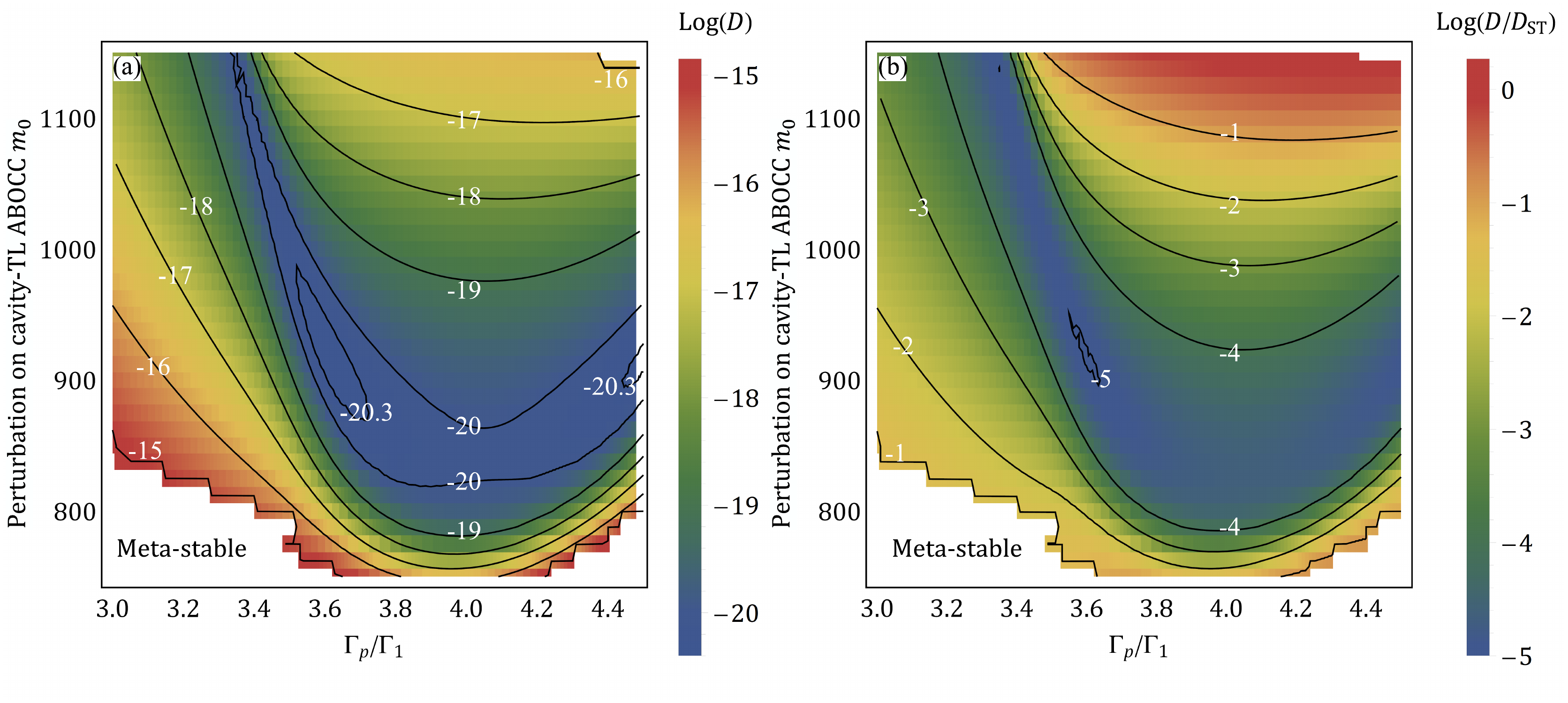}
    \caption{The Approximately Bare Operator Coupling Circuit (ABOCC) laser linewidth with perturbation on the $m_0$ value on cavity-transmission line side ABOCC operator. We plot the linewidth $D$ in (a) and linewidth ratio $D/D_\text{ST}$ in (b) as the cavity-transmission line ABOCC operator $m_0$ and the pump strength $\Gamma_p$ are tuned. Two ABOCC operators have $r_\ac = r_\ctl = 0.418$, and the atom-cavity ABOCC operator has $m_0 = 1000$. The atom-cavity coupling $g/\Gamma_1 = 1.43$.}
    \label{fig:perturb_m0}
\end{figure}

Next, we consider the case that the possible fabrication errors that cause the Eq.~\eqref{eq:za_zc_relation} not to be held, which makes the two ABOCC operators (ABOCC between the atom and the cavity, ABOCC between the cavity and the transmission line) to have different $m_0$ values. Here we assume the atom-cavity ABOCC operator always have $m_0 = 1000$ and $r_\ac = r_\ctl = 0.418$, but manually tune the cavity-transmission line ABOCC operator $m_0$. As $g/\Gamma_1$ is determined by the fabricated circuit, but $\Gamma_p$ can be tuned by the pump strength to drive the transmon qubit, we manually set the $g/\Gamma_1$ to be the optimum ratio for the ABOCC laser with two balanced ABOCC operators ($m_0 = 1000$, $r_\ac = r_\ctl = 0.418$, then $g/\Gamma_1 \sim 1.43$). In Fig.~\ref{fig:perturb_m0}, we show the ABOCC laser linewidth as we tune the cavity-transmission line ABOCC $m_0$ and the incoherent pump strength $\Gamma_p$. We notice that as we decrease the $m_0$ away from the balanced value $m_0 = 1000$, the optimum laser linewidth slightly improves. The best linewidth ratio is achieved at $\Gamma_p/\Gamma_1 = 3.59$, $m_0 = 925$ with $\text{log}(D/D_{\text{ST}}) \sim -5.02$ (see Fig.~\ref{fig:perturb_m0}b).
As we further decrease the $m_0$ value, the optimum linewidth starts to decrease. The optimum pump strength $\Gamma_p$ also shifts towards $\Gamma_p / \Gamma_1 \sim 4$. For the ABOCC operators with $m_0 \leq 888$, as we tune the pump strength away from $\Gamma_p / \Gamma_1 = 4$, the laser system starts to be tuned into a ``meta-stable'' regime, where there are multiple stable states solved from the eigen-spectrum of the super-operator of the ABOCC laser systems. This regime is labeled as ``Meta-stable'' and filled in white in Fig.~\ref{fig:perturb_m0}a and b. As we increase $m_0$ from $1000$, the best linewidth of the ABOCC laser by tuning pump strength can still reach $log(D) < -20$, which gives a decent suppression beyond the standard limit. In summary, by tuning the pump strength, the ABOCC laser can maintain the narrow linewidth with decent perturbation on ABOCC operator $m_0$ value. 

\subsection{Perturb the atom frequency}

\begin{figure}[h]
    \centering
    \includegraphics[width = 3.5 in]{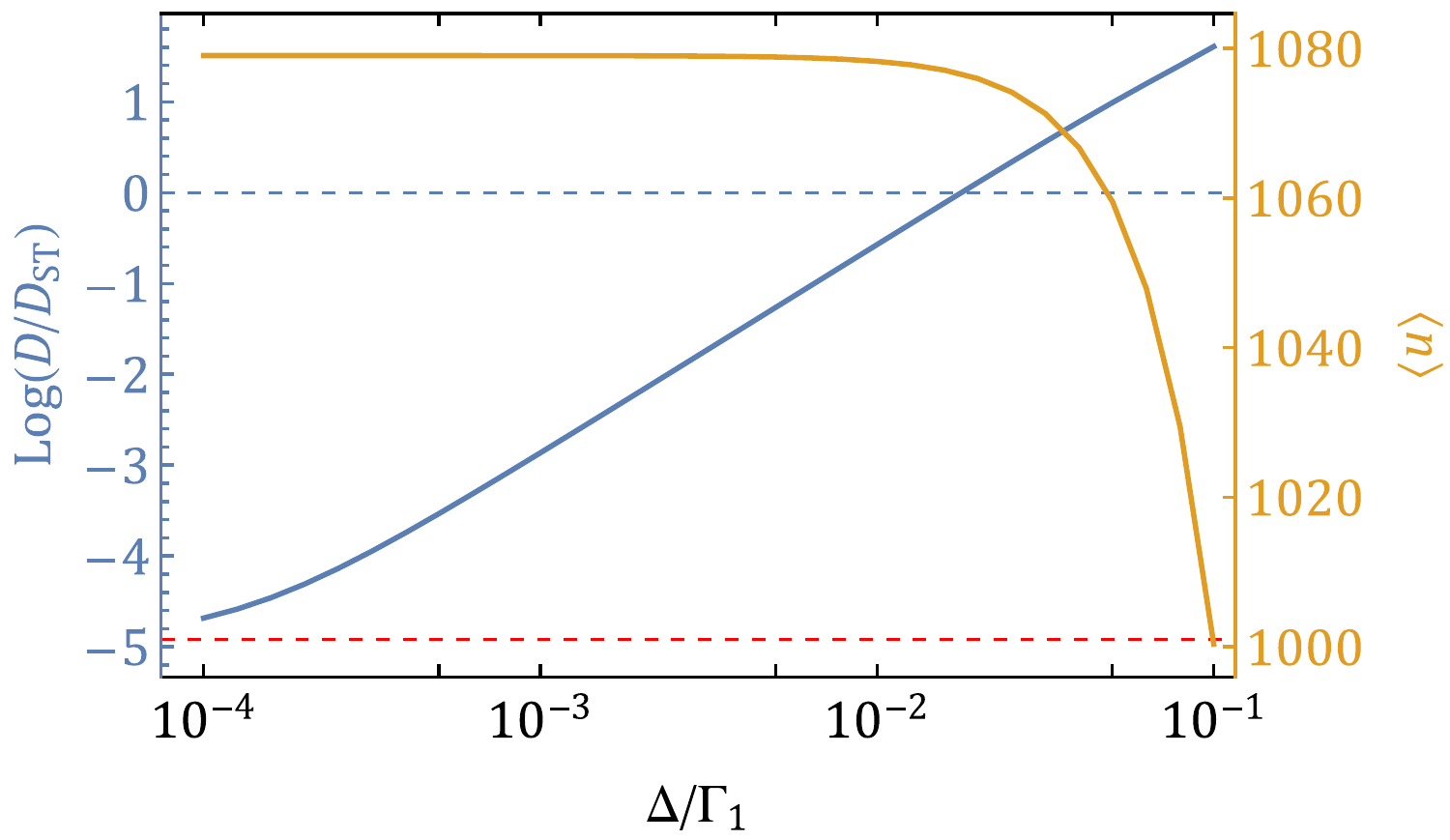}
    \caption{The ABOCC linewidth $D/D_\text{ST}$ (blue) and the mean photon number $\mean{n}$ (orange) as we shift the atom frequency to shift the cavity-atom detuning $\Delta = (\omega_a - \omega_c)$ in the unit of cavity loss rate $\Gamma_1$. For each $\Delta/\Gamma_1$ value, we tune the coupling strength $g$ and incoherent pump strength $\Gamma_p$ to optimize the laser linewidth. The red dashed line shows the laser linewidth $D/D_{\text{ST}}$ for zero detuning and the blue dashed line shows $D = D_\text{ST}$.  }
    \label{fig:detune}
\end{figure}

In this subsection, we explore the perturbation on the atomic frequency to detune it from the cavity frequency. For a typical laser system, when the cavity frequency is shifted away from the atomic frequency, the laser will be degraded. Similarly, for the ABOCC laser system, as we shift the frequency of the atom away from the cavity frequency, the laser linewidth increases, while the photon distribution does not change significantly (see Fig.~\ref{fig:detune}). As we increase the detuning $\Delta$, eventually the ABOCC laser will exceed the standard limit ($\Delta/\Gamma_1 \sim 0.0178$. In terms of fabrication of ABOCC laser, the cavity frequency needs to be fine-tuned to match the atomic frequency to achieve the ultra-narrow linewidth. This can be done by using a magnetic field tunable SQUID inside the transmon qubit, instead of the single Josephson junction, to make the frequency of the qubit to be tunable. By tuning the magnetic flux, the frequency of the transmon qubit can be manually tuned to match the cavity frequency with a relative high precision.  

\section{Properties of the ABOCC lowering operator} \label{sec:ABOCC_eigen}

In this note, we consider the properties of the ABOCC lowering operator eigenstates and demonstrate that the eigenstates which correspond to the ultranarrow linewidth laser are highly squeezed. Here, we focus on the ABOCC operator $\hat{A}_{c,m_0}$ (see Eq.~(9) in main text) that is associated with atom-cavity coupling. The operator $\hat{B}_{c,m_0}$ (see Eq.~(12) in main text), which is associated with cavity-transmission line coupling, can be made to match $\hat{A}_{c,m_0}$ by tuning the properties of the output ABOCC coupler. 

Our first goal is to obtain the eigenstates $\ket{\psi_{\alpha}}$ of $\hat{A}_{c,m_0}$. 
The eigen-equation is
\begin{equation}
    \hat{A}_{c,m_0} \ket{\psi_{\alpha}} = \alpha \ket{\psi_{\alpha}},
    \label{eq:ABOCC_eigen}
\end{equation}
where $\alpha$ is the eigenvalue. We express
\begin{align}
    \ket{\psi_{\alpha}} = \sum_{n} \psi_{\alpha}(n) \ket{n},
\end{align}
where $\ket{n}$'s are the photon number states and $\psi_{\alpha}(n)$'s are the coefficients. Since the operator $\hat{A}_{c,m_0}$ always lowers the photon number by $1$, we can write it as
\begin{equation}
    \hat{A}_{c,m_0} = \sum_{n=1}^{\infty} f(n) \ket{n-1}\bra{n},
\end{equation}
where the coefficients $f(n)$'s are determined by $m_0$ (see Fig.~4a in main text). 
The coefficients appearing in the eigenstates $\psi_{\alpha}(n)$ obey the recurrence relation
\begin{equation}
    \psi_{\alpha}(n) = \frac{\alpha}{f(n)} \psi_{\alpha}(n-1).
    \label{eq:ABOCC_eigen_state}
\end{equation}
Using this recursion relation, along with the normalization condition $\sum_n \vert \psi(n) \vert^2 = 1$, the eigenstate coefficients can be determined if the eigenvalue $\alpha$ is given.

\begin{figure}[h]
    \centering
    \includegraphics[width = 0.8 \textwidth]{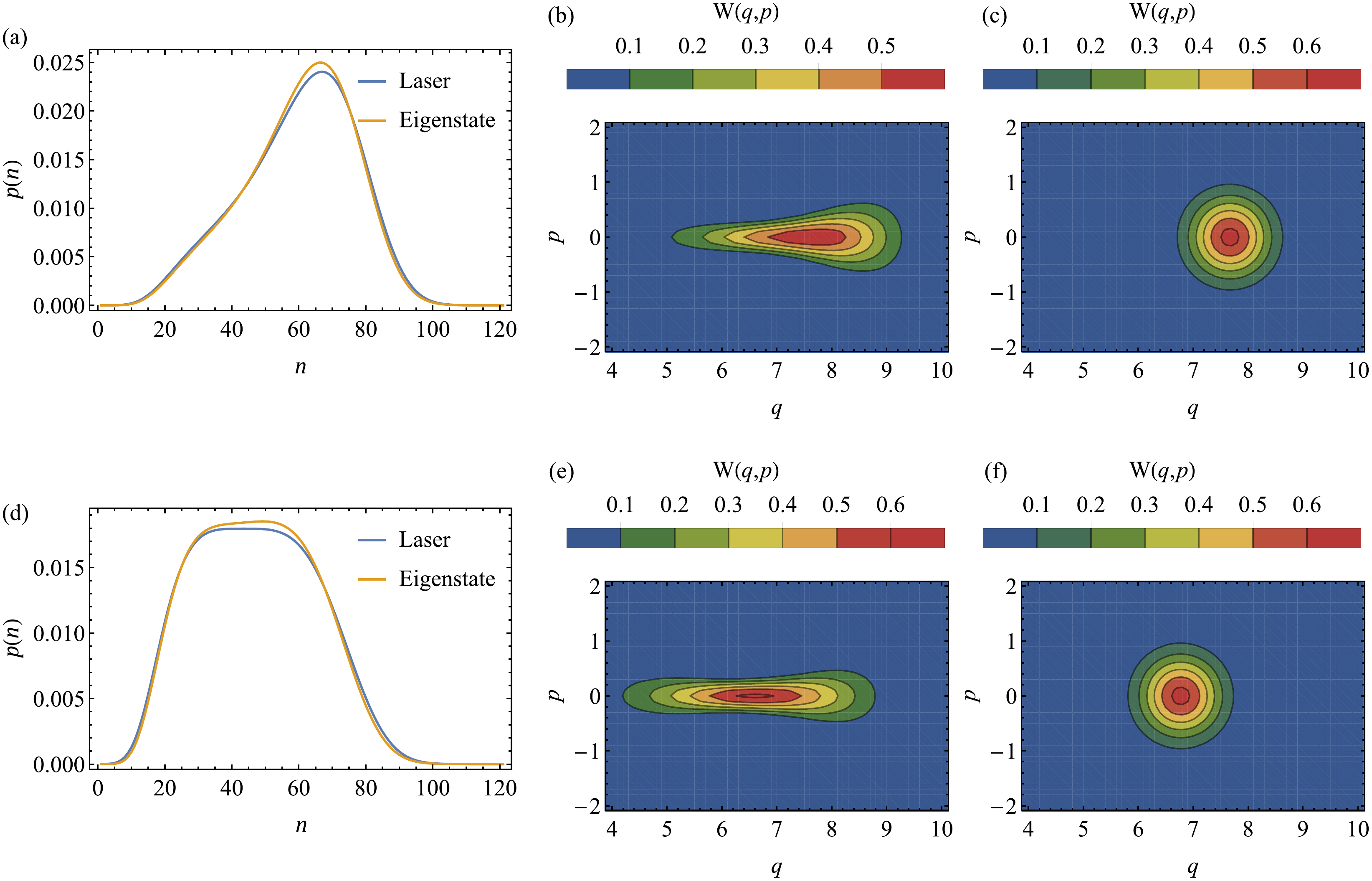}
    \caption{The photon distribution and the Wigner distribution of the corresponding eigenstates of the Approximately Bare Operator Coupling Circuit (ABOCC) operator. In (a), we show the photon distribution for the laser state that achieve the narrowest linewidth (blue) and the corresponding eigenstate of the ABOCC operator (orange). In (b) we plot the Wigner distribution of the eigenstate of the ABOCC operator whose photon distribution is shown (a). The Winger distribution for a coherent state with the same mean photon number is in (c). We also consider the laser state that utilize the most SG-operator-like regime of the ABOCC operator. The photon distribution of the laser state (blue line) and the corresponding best-match eigenstate of the ABOCC operator (orange line) are in (d). The Wigner distribution of the ABOCC eigenstate is in (e) while the coherent state with the same mean photon number is given in (f) for comparison.}
    \label{fig:compare_wigner}
\end{figure}

Now, consider a laser tuned so that the ABOCC operator for both the cavity-atom coupling and the cavity loss term are identical with $m_0 = 43$ ($Z_c = 150~\Omega$). We use the cavity loss constant $\Gamma_1$ as a frequency unit and minimize the laser linewidth to obtain $\Gamma_p/\Gamma_1 = 3.58$ and $g/\Gamma_1 = 1.468$. The corresponding photon distribution is shown in Fig.~\ref{fig:compare_wigner}a (blue line). Next, we find the eigenstate of the ABOCC operator that best matches this photon distribution by tuning $\alpha$ in Eq.~\eqref{eq:ABOCC_eigen} and using the cost function
\begin{equation}
    C(\alpha)^2 = \sum_{n} \left( \vert\psi_{\alpha}(n)\vert^2 - p(n)\right)^2,
    \label{eq:cost_fun}
\end{equation}
where $p(n)$ is the photon distribution in the laser state. In this case, we find that the optimum eigenvalue is $\vert \alpha \vert =1.0225$, where the photon distribution given by the eigenstate is shown in Fig.~\ref{fig:compare_wigner}a (orange line). Without loss of generality (as we show later) we focus on the case in which the eigenvalue $\alpha$ is a positive real number. The corresponding Wigner distribution for the eigenstate of the ABOCC operator is plotted in Fig.~\ref{fig:compare_wigner}b. For comparison, we show the Wigner distribution of a coherent state with the same phase and mean photon number in Fig.~\ref{fig:compare_wigner}c. Notice that the Wigner distribution of the eigenstate is squeezed along the phase direction (and expanded along the photon number direction) as compared to a coherent state. 

Our next goal is to quantify the amount of squeezing. Here we consider two approaches:
\begin{itemize}
    \item Fluctuations of position and momentum [$\hat{q} = \frac{1}{2}(\hat{a} + \hat{a}^{\dagger})$, $\hat{p} = \frac{1}{2i} (\hat{a} - \hat{a}^\dagger)$].
    \item Fluctuations of phase and number.
\end{itemize}
The second approach suffers from the well known problem that a Hermitian phase operator of the optical field is hard to define~\cite{Susskind1964, Carruthers1965, Carruthers1968}. The standard solution, which we adopt here, is to use the sine and cosine of the phase, which can be defined as the following Hermitian operators
\begin{equation}
    \hat{C} \equiv \cos \hat{\phi} = \frac{1}{2} \left(\hat{e} + \hat{e}^{\dagger}\right) \qquad \hat{S} \equiv \sin \hat{\phi} = \frac{1}{2 i} \left( \hat{e} - \hat{e}^{\dagger} \right).
\end{equation}
The resulting operators obey the uncertainty relation~\cite{Susskind1964, Carruthers1965, Carruthers1968}
\begin{equation}
    \frac{\Delta n^2 \Delta S^2}{\langle \hat{C} \rangle^2} \geq \frac{1}{4}.
    \label{eq:uncertainty_phase}
\end{equation}
For a state with a Wigner distribution that lies along the positive real axis we can estimate $\Delta \phi \sim \Delta S/\langle\hat{C}\rangle$. Naively, the uncertainty relation between the phase and photon number of the light field implies that $\Delta n \Delta \phi \sim \Delta n \Delta S/\vert \langle\hat{C}\rangle \vert \geq 1/2$. However, this notion does not quite hold, even for states with a Wigner distribution that lies along the positive real axis, because $\sin \hat{\phi}/\cos \hat{\phi}$ is only approximately $\hat{\phi}$. For example, coherent state $|\beta\rangle$ saturates the uncertainty bound Eq.~\ref{eq:uncertainty_phase} only as $\beta \rightarrow \infty$. 

In Table~\ref{tab:compare_eigen}, we compare the fluctuations in the narrowest linewidth eigenstate of $\hat{A}_{c,m_0}$ [column labeled Eigen-NLW] and the coherent state with the same number of photons [column labeled Coh-NLW]. Comparing the momentum and position fluctuations we observe that Coh-NLW is not squeezed while Eigen-NLW state is indeed squeezed in the momentum direction. Similarly, comparing fluctuations of $\hat{S}$ and $\hat{n}$ we observe that the Eigen-NLW state is squeezed as compared to the Coh-NLW state. Finally, we compute the uncertainties. We observe that the Coh-NLW state is a minimum uncertainty state that saturates the $\Delta p \Delta q$ uncertainty and almost saturates Eq.~\eqref{eq:uncertainty_phase} uncertainty. On the other hand, Eigen-NLW state is not a minimum uncertainty state, which has low uncertainty but does not saturate either measure of uncertainty.

To explore how much squeezing can be obtained in the laser system, we slightly decrease the atom-cavity coupling to the value $g/\Gamma_1 = 1.424$, which corresponds to the case in which the photon number distribution of the laser state has the flattest-top (see Fig.~\ref{fig:compare_wigner}d (blue line)). Again, we use the cost function in Eq.~\eqref{eq:cost_fun} to find the eigenstate of the ABOCC lowering operator that best matches the photon number distribution from the master equation and find the corresponding eigenvalue $\vert \alpha \vert \sim 1.0059$. The photon number distribution of this eigenstate is shown in Fig.~\ref{fig:compare_wigner}d (orange line). The corresponding Wigner function is plotted in Fig.~\ref{fig:compare_wigner}e and the coherent state with the same photon number is shown in Fig.~\ref{fig:compare_wigner}f. We observe that this eigenstate, which we label ``Eigen-FT'' for short, is even stronger squeezed than the Eigen-NLW state. This notion is confirmed in Table~\ref{tab:compare_eigen}.

\begin{table}[h]
\caption{\label{tab:compare_eigen} The properties of the eigenstates of the ABOCC operators and the corresponding coherent states with the same mean photon number. State ``Eigen-NLW'' is for eigenstate of the ABOCC operator that best-matches the narrowest linewidth laser state (see Fig.~\ref{fig:compare_wigner}a). State ``Eigen-FT'' is for the eigenstate of the ABOCC opreator that best-matches the flattest-top laser state (see Fig.~\ref{fig:compare_wigner}d). State ``Coh-NLW'' and ``Coh-FT'' are the corresponding coherent state with the same mean photon number.}
\begin{ruledtabular}
\begin{tabular}{c| c c c c}
 & Eigen-NLW & Coh-NLW & Eigen-FT & Coh-FT\\
\hline
$\langle n \rangle$ & $58.83$ & $58.83$ & $45.88$ & $45.88$\\
$\Delta q^2$ & $1.313$ & $0.25$ & $1.780$ & $0.25$\\
$\Delta p^2$ & $0.076$ & $0.25$ & $0.048$ & $0.25$\\
$\Delta S^2$ & $0.00116$ & $0.0043$ & $0.00122$ & $0.00548$\\
$\Delta n^2$ & $273.313$ & $58.83$ & $305.674$ & $45.88$\\
$\Delta p \Delta q$ & $0.3150$ & $0.25$ & $0.2914$ &$0.25$\\
$\Delta n^2 \Delta S^2 / \langle C \rangle^2$ & $0.3165$ & $0.2522$ & $0.3730$ & $0.2528$ \\
\end{tabular}
\end{ruledtabular}
\end{table}

\begin{figure}
    \centering
    \includegraphics[width = 0.8 \textwidth]{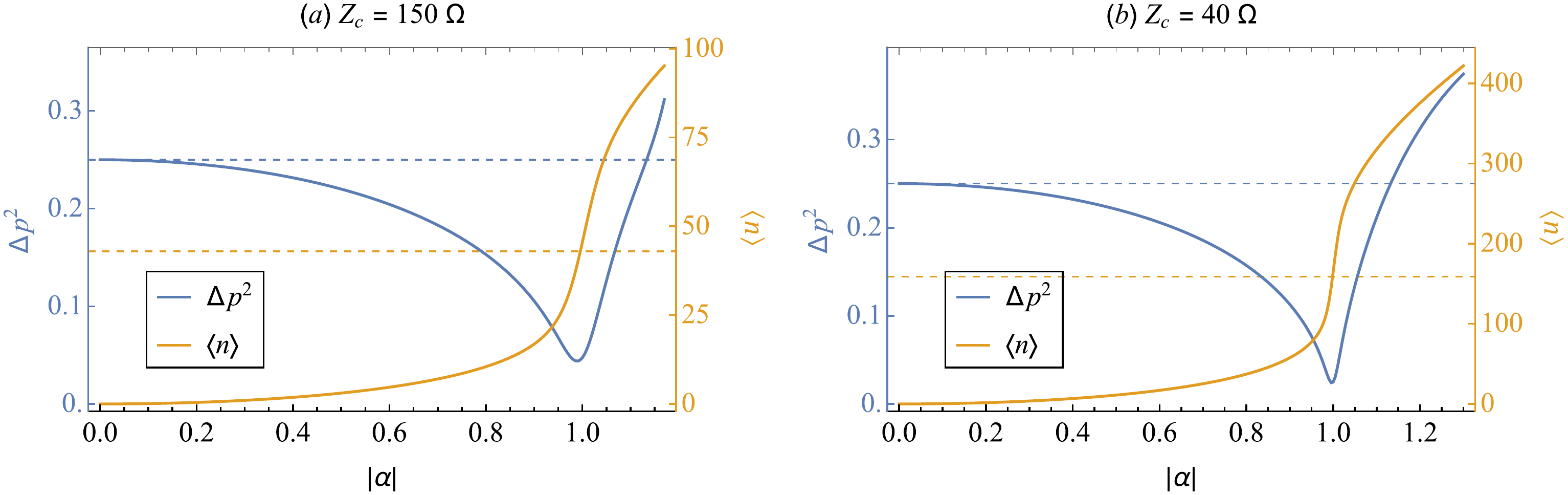}
    \caption{The mean photon number $\langle n \rangle$ (orange lines) and the noise along the $\hat{p} = \frac{1}{2 i} (\hat{a}- \hat{a}^{\dagger})$ direction (blue lines) as functions of the ABOCC operator eigenvalue $\vert \alpha \vert$. The ABOCC operators with $Z_c = 150~\Omega$ and $Z_c = 40~\Omega$ are plotted in (a) and (b), respectively. The blue dashed line show the noise limit for coherent states, while the orange dashed lines show the most flat photon number $m_0$ for the corresponding ABOCC operator.}
    \label{fig:sweep_ABOCC_noise}
\end{figure}

Having confirmed that the eigenstates of the ABOCC lowering operator are indeed squeezed, we investigate how the squeezing depends on the number of photons in the laser cavity.
In Fig.~\ref{fig:sweep_ABOCC_noise}, we plot the mean photon number $\langle n \rangle$ and the noise in $\hat{p}$ direction of the eigenstates of the ABOCC operator with two different $m_0$ values (which is controlled by the cavity effective impedance $Z_c$), as a function of the eigenvalue $\vert \alpha \vert $. In Fig.~\ref{fig:sweep_ABOCC_noise}a, we focus on the case $m_0 = 43$ ($Z_c = 150~\Omega$), while in Fig.~\ref{fig:sweep_ABOCC_noise}b, we focus on $m_0 = 159$ ($Z_c = 40~\Omega$). For both cases, the fluctuations along $\hat{p}$ direction are highly suppressed at the place where the mean photon numbers of the eigenstate is around the $m_0$ values (orange dashed lines in both plots).  

\begin{figure}[h]
    \centering
    \includegraphics[width = \textwidth]{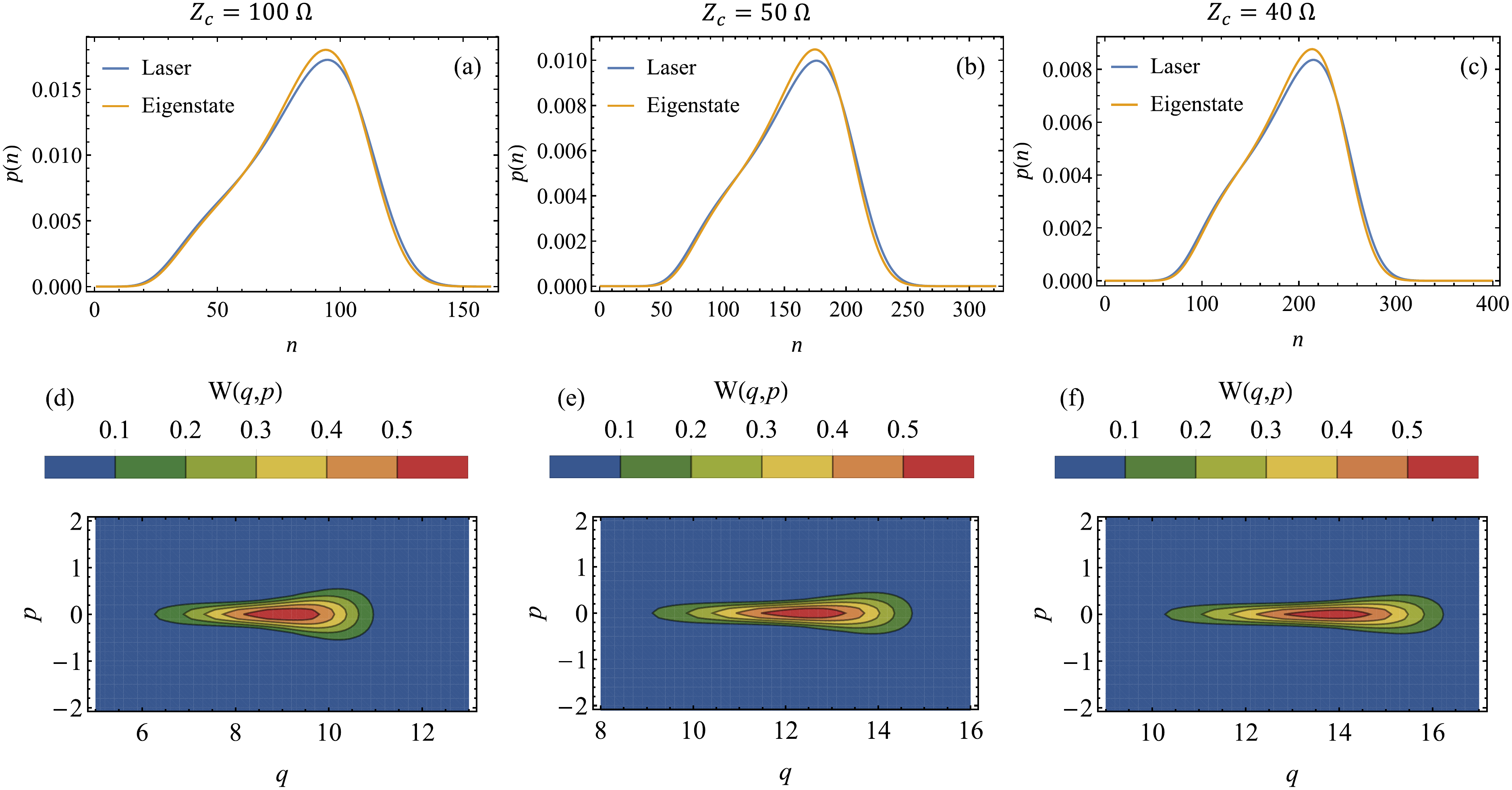}
    \caption{The photon distribution and the Wigner distribution of the corresponding eigenstates of the Approximately Bare Operator Coupling Circuit (ABOCC) operator when ABOCC laser gets the narrowest linewidth. In (a), (b) and (c), we plot the photon distribution for the narrowest linewidth ABOCC laser system and photon distribution of the corresponding best-matched ABOCC eigenstates. The Wigner distribution for the ABOCC eigenstates are plotted in (d), (e) and (f), respectively. The impedance $Z_c$ controls the central photon number $m_0$ (see Fig.~4a in main text). For (a) and (d), $Z_c = 100~\Omega$, while (b) and (e), $Z_c = 50~\Omega$, and (c), (f) are for $Z_c = 40~\Omega$.  }
    \label{fig:diff_zc_NL}
\end{figure}
\begin{figure}[h!]
    \centering
    \includegraphics[width = \textwidth]{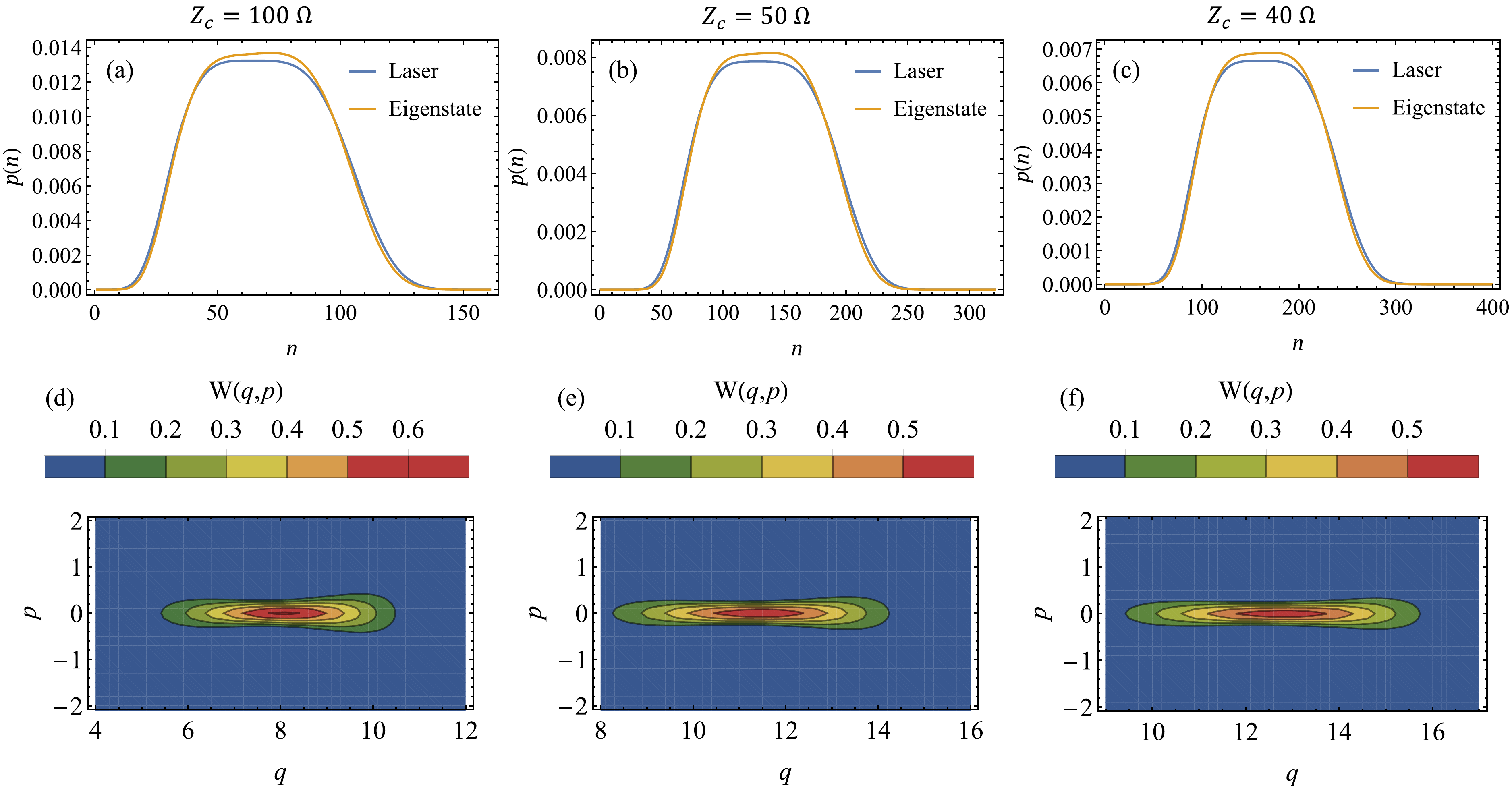}
    \caption{The photon distribution and the Wigner distribution of the corresponding eigenstates of the Approximately Bare Operator Coupling Circuit (ABOCC) operator when ABOCC laser is in flat-top states. In (a), (b) and (c), we plot the photon distribution for the flat-top ABOCC laser system and photon distribution of the corresponding best-matched ABOCC eigenstates. The Wigner distribution for the ABOCC eigenstates are plotted in (d), (e) and (f), respectively. For (a) and (d), $Z_c = 100~\Omega$, while (b) and (e), $Z_c = 50~\Omega$, and (c), (f) are for $Z_c = 40~\Omega$.  }
    \label{fig:diff_zc_FT}
\end{figure}

\begin{figure}[htbp]
    \centering
    \includegraphics[width =  \textwidth]{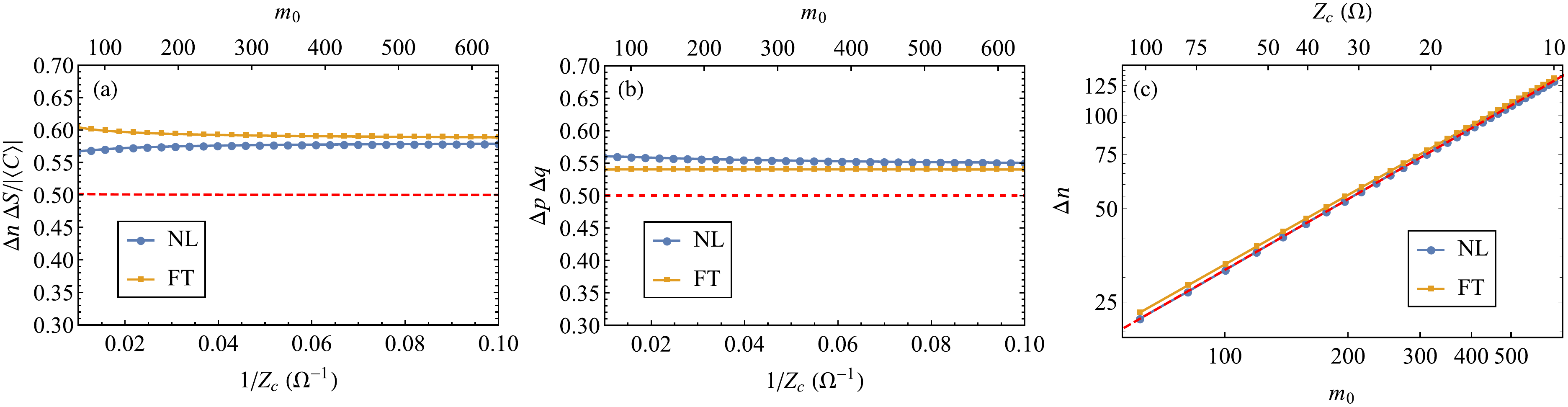}
    \caption{We plot the number-phase uncertainty relation ($\Delta n \Delta S / \vert \langle C \rangle \vert$) [in (a)] and the momentum-phase uncertainty relation $\Delta \hat{p} \Delta \hat{q}$ [in (b)] of the ABOCC operator eigenstates that best-matched to the narrowest-linewidth laser states (labeled as ``NL'') and the flat-top photon distribution (labeled as ``FT''), as we increase the $m_0$ of the ABOCC operator. The dashed lines show the minimum uncertainty relation. In (c), we calculate the photon number noise in the corresponding eigenstates of the ABOCC lowering operators as we increase the $m_0$ value. The dashed line shows the linear fit in the log-log scale, which gives a scaling parameter $0.761$.}
    \label{fig:NL_noise_sweep}
\end{figure}

In Fig.~\ref{fig:diff_zc_NL} and Fig.~\ref{fig:diff_zc_FT}, we show a few more states with different ABOCC operators. The target photon number $m_0$ of the ABOCC operators can be controlled by tuning the cavity effective impedance $Z_c$ (also see Fig.~4a in main text). In Fig.~\ref{fig:diff_zc_NL} and Fig.~\ref{fig:diff_zc_FT}, we consider three ABOCC operators: the ABOCC operators with cavity impedance $Z_c = 100~\Omega$ ($m_0 = 64$, Fig.~\ref{fig:diff_zc_NL}a and~d, Fig.~\ref{fig:diff_zc_FT}a and~d), with cavity impedance $Z_c = 50~\Omega$ ($m_0 = 128$, Fig.~\ref{fig:diff_zc_NL}b and~e, Fig.~\ref{fig:diff_zc_FT}b and~e) and with cavity impedance $Z_c = 40~\Omega$ ($m_0 = 159$, Fig.~\ref{fig:diff_zc_NL}c and~f, Fig.~\ref{fig:diff_zc_FT}c and~f). In each case we find the eigenstate of the corresponding ABOCC operator that best matches the photon number distribution of the narrowest linewidth laser state. The photon distribution of the laser states and the ABOCC eigenstates are shown in Fig.~\ref{fig:diff_zc_NL}a-c and the Wigner distribution of the corresponding ABOCC eigenstates are shown in Fig.~\ref{fig:diff_zc_NL}d-f. As the mean photon number increases with decreasing cavity impedance, the Wigner distribution width drops along the $\hat{p}$ direction, while the width along the $\hat{q}$ direction increases. This indicates that photon number noise increases while the phase noise drops. Similarly, we also calculate the ABOCC eigenstates that best match the flat-top laser states in Fig.~\ref{fig:diff_zc_FT}. We again focus on these three cavity impedance $Z_c = 100~\Omega$, $50~\Omega$ and $40~\Omega$. In this case, we observe the same trend: the Wigner distributions of the ABOCC lowering operator eigenstates become wider in the $\hat{q}$ direction and narrower along the $\hat{p}$ direction as we increase $m_0$. 

To systematically investigate the noise and the squeezing along the phase direction of the ABOCC eigenstates, we sweep the eigenstates of the ABOCC operators with different $m_0$, and calculate the number-phase uncertainty ($\Delta n \Delta S / \vert \langle \hat{C} \rangle \vert$) and the momentum-position uncertainty relation $\Delta \hat{p} \Delta \hat{q}$ in Fig.~\ref{fig:NL_noise_sweep}a and Fig.~\ref{fig:NL_noise_sweep}b. For each point, we use the ABOCC operator with the desired $m_0$ value to construct the ABOCC laser system. We solve the laser state that gives the best linewidth performance (labeled as ``NL'') and the flat-top photon distribution (labeled as ``FT''). Then we solve the ABOCC operator eigenstates and fit the eigenvalues by minimizing the cost function in Eq.~\eqref{eq:cost_fun}. With the best matched eigenstates of the ABOCC operator, we calculate the phase noise and the photon number noise, then the photon-phase uncertainty relation (in Fig.~\ref{fig:NL_noise_sweep}a) and the momentum-position uncertainty  in (in Fig.~\ref{fig:NL_noise_sweep}b). Notice that the dashed red lines in both plots are the quantum limit ($0.5$). The eigenstates of the ABOCC lowering operators under investigation are not minimum-uncertainty states. However, the uncertainty relation of both NL case and the FT cases change only slightly within the range shown in Fig.~\ref{fig:NL_noise_sweep}a and~b. Next, we consider the photon number noise for both the NL and the FT cases. Notice that the photon noise ($\Delta n$) increases as we increase $m_0$ (see Fig.~\ref{fig:NL_noise_sweep}c). We fit the scaling parameter and the dependence on the photon number is $\Delta n \sim m_0^{0.761}$ (see red dashed line in Fig.~\ref{fig:NL_noise_sweep}c). This scaling performance can be analyzed using a hypothetical ABOCC operator with the first diagonal elements in Fock basis satisfies $f(n) = 1 + \epsilon (n-m_0)^3 $ and minimize the laser performance by also tuning the small parameter $\epsilon$.~\footnote{This analysis gives a scaling parameter $3/4$.} Further, we can estimate the scaling parameter and get the relation $\Delta n \sim \langle n \rangle^{0.820}$. As the photon number-phase uncertainty relation only change slightly as the $m_0$ value is changed, the phase noise $\Delta S /\vert \langle C \rangle \vert \sim m_0 ^{-0.761} \sim \langle n \rangle^{-0.820}$. Notice that for a coherent state, the photon number noise is $\Delta n = \langle n \rangle^{0.5}$, so the corresponding phase noise should be $\sim \langle n \rangle^{-0.5}$. The eigenstates of our proposed ABOCC lowering operator will have a larger photon number noise, but squeezed along the phase direction. As we increase the $m_0$ value, the phase noise is further squeezed.

\begin{figure}[h]
    \centering
    \includegraphics[width = 3 in]{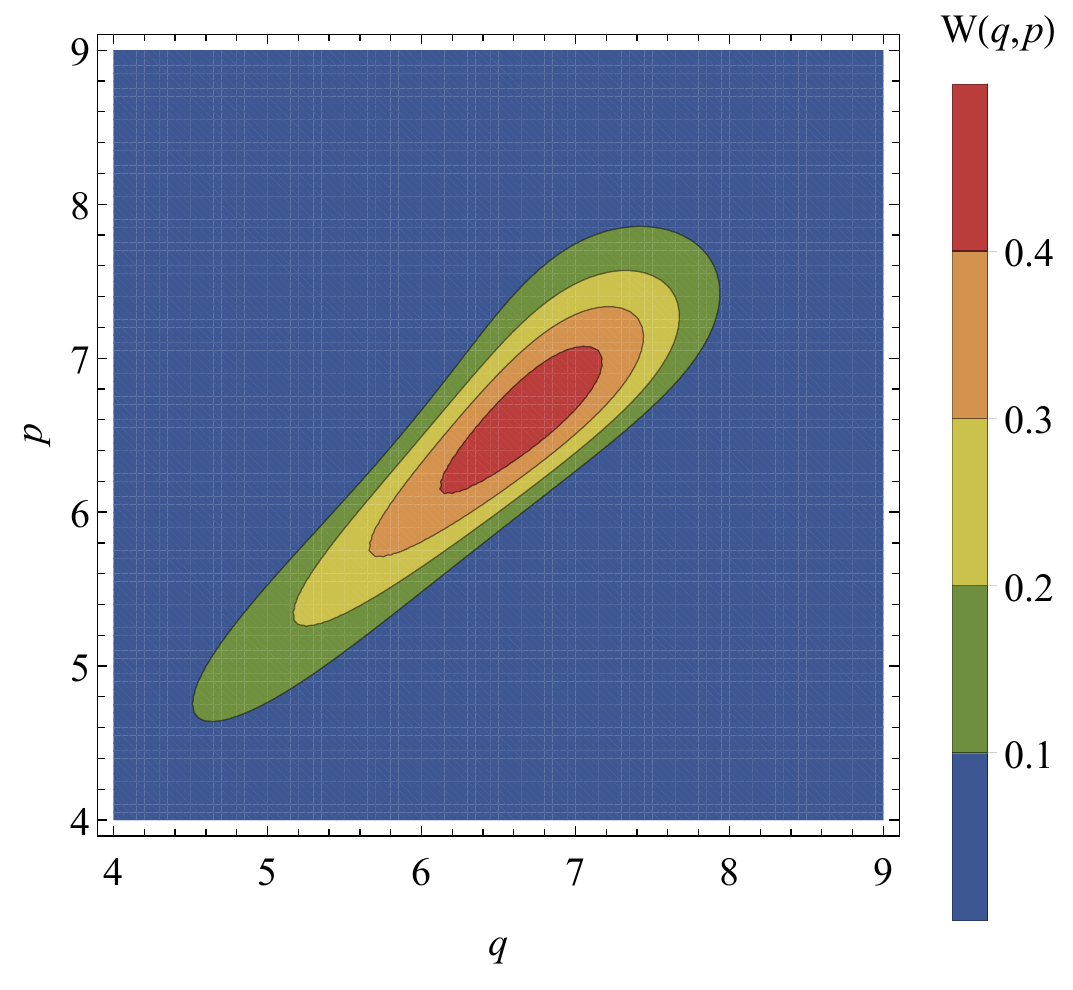}
    \caption{The Wigner distribution of the eigenstates of ABOCC operator with $Z_c = 100~\Omega$ ($m_0 = 64$) that best fit the photon distribution of the ABOCC laser state. The laser state has the optimum linewidth performance. The eigenvalue are picked to have $\pi/4$ phase, which rotate the Wigner distribution by $\pi/4$ in the phase space.}
    \label{fig:W_rotate}
\end{figure}

As a final sanity check, we verify that the squeezing direction of the eigenstates of the ABOCC lowering operator is always along the phase direction (rather than the $\hat{p}$ direction). To verify this, we construct an eigenstate with an eigenvalue $\alpha \rightarrow \alpha e^{i \theta}$. We expect that the resulting Wigner distribution is rotated around the origin $\psi(n) \rightarrow \psi(n) e^{i n \theta}$. In Fig.~\ref{fig:W_rotate}, we plot the Wigner distribution of the eigenstates of ABOCC lowering operator with $Z_c = 100~\Omega$ ($m_0 = 64$) with $\theta=\pi/4$. We observe that the Wigner distribution indeed rotated by $\pi/4$, as has the direction of squeezing, as compared to Fig.~\ref{fig:diff_zc_NL}d. 

\vspace{2 cm}

\newpage

\def\bibsection{
\section*{Supplementary References}
}
\bibliography{laser_ref}

\begin{thebibliography}{59}%
\makeatletter
\providecommand \@ifxundefined [1]{%
 \@ifx{#1\undefined}
}%
\providecommand \@ifnum [1]{%
 \ifnum #1\expandafter \@firstoftwo
 \else \expandafter \@secondoftwo
 \fi
}%
\providecommand \@ifx [1]{%
 \ifx #1\expandafter \@firstoftwo
 \else \expandafter \@secondoftwo
 \fi
}%
\providecommand \natexlab [1]{#1}%
\providecommand \enquote  [1]{``#1''}%
\providecommand \bibnamefont  [1]{#1}%
\providecommand \bibfnamefont [1]{#1}%
\providecommand \citenamefont [1]{#1}%
\providecommand \href@noop [0]{\@secondoftwo}%
\providecommand \href [0]{\begingroup \@sanitize@url \@href}%
\providecommand \@href[1]{\@@startlink{#1}\@@href}%
\providecommand \@@href[1]{\endgroup#1\@@endlink}%
\providecommand \@sanitize@url [0]{\catcode `\\12\catcode `\$12\catcode
  `\&12\catcode `\#12\catcode `\^12\catcode `\_12\catcode `\%12\relax}%
\providecommand \@@startlink[1]{}%
\providecommand \@@endlink[0]{}%
\providecommand \url  [0]{\begingroup\@sanitize@url \@url }%
\providecommand \@url [1]{\endgroup\@href {#1}{\urlprefix }}%
\providecommand \urlprefix  [0]{URL }%
\providecommand \Eprint [0]{\href }%
\providecommand \doibase [0]{http://dx.doi.org/}%
\providecommand \selectlanguage [0]{\@gobble}%
\providecommand \bibinfo  [0]{\@secondoftwo}%
\providecommand \bibfield  [0]{\@secondoftwo}%
\providecommand \translation [1]{[#1]}%
\providecommand \BibitemOpen [0]{}%
\providecommand \bibitemStop [0]{}%
\providecommand \bibitemNoStop [0]{.\EOS\space}%
\providecommand \EOS [0]{\spacefactor3000\relax}%
\providecommand \BibitemShut  [1]{\csname bibitem#1\endcsname}%
\let\auto@bib@innerbib\@empty
\bibitem [{\citenamefont {Schawlow}\ and\ \citenamefont
  {Townes}(1958)}]{Schawlow1958}%
  \BibitemOpen
  \bibfield  {author} {\bibinfo {author} {\bibfnamefont {A.~L.}\ \bibnamefont
  {Schawlow}}\ and\ \bibinfo {author} {\bibfnamefont {C.~H.}\ \bibnamefont
  {Townes}},\ }\bibfield  {title} {\enquote {\bibinfo {title} {Infrared and
  optical masers},}\ }\href {\doibase 10.1103/PhysRev.112.1940} {\bibfield
  {journal} {\bibinfo  {journal} {Phys. Rev.}\ }\textbf {\bibinfo {volume}
  {112}},\ \bibinfo {pages} {1940--1949} (\bibinfo {year} {1958})}\BibitemShut
  {NoStop}%
\bibitem [{\citenamefont {Wiseman}(1999)}]{Wiseman1999}%
  \BibitemOpen
  \bibfield  {author} {\bibinfo {author} {\bibfnamefont {H.~M.}\ \bibnamefont
  {Wiseman}},\ }\bibfield  {title} {\enquote {\bibinfo {title} {Light
  amplification without stimulated emission: Beyond the standard quantum limit
  to the laser linewidth},}\ }\href {\doibase 10.1103/PhysRevA.60.4083}
  {\bibfield  {journal} {\bibinfo  {journal} {Phys. Rev. A}\ }\textbf {\bibinfo
  {volume} {60}},\ \bibinfo {pages} {4083--4093} (\bibinfo {year}
  {1999})}\BibitemShut {NoStop}%
\bibitem [{\citenamefont {Bartalini}\ \emph {et~al.}(2010)\citenamefont
  {Bartalini}, \citenamefont {Borri}, \citenamefont {Cancio}, \citenamefont
  {Castrillo}, \citenamefont {Galli}, \citenamefont {Giusfredi}, \citenamefont
  {Mazzotti}, \citenamefont {Gianfrani},\ and\ \citenamefont
  {De~Natale}}]{Bartalini2010}%
  \BibitemOpen
  \bibfield  {author} {\bibinfo {author} {\bibfnamefont {S.}~\bibnamefont
  {Bartalini}}, \bibinfo {author} {\bibfnamefont {S.}~\bibnamefont {Borri}},
  \bibinfo {author} {\bibfnamefont {P.}~\bibnamefont {Cancio}}, \bibinfo
  {author} {\bibfnamefont {A.}~\bibnamefont {Castrillo}}, \bibinfo {author}
  {\bibfnamefont {I.}~\bibnamefont {Galli}}, \bibinfo {author} {\bibfnamefont
  {G.}~\bibnamefont {Giusfredi}}, \bibinfo {author} {\bibfnamefont
  {D.}~\bibnamefont {Mazzotti}}, \bibinfo {author} {\bibfnamefont
  {L.}~\bibnamefont {Gianfrani}}, \ and\ \bibinfo {author} {\bibfnamefont
  {P.}~\bibnamefont {De~Natale}},\ }\bibfield  {title} {\enquote {\bibinfo
  {title} {Observing the intrinsic linewidth of a quantum-cascade laser: Beyond
  the schawlow-townes limit},}\ }\href {\doibase
  10.1103/PhysRevLett.104.083904} {\bibfield  {journal} {\bibinfo  {journal}
  {Phys. Rev. Lett.}\ }\textbf {\bibinfo {volume} {104}},\ \bibinfo {pages}
  {083904} (\bibinfo {year} {2010})}\BibitemShut {NoStop}%
\bibitem [{\citenamefont {Bohnet}\ \emph {et~al.}(2012)\citenamefont {Bohnet},
  \citenamefont {Chen}, \citenamefont {Weiner}, \citenamefont {Meiser},
  \citenamefont {Holland},\ and\ \citenamefont {Thompson}}]{Bohnet2012}%
  \BibitemOpen
  \bibfield  {author} {\bibinfo {author} {\bibfnamefont {Justin~G}\
  \bibnamefont {Bohnet}}, \bibinfo {author} {\bibfnamefont {Zilong}\
  \bibnamefont {Chen}}, \bibinfo {author} {\bibfnamefont {Joshua~M}\
  \bibnamefont {Weiner}}, \bibinfo {author} {\bibfnamefont {Dominic}\
  \bibnamefont {Meiser}}, \bibinfo {author} {\bibfnamefont {Murray~J}\
  \bibnamefont {Holland}}, \ and\ \bibinfo {author} {\bibfnamefont {James~K}\
  \bibnamefont {Thompson}},\ }\bibfield  {title} {\enquote {\bibinfo {title} {A
  steady-state superradiant laser with less than one intracavity photon},}\
  }\href@noop {} {\bibfield  {journal} {\bibinfo  {journal} {Nature}\ }\textbf
  {\bibinfo {volume} {484}},\ \bibinfo {pages} {78--81} (\bibinfo {year}
  {2012})}\BibitemShut {NoStop}%
\bibitem [{\citenamefont {Yamamoto}\ \emph {et~al.}(1986)\citenamefont
  {Yamamoto}, \citenamefont {Machida},\ and\ \citenamefont
  {Nilsson}}]{Yamamoto1986}%
  \BibitemOpen
  \bibfield  {author} {\bibinfo {author} {\bibfnamefont {Y.}~\bibnamefont
  {Yamamoto}}, \bibinfo {author} {\bibfnamefont {S.}~\bibnamefont {Machida}}, \
  and\ \bibinfo {author} {\bibfnamefont {O.}~\bibnamefont {Nilsson}},\
  }\bibfield  {title} {\enquote {\bibinfo {title} {Amplitude squeezing in a
  pump-noise-suppressed laser oscillator},}\ }\href {\doibase
  10.1103/PhysRevA.34.4025} {\bibfield  {journal} {\bibinfo  {journal} {Phys.
  Rev. A}\ }\textbf {\bibinfo {volume} {34}},\ \bibinfo {pages} {4025--4042}
  (\bibinfo {year} {1986})}\BibitemShut {NoStop}%
\bibitem [{\citenamefont {Susskind}\ and\ \citenamefont
  {Glogower}(1964)}]{Susskind1964}%
  \BibitemOpen
  \bibfield  {author} {\bibinfo {author} {\bibfnamefont {Leonard}\ \bibnamefont
  {Susskind}}\ and\ \bibinfo {author} {\bibfnamefont {Jonathan}\ \bibnamefont
  {Glogower}},\ }\bibfield  {title} {\enquote {\bibinfo {title} {Quantum
  mechanical phase and time operator},}\ }\href {\doibase
  10.1103/PhysicsPhysiqueFizika.1.49} {\bibfield  {journal} {\bibinfo
  {journal} {Physics Physique Fizika}\ }\textbf {\bibinfo {volume} {1}},\
  \bibinfo {pages} {49--61} (\bibinfo {year} {1964})}\BibitemShut {NoStop}%
\bibitem [{\citenamefont {Barnett}\ \emph {et~al.}(1989)\citenamefont
  {Barnett}, \citenamefont {Stenholm},\ and\ \citenamefont
  {Pegg}}]{Barnet1989}%
  \BibitemOpen
  \bibfield  {author} {\bibinfo {author} {\bibfnamefont {S.M.}\ \bibnamefont
  {Barnett}}, \bibinfo {author} {\bibfnamefont {S.}~\bibnamefont {Stenholm}}, \
  and\ \bibinfo {author} {\bibfnamefont {D.T.}\ \bibnamefont {Pegg}},\
  }\bibfield  {title} {\enquote {\bibinfo {title} {A new approach to optical
  phase diffusion},}\ }\href {\doibase
  https://doi.org/10.1016/0030-4018(89)90224-1} {\bibfield  {journal} {\bibinfo
   {journal} {Optics Communications}\ }\textbf {\bibinfo {volume} {73}},\
  \bibinfo {pages} {314 -- 318} (\bibinfo {year} {1989})}\BibitemShut {NoStop}%
\bibitem [{\citenamefont {Kimble}(1998)}]{Kimble1998}%
  \BibitemOpen
  \bibfield  {author} {\bibinfo {author} {\bibfnamefont {H.~J.}\ \bibnamefont
  {Kimble}},\ }\bibfield  {title} {\enquote {\bibinfo {title} {Strong
  interactions of single atoms and photons in cavity {QED}},}\ }\href {\doibase
  10.1238/physica.topical.076a00127} {\bibfield  {journal} {\bibinfo  {journal}
  {Physica Scripta}\ }\textbf {\bibinfo {volume} {T76}},\ \bibinfo {pages}
  {127} (\bibinfo {year} {1998})}\BibitemShut {NoStop}%
\bibitem [{\citenamefont {Mabuchi}\ and\ \citenamefont
  {Doherty}(2002)}]{Mabuchi2002}%
  \BibitemOpen
  \bibfield  {author} {\bibinfo {author} {\bibfnamefont {H.}~\bibnamefont
  {Mabuchi}}\ and\ \bibinfo {author} {\bibfnamefont {A.~C.}\ \bibnamefont
  {Doherty}},\ }\bibfield  {title} {\enquote {\bibinfo {title} {Cavity quantum
  electrodynamics: Coherence in context},}\ }\href {\doibase
  10.1126/science.1078446} {\bibfield  {journal} {\bibinfo  {journal}
  {Science}\ }\textbf {\bibinfo {volume} {298}},\ \bibinfo {pages} {1372--1377}
  (\bibinfo {year} {2002})}\BibitemShut {NoStop}%
\bibitem [{\citenamefont {McKeever}\ \emph {et~al.}(2003)\citenamefont
  {McKeever}, \citenamefont {Boca}, \citenamefont {Boozer}, \citenamefont
  {Buck},\ and\ \citenamefont {Kimble}}]{McKeever2003}%
  \BibitemOpen
  \bibfield  {author} {\bibinfo {author} {\bibfnamefont {J.}~\bibnamefont
  {McKeever}}, \bibinfo {author} {\bibfnamefont {A.}~\bibnamefont {Boca}},
  \bibinfo {author} {\bibfnamefont {A.~D.}\ \bibnamefont {Boozer}}, \bibinfo
  {author} {\bibfnamefont {J.~R.}\ \bibnamefont {Buck}}, \ and\ \bibinfo
  {author} {\bibfnamefont {H.~J.}\ \bibnamefont {Kimble}},\ }\bibfield  {title}
  {\enquote {\bibinfo {title} {Experimental realization of a one-atom laser in
  the regime of strong coupling},}\ }\href {\doibase 10.1038/nature01974}
  {\bibfield  {journal} {\bibinfo  {journal} {Nature}\ }\textbf {\bibinfo
  {volume} {425}},\ \bibinfo {pages} {268--271} (\bibinfo {year}
  {2003})}\BibitemShut {NoStop}%
\bibitem [{\citenamefont {Walther}\ \emph {et~al.}(2006)\citenamefont
  {Walther}, \citenamefont {Varcoe}, \citenamefont {Englert},\ and\
  \citenamefont {Becker}}]{Walther2006}%
  \BibitemOpen
  \bibfield  {author} {\bibinfo {author} {\bibfnamefont {Herbert}\ \bibnamefont
  {Walther}}, \bibinfo {author} {\bibfnamefont {Benjamin~TH}\ \bibnamefont
  {Varcoe}}, \bibinfo {author} {\bibfnamefont {Berthold-Georg}\ \bibnamefont
  {Englert}}, \ and\ \bibinfo {author} {\bibfnamefont {Thomas}\ \bibnamefont
  {Becker}},\ }\bibfield  {title} {\enquote {\bibinfo {title} {Cavity quantum
  electrodynamics},}\ }\href@noop {} {\bibfield  {journal} {\bibinfo  {journal}
  {Reports on Progress in Physics}\ }\textbf {\bibinfo {volume} {69}},\
  \bibinfo {pages} {1325} (\bibinfo {year} {2006})}\BibitemShut {NoStop}%
\bibitem [{\citenamefont {Girvin}(2014)}]{Girvin2014}%
  \BibitemOpen
  \bibfield  {author} {\bibinfo {author} {\bibfnamefont {S.~M.}\ \bibnamefont
  {Girvin}},\ }\href
  {https://www.oxfordscholarship.com/view/10.1093/acprof:oso/9780199681181.001.0001/acprof-9780199681181-chapter-3}
  {\emph {\bibinfo {title} {Circuit {QED}: superconducting qubits coupled to
  microwave photons}}}\ (\bibinfo  {publisher} {Oxford University Press},\
  \bibinfo {year} {2014})\BibitemShut {NoStop}%
\bibitem [{\citenamefont {Manucharyan}\ \emph {et~al.}(2009)\citenamefont
  {Manucharyan}, \citenamefont {Koch}, \citenamefont {Glazman},\ and\
  \citenamefont {Devoret}}]{Manucharyan2009}%
  \BibitemOpen
  \bibfield  {author} {\bibinfo {author} {\bibfnamefont {Vladimir~E.}\
  \bibnamefont {Manucharyan}}, \bibinfo {author} {\bibfnamefont {Jens}\
  \bibnamefont {Koch}}, \bibinfo {author} {\bibfnamefont {Leonid~I.}\
  \bibnamefont {Glazman}}, \ and\ \bibinfo {author} {\bibfnamefont {Michel~H.}\
  \bibnamefont {Devoret}},\ }\bibfield  {title} {\enquote {\bibinfo {title}
  {Fluxonium: Single cooper-pair circuit free of charge offsets},}\ }\href
  {\doibase 10.1126/science.1175552} {\bibfield  {journal} {\bibinfo  {journal}
  {Science}\ }\textbf {\bibinfo {volume} {326}},\ \bibinfo {pages} {113--116}
  (\bibinfo {year} {2009})}\BibitemShut {NoStop}%
\bibitem [{\citenamefont {Koch}\ \emph {et~al.}(2007)\citenamefont {Koch},
  \citenamefont {Yu}, \citenamefont {Gambetta}, \citenamefont {Houck},
  \citenamefont {Schuster}, \citenamefont {Majer}, \citenamefont {Blais},
  \citenamefont {Devoret}, \citenamefont {Girvin},\ and\ \citenamefont
  {Schoelkopf}}]{Koch2007}%
  \BibitemOpen
  \bibfield  {author} {\bibinfo {author} {\bibfnamefont {Jens}\ \bibnamefont
  {Koch}}, \bibinfo {author} {\bibfnamefont {Terri~M.}\ \bibnamefont {Yu}},
  \bibinfo {author} {\bibfnamefont {Jay}\ \bibnamefont {Gambetta}}, \bibinfo
  {author} {\bibfnamefont {A.~A.}\ \bibnamefont {Houck}}, \bibinfo {author}
  {\bibfnamefont {D.~I.}\ \bibnamefont {Schuster}}, \bibinfo {author}
  {\bibfnamefont {J.}~\bibnamefont {Majer}}, \bibinfo {author} {\bibfnamefont
  {Alexandre}\ \bibnamefont {Blais}}, \bibinfo {author} {\bibfnamefont {M.~H.}\
  \bibnamefont {Devoret}}, \bibinfo {author} {\bibfnamefont {S.~M.}\
  \bibnamefont {Girvin}}, \ and\ \bibinfo {author} {\bibfnamefont {R.~J.}\
  \bibnamefont {Schoelkopf}},\ }\bibfield  {title} {\enquote {\bibinfo {title}
  {Charge-insensitive qubit design derived from the cooper pair box},}\ }\href
  {\doibase 10.1103/PhysRevA.76.042319} {\bibfield  {journal} {\bibinfo
  {journal} {Phys. Rev. A}\ }\textbf {\bibinfo {volume} {76}},\ \bibinfo
  {pages} {042319} (\bibinfo {year} {2007})}\BibitemShut {NoStop}%
\bibitem [{\citenamefont {Schreier}\ \emph {et~al.}(2008)\citenamefont
  {Schreier}, \citenamefont {Houck}, \citenamefont {Koch}, \citenamefont
  {Schuster}, \citenamefont {Johnson}, \citenamefont {Chow}, \citenamefont
  {Gambetta}, \citenamefont {Majer}, \citenamefont {Frunzio}, \citenamefont
  {Devoret}, \citenamefont {Girvin},\ and\ \citenamefont
  {Schoelkopf}}]{Schreier2008}%
  \BibitemOpen
  \bibfield  {author} {\bibinfo {author} {\bibfnamefont {J.~A.}\ \bibnamefont
  {Schreier}}, \bibinfo {author} {\bibfnamefont {A.~A.}\ \bibnamefont {Houck}},
  \bibinfo {author} {\bibfnamefont {Jens}\ \bibnamefont {Koch}}, \bibinfo
  {author} {\bibfnamefont {D.~I.}\ \bibnamefont {Schuster}}, \bibinfo {author}
  {\bibfnamefont {B.~R.}\ \bibnamefont {Johnson}}, \bibinfo {author}
  {\bibfnamefont {J.~M.}\ \bibnamefont {Chow}}, \bibinfo {author}
  {\bibfnamefont {J.~M.}\ \bibnamefont {Gambetta}}, \bibinfo {author}
  {\bibfnamefont {J.}~\bibnamefont {Majer}}, \bibinfo {author} {\bibfnamefont
  {L.}~\bibnamefont {Frunzio}}, \bibinfo {author} {\bibfnamefont {M.~H.}\
  \bibnamefont {Devoret}}, \bibinfo {author} {\bibfnamefont {S.~M.}\
  \bibnamefont {Girvin}}, \ and\ \bibinfo {author} {\bibfnamefont {R.~J.}\
  \bibnamefont {Schoelkopf}},\ }\bibfield  {title} {\enquote {\bibinfo {title}
  {Suppressing charge noise decoherence in superconducting charge qubits},}\
  }\href {\doibase 10.1103/PhysRevB.77.180502} {\bibfield  {journal} {\bibinfo
  {journal} {Phys. Rev. B}\ }\textbf {\bibinfo {volume} {77}},\ \bibinfo
  {pages} {180502} (\bibinfo {year} {2008})}\BibitemShut {NoStop}%
\bibitem [{\citenamefont {Clerk}\ \emph {et~al.}(2010)\citenamefont {Clerk},
  \citenamefont {Devoret}, \citenamefont {Girvin}, \citenamefont {Marquardt},\
  and\ \citenamefont {Schoelkopf}}]{ClerkRMP2008}%
  \BibitemOpen
  \bibfield  {author} {\bibinfo {author} {\bibfnamefont {A.~A.}\ \bibnamefont
  {Clerk}}, \bibinfo {author} {\bibfnamefont {M.~H.}\ \bibnamefont {Devoret}},
  \bibinfo {author} {\bibfnamefont {S.~M.}\ \bibnamefont {Girvin}}, \bibinfo
  {author} {\bibfnamefont {Florian}\ \bibnamefont {Marquardt}}, \ and\ \bibinfo
  {author} {\bibfnamefont {R.~J.}\ \bibnamefont {Schoelkopf}},\ }\bibfield
  {title} {\enquote {\bibinfo {title} {Introduction to quantum noise,
  measurement, and amplification},}\ }\href {\doibase
  10.1103/RevModPhys.82.1155} {\bibfield  {journal} {\bibinfo  {journal} {Rev.
  Mod. Phys.}\ }\textbf {\bibinfo {volume} {82}},\ \bibinfo {pages}
  {1155--1208} (\bibinfo {year} {2010})}\BibitemShut {NoStop}%
\bibitem [{\citenamefont {Bergeal}\ \emph {et~al.}(2010)\citenamefont
  {Bergeal}, \citenamefont {Schackert}, \citenamefont {Metcalfe}, \citenamefont
  {Vijay}, \citenamefont {Manucharyan}, \citenamefont {Frunzio}, \citenamefont
  {Prober}, \citenamefont {Schoelkopf}, \citenamefont {Girvin},\ and\
  \citenamefont {Devoret}}]{Bergeal2010}%
  \BibitemOpen
  \bibfield  {author} {\bibinfo {author} {\bibfnamefont {N}~\bibnamefont
  {Bergeal}}, \bibinfo {author} {\bibfnamefont {F}~\bibnamefont {Schackert}},
  \bibinfo {author} {\bibfnamefont {M}~\bibnamefont {Metcalfe}}, \bibinfo
  {author} {\bibfnamefont {R}~\bibnamefont {Vijay}}, \bibinfo {author}
  {\bibfnamefont {VE}~\bibnamefont {Manucharyan}}, \bibinfo {author}
  {\bibfnamefont {L}~\bibnamefont {Frunzio}}, \bibinfo {author} {\bibfnamefont
  {DE}~\bibnamefont {Prober}}, \bibinfo {author} {\bibfnamefont
  {RJ}~\bibnamefont {Schoelkopf}}, \bibinfo {author} {\bibfnamefont
  {SM}~\bibnamefont {Girvin}}, \ and\ \bibinfo {author} {\bibfnamefont
  {MH}~\bibnamefont {Devoret}},\ }\bibfield  {title} {\enquote {\bibinfo
  {title} {Phase-preserving amplification near the quantum limit with a
  josephson ring modulator},}\ }\href@noop {} {\bibfield  {journal} {\bibinfo
  {journal} {Nature}\ }\textbf {\bibinfo {volume} {465}},\ \bibinfo {pages}
  {64--68} (\bibinfo {year} {2010})}\BibitemShut {NoStop}%
\bibitem [{\citenamefont {Astafiev}\ \emph {et~al.}(2007)\citenamefont
  {Astafiev}, \citenamefont {Inomata}, \citenamefont {Niskanen}, \citenamefont
  {Yamamoto}, \citenamefont {Pashkin}, \citenamefont {Nakamura},\ and\
  \citenamefont {Tsai}}]{Astafiev2007}%
  \BibitemOpen
  \bibfield  {author} {\bibinfo {author} {\bibfnamefont {O.}~\bibnamefont
  {Astafiev}}, \bibinfo {author} {\bibfnamefont {K.}~\bibnamefont {Inomata}},
  \bibinfo {author} {\bibfnamefont {A.~O.}\ \bibnamefont {Niskanen}}, \bibinfo
  {author} {\bibfnamefont {T.}~\bibnamefont {Yamamoto}}, \bibinfo {author}
  {\bibfnamefont {Yu.~A.}\ \bibnamefont {Pashkin}}, \bibinfo {author}
  {\bibfnamefont {Y.}~\bibnamefont {Nakamura}}, \ and\ \bibinfo {author}
  {\bibfnamefont {J.~S.}\ \bibnamefont {Tsai}},\ }\bibfield  {title} {\enquote
  {\bibinfo {title} {Single artificial-atom lasing},}\ }\href {\doibase
  10.1038/nature06141} {\bibfield  {journal} {\bibinfo  {journal} {Nature}\
  }\textbf {\bibinfo {volume} {449}},\ \bibinfo {pages} {588--590} (\bibinfo
  {year} {2007})}\BibitemShut {NoStop}%
\bibitem [{\citenamefont {Chen}\ \emph {et~al.}(2014)\citenamefont {Chen},
  \citenamefont {Li}, \citenamefont {Armour}, \citenamefont {Brahimi},
  \citenamefont {Stettenheim}, \citenamefont {Sirois}, \citenamefont
  {Simmonds}, \citenamefont {Blencowe},\ and\ \citenamefont
  {Rimberg}}]{Chen2014}%
  \BibitemOpen
  \bibfield  {author} {\bibinfo {author} {\bibfnamefont {Fei}\ \bibnamefont
  {Chen}}, \bibinfo {author} {\bibfnamefont {Juliang}\ \bibnamefont {Li}},
  \bibinfo {author} {\bibfnamefont {A.~D.}\ \bibnamefont {Armour}}, \bibinfo
  {author} {\bibfnamefont {E.}~\bibnamefont {Brahimi}}, \bibinfo {author}
  {\bibfnamefont {Joel}\ \bibnamefont {Stettenheim}}, \bibinfo {author}
  {\bibfnamefont {A.~J.}\ \bibnamefont {Sirois}}, \bibinfo {author}
  {\bibfnamefont {R.~W.}\ \bibnamefont {Simmonds}}, \bibinfo {author}
  {\bibfnamefont {M.~P.}\ \bibnamefont {Blencowe}}, \ and\ \bibinfo {author}
  {\bibfnamefont {A.~J.}\ \bibnamefont {Rimberg}},\ }\bibfield  {title}
  {\enquote {\bibinfo {title} {Realization of a single-cooper-pair josephson
  laser},}\ }\href {\doibase 10.1103/PhysRevB.90.020506} {\bibfield  {journal}
  {\bibinfo  {journal} {Phys. Rev. B}\ }\textbf {\bibinfo {volume} {90}},\
  \bibinfo {pages} {020506} (\bibinfo {year} {2014})}\BibitemShut {NoStop}%
\bibitem [{\citenamefont {Rolland}\ \emph {et~al.}(2019)\citenamefont
  {Rolland}, \citenamefont {Peugeot}, \citenamefont {Dambach}, \citenamefont
  {Westig}, \citenamefont {Kubala}, \citenamefont {Mukharsky}, \citenamefont
  {Altimiras}, \citenamefont {le~Sueur}, \citenamefont {Joyez}, \citenamefont
  {Vion}, \citenamefont {Roche}, \citenamefont {Esteve}, \citenamefont
  {Ankerhold},\ and\ \citenamefont {Portier}}]{Rolland2019}%
  \BibitemOpen
  \bibfield  {author} {\bibinfo {author} {\bibfnamefont {C.}~\bibnamefont
  {Rolland}}, \bibinfo {author} {\bibfnamefont {A.}~\bibnamefont {Peugeot}},
  \bibinfo {author} {\bibfnamefont {S.}~\bibnamefont {Dambach}}, \bibinfo
  {author} {\bibfnamefont {M.}~\bibnamefont {Westig}}, \bibinfo {author}
  {\bibfnamefont {B.}~\bibnamefont {Kubala}}, \bibinfo {author} {\bibfnamefont
  {Y.}~\bibnamefont {Mukharsky}}, \bibinfo {author} {\bibfnamefont
  {C.}~\bibnamefont {Altimiras}}, \bibinfo {author} {\bibfnamefont
  {H.}~\bibnamefont {le~Sueur}}, \bibinfo {author} {\bibfnamefont
  {P.}~\bibnamefont {Joyez}}, \bibinfo {author} {\bibfnamefont
  {D.}~\bibnamefont {Vion}}, \bibinfo {author} {\bibfnamefont {P.}~\bibnamefont
  {Roche}}, \bibinfo {author} {\bibfnamefont {D.}~\bibnamefont {Esteve}},
  \bibinfo {author} {\bibfnamefont {J.}~\bibnamefont {Ankerhold}}, \ and\
  \bibinfo {author} {\bibfnamefont {F.}~\bibnamefont {Portier}},\ }\bibfield
  {title} {\enquote {\bibinfo {title} {Antibunched photons emitted by a
  dc-biased josephson junction},}\ }\href {\doibase
  10.1103/PhysRevLett.122.186804} {\bibfield  {journal} {\bibinfo  {journal}
  {Phys. Rev. Lett.}\ }\textbf {\bibinfo {volume} {122}},\ \bibinfo {pages}
  {186804} (\bibinfo {year} {2019})}\BibitemShut {NoStop}%
\bibitem [{\citenamefont {Cassidy}\ \emph {et~al.}(2017)\citenamefont
  {Cassidy}, \citenamefont {Bruno}, \citenamefont {Rubbert}, \citenamefont
  {Irfan}, \citenamefont {Kammhuber}, \citenamefont {Schouten}, \citenamefont
  {Akhmerov},\ and\ \citenamefont {Kouwenhoven}}]{Cassidy2017}%
  \BibitemOpen
  \bibfield  {author} {\bibinfo {author} {\bibfnamefont {M.~C.}\ \bibnamefont
  {Cassidy}}, \bibinfo {author} {\bibfnamefont {A.}~\bibnamefont {Bruno}},
  \bibinfo {author} {\bibfnamefont {S.}~\bibnamefont {Rubbert}}, \bibinfo
  {author} {\bibfnamefont {M.}~\bibnamefont {Irfan}}, \bibinfo {author}
  {\bibfnamefont {J.}~\bibnamefont {Kammhuber}}, \bibinfo {author}
  {\bibfnamefont {R.~N.}\ \bibnamefont {Schouten}}, \bibinfo {author}
  {\bibfnamefont {A.~R.}\ \bibnamefont {Akhmerov}}, \ and\ \bibinfo {author}
  {\bibfnamefont {L.~P.}\ \bibnamefont {Kouwenhoven}},\ }\bibfield  {title}
  {\enquote {\bibinfo {title} {Demonstration of an ac josephson junction
  laser},}\ }\href {\doibase 10.1126/science.aah6640} {\bibfield  {journal}
  {\bibinfo  {journal} {Science}\ }\textbf {\bibinfo {volume} {355}},\ \bibinfo
  {pages} {939--942} (\bibinfo {year} {2017})},\ \Eprint
  {http://arxiv.org/abs/https://science.sciencemag.org/content/355/6328/939.full.pdf}
  {https://science.sciencemag.org/content/355/6328/939.full.pdf} \BibitemShut
  {NoStop}%
\bibitem [{\citenamefont {Simon}\ and\ \citenamefont
  {Cooper}(2018)}]{Simon2018}%
  \BibitemOpen
  \bibfield  {author} {\bibinfo {author} {\bibfnamefont {Steven~H.}\
  \bibnamefont {Simon}}\ and\ \bibinfo {author} {\bibfnamefont {Nigel~R.}\
  \bibnamefont {Cooper}},\ }\bibfield  {title} {\enquote {\bibinfo {title}
  {Theory of the josephson junction laser},}\ }\href {\doibase
  10.1103/PhysRevLett.121.027004} {\bibfield  {journal} {\bibinfo  {journal}
  {Phys. Rev. Lett.}\ }\textbf {\bibinfo {volume} {121}},\ \bibinfo {pages}
  {027004} (\bibinfo {year} {2018})}\BibitemShut {NoStop}%
\bibitem [{\citenamefont {Baker}\ \emph {et~al.}(2020)\citenamefont {Baker},
  \citenamefont {Saadatmand}, \citenamefont {Berry},\ and\ \citenamefont
  {Wiseman}}]{Wiseman2020arXiv}%
  \BibitemOpen
  \bibfield  {author} {\bibinfo {author} {\bibfnamefont {Travis~J.}\
  \bibnamefont {Baker}}, \bibinfo {author} {\bibfnamefont {Seyed~N.}\
  \bibnamefont {Saadatmand}}, \bibinfo {author} {\bibfnamefont {Dominic~W.}\
  \bibnamefont {Berry}}, \ and\ \bibinfo {author} {\bibfnamefont {Howard~M.}\
  \bibnamefont {Wiseman}},\ }\bibfield  {title} {\enquote {\bibinfo {title}
  {The heisenberg limit for laser coherence},}\ }\href {\doibase
  10.1038/s41567-020-01049-3} {\bibfield  {journal} {\bibinfo  {journal}
  {Nature Physics}\ } (\bibinfo {year} {2020}),\
  10.1038/s41567-020-01049-3}\BibitemShut {NoStop}%
\bibitem [{\citenamefont {Scully}\ and\ \citenamefont
  {Zubairy}(1997)}]{Scully1997}%
  \BibitemOpen
  \bibfield  {author} {\bibinfo {author} {\bibfnamefont {Marlan~O.}\
  \bibnamefont {Scully}}\ and\ \bibinfo {author} {\bibfnamefont {M.~Suhail}\
  \bibnamefont {Zubairy}},\ }\href {\doibase 10.1017/CBO9780511813993} {\emph
  {\bibinfo {title} {Quantum Optics}}}\ (\bibinfo  {publisher} {Cambridge
  University Press},\ \bibinfo {year} {1997})\BibitemShut {NoStop}%
\bibitem [{\citenamefont {Oi}\ \emph {et~al.}(2013)\citenamefont {Oi},
  \citenamefont {Poto\ifmmode~\check{c}\else \v{c}\fi{}ek},\ and\ \citenamefont
  {Jeffers}}]{Oi2013}%
  \BibitemOpen
  \bibfield  {author} {\bibinfo {author} {\bibfnamefont {Daniel K.~L.}\
  \bibnamefont {Oi}}, \bibinfo {author} {\bibfnamefont {V\'aclav}\ \bibnamefont
  {Poto\ifmmode~\check{c}\else \v{c}\fi{}ek}}, \ and\ \bibinfo {author}
  {\bibfnamefont {John}\ \bibnamefont {Jeffers}},\ }\bibfield  {title}
  {\enquote {\bibinfo {title} {Nondemolition measurement of the vacuum state or
  its complement},}\ }\href {\doibase 10.1103/PhysRevLett.110.210504}
  {\bibfield  {journal} {\bibinfo  {journal} {Phys. Rev. Lett.}\ }\textbf
  {\bibinfo {volume} {110}},\ \bibinfo {pages} {210504} (\bibinfo {year}
  {2013})}\BibitemShut {NoStop}%
\bibitem [{\citenamefont {Govia}\ \emph {et~al.}(2014)\citenamefont {Govia},
  \citenamefont {Pritchett},\ and\ \citenamefont {Wilhelm}}]{Govia2014}%
  \BibitemOpen
  \bibfield  {author} {\bibinfo {author} {\bibfnamefont {Luke C~G}\
  \bibnamefont {Govia}}, \bibinfo {author} {\bibfnamefont {Emily~J}\
  \bibnamefont {Pritchett}}, \ and\ \bibinfo {author} {\bibfnamefont {Frank~K}\
  \bibnamefont {Wilhelm}},\ }\bibfield  {title} {\enquote {\bibinfo {title}
  {Generating nonclassical states from classical radiation by subtraction
  measurements},}\ }\href {\doibase 10.1088/1367-2630/16/4/045011} {\bibfield
  {journal} {\bibinfo  {journal} {New Journal of Physics}\ }\textbf {\bibinfo
  {volume} {16}},\ \bibinfo {pages} {045011} (\bibinfo {year}
  {2014})}\BibitemShut {NoStop}%
\bibitem [{\citenamefont {Rosenblum}\ \emph {et~al.}(2016)\citenamefont
  {Rosenblum}, \citenamefont {Bechler}, \citenamefont {Shomroni}, \citenamefont
  {Lovsky}, \citenamefont {Guendelman},\ and\ \citenamefont
  {Dayan}}]{Rosenblum2016}%
  \BibitemOpen
  \bibfield  {author} {\bibinfo {author} {\bibfnamefont {Serge}\ \bibnamefont
  {Rosenblum}}, \bibinfo {author} {\bibfnamefont {Orel}\ \bibnamefont
  {Bechler}}, \bibinfo {author} {\bibfnamefont {Itay}\ \bibnamefont
  {Shomroni}}, \bibinfo {author} {\bibfnamefont {Yulia}\ \bibnamefont
  {Lovsky}}, \bibinfo {author} {\bibfnamefont {Gabriel}\ \bibnamefont
  {Guendelman}}, \ and\ \bibinfo {author} {\bibfnamefont {Barak}\ \bibnamefont
  {Dayan}},\ }\bibfield  {title} {\enquote {\bibinfo {title} {Extraction of a
  single photon from an optical pulse},}\ }\href {\doibase
  10.1038/nphoton.2015.227} {\bibfield  {journal} {\bibinfo  {journal} {Nature
  Photonics}\ }\textbf {\bibinfo {volume} {10}},\ \bibinfo {pages} {19--22}
  (\bibinfo {year} {2016})}\BibitemShut {NoStop}%
\bibitem [{\citenamefont {Um}\ \emph {et~al.}(2016)\citenamefont {Um},
  \citenamefont {Zhang}, \citenamefont {Lv}, \citenamefont {Lu}, \citenamefont
  {An}, \citenamefont {Zhang}, \citenamefont {Nha}, \citenamefont {Kim},\ and\
  \citenamefont {Kim}}]{Um2016}%
  \BibitemOpen
  \bibfield  {author} {\bibinfo {author} {\bibfnamefont {Mark}\ \bibnamefont
  {Um}}, \bibinfo {author} {\bibfnamefont {Junhua}\ \bibnamefont {Zhang}},
  \bibinfo {author} {\bibfnamefont {Dingshun}\ \bibnamefont {Lv}}, \bibinfo
  {author} {\bibfnamefont {Yao}\ \bibnamefont {Lu}}, \bibinfo {author}
  {\bibfnamefont {Shuoming}\ \bibnamefont {An}}, \bibinfo {author}
  {\bibfnamefont {Jing-Ning}\ \bibnamefont {Zhang}}, \bibinfo {author}
  {\bibfnamefont {Hyunchul}\ \bibnamefont {Nha}}, \bibinfo {author}
  {\bibfnamefont {M.~S.}\ \bibnamefont {Kim}}, \ and\ \bibinfo {author}
  {\bibfnamefont {Kihwan}\ \bibnamefont {Kim}},\ }\bibfield  {title} {\enquote
  {\bibinfo {title} {Phonon arithmetic in a trapped ion system},}\ }\href
  {\doibase 10.1038/ncomms11410} {\bibfield  {journal} {\bibinfo  {journal}
  {Nature Communications}\ }\textbf {\bibinfo {volume} {7}},\ \bibinfo {pages}
  {11410} (\bibinfo {year} {2016})}\BibitemShut {NoStop}%
\bibitem [{\citenamefont {Radtke}\ \emph {et~al.}(2017)\citenamefont {Radtke},
  \citenamefont {Oi},\ and\ \citenamefont {Jeffers}}]{Radtke2017}%
  \BibitemOpen
  \bibfield  {author} {\bibinfo {author} {\bibfnamefont {Jennifer C~J}\
  \bibnamefont {Radtke}}, \bibinfo {author} {\bibfnamefont {Daniel K~L}\
  \bibnamefont {Oi}}, \ and\ \bibinfo {author} {\bibfnamefont {John}\
  \bibnamefont {Jeffers}},\ }\bibfield  {title} {\enquote {\bibinfo {title}
  {Linear quantum optical bare raising operator},}\ }\href {\doibase
  10.1088/1361-6455/aa8e69} {\bibfield  {journal} {\bibinfo  {journal} {Journal
  of Physics B: Atomic, Molecular and Optical Physics}\ }\textbf {\bibinfo
  {volume} {50}},\ \bibinfo {pages} {215501} (\bibinfo {year}
  {2017})}\BibitemShut {NoStop}%
\bibitem [{\citenamefont {Ma}\ \emph {et~al.}(2019)\citenamefont {Ma},
  \citenamefont {Saxberg}, \citenamefont {Owens}, \citenamefont {Leung},
  \citenamefont {Lu}, \citenamefont {Simon},\ and\ \citenamefont
  {Schuster}}]{Ma2019}%
  \BibitemOpen
  \bibfield  {author} {\bibinfo {author} {\bibfnamefont {Ruichao}\ \bibnamefont
  {Ma}}, \bibinfo {author} {\bibfnamefont {Brendan}\ \bibnamefont {Saxberg}},
  \bibinfo {author} {\bibfnamefont {Clai}\ \bibnamefont {Owens}}, \bibinfo
  {author} {\bibfnamefont {Nelson}\ \bibnamefont {Leung}}, \bibinfo {author}
  {\bibfnamefont {Yao}\ \bibnamefont {Lu}}, \bibinfo {author} {\bibfnamefont
  {Jonathan}\ \bibnamefont {Simon}}, \ and\ \bibinfo {author} {\bibfnamefont
  {David~I.}\ \bibnamefont {Schuster}},\ }\bibfield  {title} {\enquote
  {\bibinfo {title} {Author correction: A dissipatively stabilized mott
  insulator of photons},}\ }\href {\doibase 10.1038/s41586-019-1221-4}
  {\bibfield  {journal} {\bibinfo  {journal} {Nature}\ }\textbf {\bibinfo
  {volume} {570}},\ \bibinfo {pages} {E52--E52} (\bibinfo {year}
  {2019})}\BibitemShut {NoStop}%
\bibitem [{\citenamefont {Mucci}\ \emph {et~al.}(2020)\citenamefont {Mucci},
  \citenamefont {Cao}, \citenamefont {Liu}, \citenamefont {Kaufman},
  \citenamefont {Pekker},\ and\ \citenamefont {Hatridge}}]{Mucci2020}%
  \BibitemOpen
  \bibfield  {author} {\bibinfo {author} {\bibfnamefont {Maria}\ \bibnamefont
  {Mucci}}, \bibinfo {author} {\bibfnamefont {Xi}~\bibnamefont {Cao}}, \bibinfo
  {author} {\bibfnamefont {Chenxu}\ \bibnamefont {Liu}}, \bibinfo {author}
  {\bibfnamefont {Ryan}\ \bibnamefont {Kaufman}}, \bibinfo {author}
  {\bibfnamefont {David}\ \bibnamefont {Pekker}}, \ and\ \bibinfo {author}
  {\bibfnamefont {Michael}\ \bibnamefont {Hatridge}},\ }\bibfield  {title}
  {\enquote {\bibinfo {title} {A josephson maser via three-wave coupling},}\ \
  }(\bibinfo  {publisher} {APS},\ \bibinfo {year} {2020})\BibitemShut {NoStop}%
\bibitem [{\citenamefont {Meschede}\ \emph {et~al.}(1985)\citenamefont
  {Meschede}, \citenamefont {Walther},\ and\ \citenamefont
  {M\"uller}}]{Meschede1985}%
  \BibitemOpen
  \bibfield  {author} {\bibinfo {author} {\bibfnamefont {D.}~\bibnamefont
  {Meschede}}, \bibinfo {author} {\bibfnamefont {H.}~\bibnamefont {Walther}}, \
  and\ \bibinfo {author} {\bibfnamefont {G.}~\bibnamefont {M\"uller}},\
  }\bibfield  {title} {\enquote {\bibinfo {title} {One-atom maser},}\ }\href
  {\doibase 10.1103/PhysRevLett.54.551} {\bibfield  {journal} {\bibinfo
  {journal} {Phys. Rev. Lett.}\ }\textbf {\bibinfo {volume} {54}},\ \bibinfo
  {pages} {551--554} (\bibinfo {year} {1985})}\BibitemShut {NoStop}%
\bibitem [{\citenamefont {Filipowicz}\ \emph {et~al.}(1986)\citenamefont
  {Filipowicz}, \citenamefont {Javanainen},\ and\ \citenamefont
  {Meystre}}]{Filipowicz1986}%
  \BibitemOpen
  \bibfield  {author} {\bibinfo {author} {\bibfnamefont {P.}~\bibnamefont
  {Filipowicz}}, \bibinfo {author} {\bibfnamefont {J.}~\bibnamefont
  {Javanainen}}, \ and\ \bibinfo {author} {\bibfnamefont {P.}~\bibnamefont
  {Meystre}},\ }\bibfield  {title} {\enquote {\bibinfo {title} {Theory of a
  microscopic maser},}\ }\href {\doibase 10.1103/PhysRevA.34.3077} {\bibfield
  {journal} {\bibinfo  {journal} {Phys. Rev. A}\ }\textbf {\bibinfo {volume}
  {34}},\ \bibinfo {pages} {3077--3087} (\bibinfo {year} {1986})}\BibitemShut
  {NoStop}%
\bibitem [{\citenamefont {Lugiato}\ \emph {et~al.}(1987)\citenamefont
  {Lugiato}, \citenamefont {Scully},\ and\ \citenamefont
  {Walther}}]{Lugiato1987}%
  \BibitemOpen
  \bibfield  {author} {\bibinfo {author} {\bibfnamefont {L.~A.}\ \bibnamefont
  {Lugiato}}, \bibinfo {author} {\bibfnamefont {M.~O.}\ \bibnamefont {Scully}},
  \ and\ \bibinfo {author} {\bibfnamefont {H.}~\bibnamefont {Walther}},\
  }\bibfield  {title} {\enquote {\bibinfo {title} {Connection between
  microscopic and macroscopic maser theory},}\ }\href {\doibase
  10.1103/PhysRevA.36.740} {\bibfield  {journal} {\bibinfo  {journal} {Phys.
  Rev. A}\ }\textbf {\bibinfo {volume} {36}},\ \bibinfo {pages} {740--743}
  (\bibinfo {year} {1987})}\BibitemShut {NoStop}%
\bibitem [{\citenamefont {Mu}\ and\ \citenamefont {Savage}(1992)}]{Mu1992}%
  \BibitemOpen
  \bibfield  {author} {\bibinfo {author} {\bibfnamefont {Yi}~\bibnamefont
  {Mu}}\ and\ \bibinfo {author} {\bibfnamefont {C.~M.}\ \bibnamefont
  {Savage}},\ }\bibfield  {title} {\enquote {\bibinfo {title} {One-atom
  lasers},}\ }\href {\doibase 10.1103/PhysRevA.46.5944} {\bibfield  {journal}
  {\bibinfo  {journal} {Phys. Rev. A}\ }\textbf {\bibinfo {volume} {46}},\
  \bibinfo {pages} {5944--5954} (\bibinfo {year} {1992})}\BibitemShut {NoStop}%
\bibitem [{\citenamefont {Bj\"ork}\ \emph {et~al.}(1994)\citenamefont
  {Bj\"ork}, \citenamefont {Karlsson},\ and\ \citenamefont
  {Yamamoto}}]{Gunnar1994}%
  \BibitemOpen
  \bibfield  {author} {\bibinfo {author} {\bibfnamefont {Gunnar}\ \bibnamefont
  {Bj\"ork}}, \bibinfo {author} {\bibfnamefont {Anders}\ \bibnamefont
  {Karlsson}}, \ and\ \bibinfo {author} {\bibfnamefont {Yoshihisa}\
  \bibnamefont {Yamamoto}},\ }\bibfield  {title} {\enquote {\bibinfo {title}
  {Definition of a laser threshold},}\ }\href {\doibase
  10.1103/PhysRevA.50.1675} {\bibfield  {journal} {\bibinfo  {journal} {Phys.
  Rev. A}\ }\textbf {\bibinfo {volume} {50}},\ \bibinfo {pages} {1675--1680}
  (\bibinfo {year} {1994})}\BibitemShut {NoStop}%
\bibitem [{\citenamefont {Wiseman}(1997)}]{Wiseman1997}%
  \BibitemOpen
  \bibfield  {author} {\bibinfo {author} {\bibfnamefont {H.~M.}\ \bibnamefont
  {Wiseman}},\ }\bibfield  {title} {\enquote {\bibinfo {title} {Defining the
  (atom) laser},}\ }\href {\doibase 10.1103/PhysRevA.56.2068} {\bibfield
  {journal} {\bibinfo  {journal} {Phys. Rev. A}\ }\textbf {\bibinfo {volume}
  {56}},\ \bibinfo {pages} {2068--2084} (\bibinfo {year} {1997})}\BibitemShut
  {NoStop}%
\bibitem [{\citenamefont {Boozer}\ \emph {et~al.}(2004)\citenamefont {Boozer},
  \citenamefont {Boca}, \citenamefont {Buck}, \citenamefont {McKeever},\ and\
  \citenamefont {Kimble}}]{Boozer2004}%
  \BibitemOpen
  \bibfield  {author} {\bibinfo {author} {\bibfnamefont {A.~D.}\ \bibnamefont
  {Boozer}}, \bibinfo {author} {\bibfnamefont {A.}~\bibnamefont {Boca}},
  \bibinfo {author} {\bibfnamefont {J.~R.}\ \bibnamefont {Buck}}, \bibinfo
  {author} {\bibfnamefont {J.}~\bibnamefont {McKeever}}, \ and\ \bibinfo
  {author} {\bibfnamefont {H.~J.}\ \bibnamefont {Kimble}},\ }\bibfield  {title}
  {\enquote {\bibinfo {title} {Comparison of theory and experiment for a
  one-atom laser in a regime of strong coupling},}\ }\href {\doibase
  10.1103/PhysRevA.70.023814} {\bibfield  {journal} {\bibinfo  {journal} {Phys.
  Rev. A}\ }\textbf {\bibinfo {volume} {70}},\ \bibinfo {pages} {023814}
  (\bibinfo {year} {2004})}\BibitemShut {NoStop}%
\bibitem [{\citenamefont {Wiseman}\ \emph {et~al.}(2019)\citenamefont
  {Wiseman}, \citenamefont {Saadatmand}, \citenamefont {Baker},\ and\
  \citenamefont {Berry}}]{Wiseman2019}%
  \BibitemOpen
  \bibfield  {author} {\bibinfo {author} {\bibfnamefont {Howard~M.}\
  \bibnamefont {Wiseman}}, \bibinfo {author} {\bibfnamefont {S.~Nariman}\
  \bibnamefont {Saadatmand}}, \bibinfo {author} {\bibfnamefont {Travis~J.}\
  \bibnamefont {Baker}}, \ and\ \bibinfo {author} {\bibfnamefont {Dominic~W.}\
  \bibnamefont {Berry}},\ }\bibfield  {title} {\enquote {\bibinfo {title} {The
  heisenberg limit for laser coherence},}\ }in\ \href {\doibase
  10.1364/CQO.2019.M3A.1} {\emph {\bibinfo {booktitle} {Rochester Conference on
  Coherence and Quantum Optics (CQO-11)}}}\ (\bibinfo  {publisher} {Optical
  Society of America},\ \bibinfo {year} {2019})\ p.\ \bibinfo {pages}
  {M3A.1}\BibitemShut {NoStop}%
\bibitem [{\citenamefont {Gottesman}\ \emph {et~al.}(2001)\citenamefont
  {Gottesman}, \citenamefont {Kitaev},\ and\ \citenamefont
  {Preskill}}]{Gottesman2001}%
  \BibitemOpen
  \bibfield  {author} {\bibinfo {author} {\bibfnamefont {Daniel}\ \bibnamefont
  {Gottesman}}, \bibinfo {author} {\bibfnamefont {Alexei}\ \bibnamefont
  {Kitaev}}, \ and\ \bibinfo {author} {\bibfnamefont {John}\ \bibnamefont
  {Preskill}},\ }\bibfield  {title} {\enquote {\bibinfo {title} {Encoding a
  qubit in an oscillator},}\ }\href {\doibase 10.1103/PhysRevA.64.012310}
  {\bibfield  {journal} {\bibinfo  {journal} {Phys. Rev. A}\ }\textbf {\bibinfo
  {volume} {64}},\ \bibinfo {pages} {012310} (\bibinfo {year}
  {2001})}\BibitemShut {NoStop}%
\bibitem [{\citenamefont {Menicucci}(2014)}]{Menicucci2014}%
  \BibitemOpen
  \bibfield  {author} {\bibinfo {author} {\bibfnamefont {Nicolas~C.}\
  \bibnamefont {Menicucci}},\ }\bibfield  {title} {\enquote {\bibinfo {title}
  {Fault-tolerant measurement-based quantum computing with continuous-variable
  cluster states},}\ }\href {\doibase 10.1103/PhysRevLett.112.120504}
  {\bibfield  {journal} {\bibinfo  {journal} {Phys. Rev. Lett.}\ }\textbf
  {\bibinfo {volume} {112}},\ \bibinfo {pages} {120504} (\bibinfo {year}
  {2014})}\BibitemShut {NoStop}%
\bibitem [{\citenamefont {Marshall}\ \emph {et~al.}(2016)\citenamefont
  {Marshall}, \citenamefont {Jacobsen}, \citenamefont {Sch{\"a}fermeier},
  \citenamefont {Gehring}, \citenamefont {Weedbrook},\ and\ \citenamefont
  {Andersen}}]{Marshall2016}%
  \BibitemOpen
  \bibfield  {author} {\bibinfo {author} {\bibfnamefont {Kevin}\ \bibnamefont
  {Marshall}}, \bibinfo {author} {\bibfnamefont {Christian~S.}\ \bibnamefont
  {Jacobsen}}, \bibinfo {author} {\bibfnamefont {Clemens}\ \bibnamefont
  {Sch{\"a}fermeier}}, \bibinfo {author} {\bibfnamefont {Tobias}\ \bibnamefont
  {Gehring}}, \bibinfo {author} {\bibfnamefont {Christian}\ \bibnamefont
  {Weedbrook}}, \ and\ \bibinfo {author} {\bibfnamefont {Ulrik~L.}\
  \bibnamefont {Andersen}},\ }\bibfield  {title} {\enquote {\bibinfo {title}
  {Continuous-variable quantum computing on encrypted data},}\ }\href {\doibase
  10.1038/ncomms13795} {\bibfield  {journal} {\bibinfo  {journal} {Nature
  Communications}\ }\textbf {\bibinfo {volume} {7}},\ \bibinfo {pages} {13795}
  (\bibinfo {year} {2016})}\BibitemShut {NoStop}%
\bibitem [{\citenamefont {Didier}\ \emph {et~al.}(2015)\citenamefont {Didier},
  \citenamefont {Kamal}, \citenamefont {Oliver}, \citenamefont {Blais},\ and\
  \citenamefont {Clerk}}]{Didier2015}%
  \BibitemOpen
  \bibfield  {author} {\bibinfo {author} {\bibfnamefont {Nicolas}\ \bibnamefont
  {Didier}}, \bibinfo {author} {\bibfnamefont {Archana}\ \bibnamefont {Kamal}},
  \bibinfo {author} {\bibfnamefont {William~D.}\ \bibnamefont {Oliver}},
  \bibinfo {author} {\bibfnamefont {Alexandre}\ \bibnamefont {Blais}}, \ and\
  \bibinfo {author} {\bibfnamefont {Aashish~A.}\ \bibnamefont {Clerk}},\
  }\bibfield  {title} {\enquote {\bibinfo {title} {Heisenberg-limited qubit
  read-out with two-mode squeezed light},}\ }\href {\doibase
  10.1103/PhysRevLett.115.093604} {\bibfield  {journal} {\bibinfo  {journal}
  {Phys. Rev. Lett.}\ }\textbf {\bibinfo {volume} {115}},\ \bibinfo {pages}
  {093604} (\bibinfo {year} {2015})}\BibitemShut {NoStop}%
\bibitem [{\citenamefont {Eddins}\ \emph {et~al.}(2018)\citenamefont {Eddins},
  \citenamefont {Schreppler}, \citenamefont {Toyli}, \citenamefont {Martin},
  \citenamefont {Hacohen-Gourgy}, \citenamefont {Govia}, \citenamefont
  {Ribeiro}, \citenamefont {Clerk},\ and\ \citenamefont
  {Siddiqi}}]{Eddins2018}%
  \BibitemOpen
  \bibfield  {author} {\bibinfo {author} {\bibfnamefont {A.}~\bibnamefont
  {Eddins}}, \bibinfo {author} {\bibfnamefont {S.}~\bibnamefont {Schreppler}},
  \bibinfo {author} {\bibfnamefont {D.~M.}\ \bibnamefont {Toyli}}, \bibinfo
  {author} {\bibfnamefont {L.~S.}\ \bibnamefont {Martin}}, \bibinfo {author}
  {\bibfnamefont {S.}~\bibnamefont {Hacohen-Gourgy}}, \bibinfo {author}
  {\bibfnamefont {L.~C.~G.}\ \bibnamefont {Govia}}, \bibinfo {author}
  {\bibfnamefont {H.}~\bibnamefont {Ribeiro}}, \bibinfo {author} {\bibfnamefont
  {A.~A.}\ \bibnamefont {Clerk}}, \ and\ \bibinfo {author} {\bibfnamefont
  {I.}~\bibnamefont {Siddiqi}},\ }\bibfield  {title} {\enquote {\bibinfo
  {title} {Stroboscopic qubit measurement with squeezed illumination},}\ }\href
  {\doibase 10.1103/PhysRevLett.120.040505} {\bibfield  {journal} {\bibinfo
  {journal} {Phys. Rev. Lett.}\ }\textbf {\bibinfo {volume} {120}},\ \bibinfo
  {pages} {040505} (\bibinfo {year} {2018})}\BibitemShut {NoStop}%
\bibitem [{\citenamefont {Caves}(1981)}]{caves1981}%
  \BibitemOpen
  \bibfield  {author} {\bibinfo {author} {\bibfnamefont {Carlton~M.}\
  \bibnamefont {Caves}},\ }\bibfield  {title} {\enquote {\bibinfo {title}
  {Quantum-mechanical noise in an interferometer},}\ }\href {\doibase
  10.1103/PhysRevD.23.1693} {\bibfield  {journal} {\bibinfo  {journal} {Phys.
  Rev. D}\ }\textbf {\bibinfo {volume} {23}},\ \bibinfo {pages} {1693--1708}
  (\bibinfo {year} {1981})}\BibitemShut {NoStop}%
\bibitem [{\citenamefont {Bondurant}\ and\ \citenamefont
  {Shapiro}(1984)}]{Bondurant1984}%
  \BibitemOpen
  \bibfield  {author} {\bibinfo {author} {\bibfnamefont {Roy~S.}\ \bibnamefont
  {Bondurant}}\ and\ \bibinfo {author} {\bibfnamefont {Jeffrey~H.}\
  \bibnamefont {Shapiro}},\ }\bibfield  {title} {\enquote {\bibinfo {title}
  {Squeezed states in phase-sensing interferometers},}\ }\href {\doibase
  10.1103/PhysRevD.30.2548} {\bibfield  {journal} {\bibinfo  {journal} {Phys.
  Rev. D}\ }\textbf {\bibinfo {volume} {30}},\ \bibinfo {pages} {2548--2556}
  (\bibinfo {year} {1984})}\BibitemShut {NoStop}%
\bibitem [{\citenamefont {Ma}\ \emph {et~al.}(2017)\citenamefont {Ma},
  \citenamefont {Miao}, \citenamefont {Pang}, \citenamefont {Evans},
  \citenamefont {Zhao}, \citenamefont {Harms}, \citenamefont {Schnabel},\ and\
  \citenamefont {Chen}}]{Ma2017}%
  \BibitemOpen
  \bibfield  {author} {\bibinfo {author} {\bibfnamefont {Yiqiu}\ \bibnamefont
  {Ma}}, \bibinfo {author} {\bibfnamefont {Haixing}\ \bibnamefont {Miao}},
  \bibinfo {author} {\bibfnamefont {Belinda~Heyun}\ \bibnamefont {Pang}},
  \bibinfo {author} {\bibfnamefont {Matthew}\ \bibnamefont {Evans}}, \bibinfo
  {author} {\bibfnamefont {Chunnong}\ \bibnamefont {Zhao}}, \bibinfo {author}
  {\bibfnamefont {Jan}\ \bibnamefont {Harms}}, \bibinfo {author} {\bibfnamefont
  {Roman}\ \bibnamefont {Schnabel}}, \ and\ \bibinfo {author} {\bibfnamefont
  {Yanbei}\ \bibnamefont {Chen}},\ }\bibfield  {title} {\enquote {\bibinfo
  {title} {Proposal for gravitational-wave detection beyond the standard
  quantum limit through epr entanglement},}\ }\href {\doibase
  10.1038/nphys4118} {\bibfield  {journal} {\bibinfo  {journal} {Nature
  Physics}\ }\textbf {\bibinfo {volume} {13}},\ \bibinfo {pages} {776--780}
  (\bibinfo {year} {2017})}\BibitemShut {NoStop}%
\bibitem [{\citenamefont {Zuo}\ \emph {et~al.}(2020)\citenamefont {Zuo},
  \citenamefont {Yan}, \citenamefont {Feng}, \citenamefont {Ma}, \citenamefont
  {Jia}, \citenamefont {Xie},\ and\ \citenamefont {Peng}}]{XJZuo2020}%
  \BibitemOpen
  \bibfield  {author} {\bibinfo {author} {\bibfnamefont {Xiaojie}\ \bibnamefont
  {Zuo}}, \bibinfo {author} {\bibfnamefont {Zhihui}\ \bibnamefont {Yan}},
  \bibinfo {author} {\bibfnamefont {Yanni}\ \bibnamefont {Feng}}, \bibinfo
  {author} {\bibfnamefont {Jingxu}\ \bibnamefont {Ma}}, \bibinfo {author}
  {\bibfnamefont {Xiaojun}\ \bibnamefont {Jia}}, \bibinfo {author}
  {\bibfnamefont {Changde}\ \bibnamefont {Xie}}, \ and\ \bibinfo {author}
  {\bibfnamefont {Kunchi}\ \bibnamefont {Peng}},\ }\bibfield  {title} {\enquote
  {\bibinfo {title} {Quantum interferometer combining squeezing and parametric
  amplification},}\ }\href {\doibase 10.1103/PhysRevLett.124.173602} {\bibfield
   {journal} {\bibinfo  {journal} {Phys. Rev. Lett.}\ }\textbf {\bibinfo
  {volume} {124}},\ \bibinfo {pages} {173602} (\bibinfo {year}
  {2020})}\BibitemShut {NoStop}%
\bibitem [{\citenamefont {Yuen}\ and\ \citenamefont
  {Shapiro}(1978)}]{Yuen1978}%
  \BibitemOpen
  \bibfield  {author} {\bibinfo {author} {\bibfnamefont {H.}~\bibnamefont
  {Yuen}}\ and\ \bibinfo {author} {\bibfnamefont {J.}~\bibnamefont {Shapiro}},\
  }\bibfield  {title} {\enquote {\bibinfo {title} {Optical communication with
  two-photon coherent states--part i: Quantum-state propagation and
  quantum-noise},}\ }\href@noop {} {\bibfield  {journal} {\bibinfo  {journal}
  {IEEE Transactions on Information Theory}\ }\textbf {\bibinfo {volume}
  {24}},\ \bibinfo {pages} {657--668} (\bibinfo {year} {1978})}\BibitemShut
  {NoStop}%
\bibitem [{\citenamefont {Slusher}\ and\ \citenamefont
  {Yurke}(1990)}]{Slusher1990}%
  \BibitemOpen
  \bibfield  {author} {\bibinfo {author} {\bibfnamefont {R.~E.}\ \bibnamefont
  {Slusher}}\ and\ \bibinfo {author} {\bibfnamefont {B.}~\bibnamefont
  {Yurke}},\ }\bibfield  {title} {\enquote {\bibinfo {title} {Squeezed light
  for coherent communications},}\ }\href@noop {} {\bibfield  {journal}
  {\bibinfo  {journal} {Journal of Lightwave Technology}\ }\textbf {\bibinfo
  {volume} {8}},\ \bibinfo {pages} {466--477} (\bibinfo {year}
  {1990})}\BibitemShut {NoStop}%
\bibitem [{\citenamefont {Gottesman}\ and\ \citenamefont
  {Preskill}(2003)}]{Gottesman2003}%
  \BibitemOpen
  \bibfield  {author} {\bibinfo {author} {\bibfnamefont {Daniel}\ \bibnamefont
  {Gottesman}}\ and\ \bibinfo {author} {\bibfnamefont {John}\ \bibnamefont
  {Preskill}},\ }\enquote {\bibinfo {title} {Secure quantum key distribution
  using squeezed states},}\ in\ \href {\doibase 10.1007/978-94-015-1258-9_22}
  {\emph {\bibinfo {booktitle} {Quantum Information with Continuous
  Variables}}},\ \bibinfo {editor} {edited by\ \bibinfo {editor} {\bibfnamefont
  {Samuel~L.}\ \bibnamefont {Braunstein}}\ and\ \bibinfo {editor}
  {\bibfnamefont {Arun~K.}\ \bibnamefont {Pati}}}\ (\bibinfo  {publisher}
  {Springer Netherlands},\ \bibinfo {address} {Dordrecht},\ \bibinfo {year}
  {2003})\ pp.\ \bibinfo {pages} {317--356}\BibitemShut {NoStop}%
\bibitem [{\citenamefont {Vahlbruch}\ \emph {et~al.}(2008)\citenamefont
  {Vahlbruch}, \citenamefont {Mehmet}, \citenamefont {Chelkowski},
  \citenamefont {Hage}, \citenamefont {Franzen}, \citenamefont {Lastzka},
  \citenamefont {Go\ss{}ler}, \citenamefont {Danzmann},\ and\ \citenamefont
  {Schnabel}}]{Vahlbruch2008}%
  \BibitemOpen
  \bibfield  {author} {\bibinfo {author} {\bibfnamefont {Henning}\ \bibnamefont
  {Vahlbruch}}, \bibinfo {author} {\bibfnamefont {Moritz}\ \bibnamefont
  {Mehmet}}, \bibinfo {author} {\bibfnamefont {Simon}\ \bibnamefont
  {Chelkowski}}, \bibinfo {author} {\bibfnamefont {Boris}\ \bibnamefont
  {Hage}}, \bibinfo {author} {\bibfnamefont {Alexander}\ \bibnamefont
  {Franzen}}, \bibinfo {author} {\bibfnamefont {Nico}\ \bibnamefont {Lastzka}},
  \bibinfo {author} {\bibfnamefont {Stefan}\ \bibnamefont {Go\ss{}ler}},
  \bibinfo {author} {\bibfnamefont {Karsten}\ \bibnamefont {Danzmann}}, \ and\
  \bibinfo {author} {\bibfnamefont {Roman}\ \bibnamefont {Schnabel}},\
  }\bibfield  {title} {\enquote {\bibinfo {title} {Observation of squeezed
  light with 10-db quantum-noise reduction},}\ }\href {\doibase
  10.1103/PhysRevLett.100.033602} {\bibfield  {journal} {\bibinfo  {journal}
  {Phys. Rev. Lett.}\ }\textbf {\bibinfo {volume} {100}},\ \bibinfo {pages}
  {033602} (\bibinfo {year} {2008})}\BibitemShut {NoStop}%
\bibitem [{\citenamefont {Liu}\ \emph {et~al.}(2021)\citenamefont {Liu},
  \citenamefont {Mucci}, \citenamefont {Cao}, \citenamefont {Dutt},
  \citenamefont {Hatridge},\ and\ \citenamefont {Pekker}}]{code_DOI}%
  \BibitemOpen
  \bibfield  {author} {\bibinfo {author} {\bibfnamefont {Chenxu}\ \bibnamefont
  {Liu}}, \bibinfo {author} {\bibfnamefont {Maria}\ \bibnamefont {Mucci}},
  \bibinfo {author} {\bibfnamefont {Xi}~\bibnamefont {Cao}}, \bibinfo {author}
  {\bibfnamefont {M.~V.~Gurudev}\ \bibnamefont {Dutt}}, \bibinfo {author}
  {\bibfnamefont {Michael}\ \bibnamefont {Hatridge}}, \ and\ \bibinfo {author}
  {\bibfnamefont {David}\ \bibnamefont {Pekker}},\ }\href {\doibase
  10.5281/zenodo.5016168} {\enquote {\bibinfo {title} {Proposal for a
  continuous wave laser with linewidth well below the standard quantum limit:
  Calculation code},}\ } (\bibinfo {year} {2021}),\ \bibinfo {note} {doi:
  \url{10.5281/zenodo.5016168}}\BibitemShut {NoStop}%
\bibitem [{\citenamefont {Gardiner}\ \emph {et~al.}(2004)\citenamefont
  {Gardiner}, \citenamefont {Zoller},\ and\ \citenamefont
  {Zoller}}]{GardinerQNbook}%
  \BibitemOpen
  \bibfield  {author} {\bibinfo {author} {\bibfnamefont {C.}~\bibnamefont
  {Gardiner}}, \bibinfo {author} {\bibfnamefont {P.}~\bibnamefont {Zoller}}, \
  and\ \bibinfo {author} {\bibfnamefont {P.}~\bibnamefont {Zoller}},\
  }\href@noop {} {\emph {\bibinfo {title} {Quantum Noise: A Handbook of
  Markovian and Non-Markovian Quantum Stochastic Methods with Applications to
  Quantum Optics}}},\ Springer Series in Synergetics\ (\bibinfo  {publisher}
  {Springer},\ \bibinfo {year} {2004})\BibitemShut {NoStop}%
\bibitem [{\citenamefont {Briegel}\ and\ \citenamefont
  {Englert}(1993)}]{Briegel1993}%
  \BibitemOpen
  \bibfield  {author} {\bibinfo {author} {\bibfnamefont {Hans-J\"urgen}\
  \bibnamefont {Briegel}}\ and\ \bibinfo {author} {\bibfnamefont
  {Berthold-Georg}\ \bibnamefont {Englert}},\ }\bibfield  {title} {\enquote
  {\bibinfo {title} {Quantum optical master equations: The use of damping
  bases},}\ }\href {\doibase 10.1103/PhysRevA.47.3311} {\bibfield  {journal}
  {\bibinfo  {journal} {Phys. Rev. A}\ }\textbf {\bibinfo {volume} {47}},\
  \bibinfo {pages} {3311--3329} (\bibinfo {year} {1993})}\BibitemShut {NoStop}%
\bibitem [{\citenamefont {Ginzel}\ \emph {et~al.}(1993)\citenamefont {Ginzel},
  \citenamefont {Briegel}, \citenamefont {Martini}, \citenamefont {Englert},\
  and\ \citenamefont {Schenzle}}]{Ginzel1993}%
  \BibitemOpen
  \bibfield  {author} {\bibinfo {author} {\bibfnamefont {Christian}\
  \bibnamefont {Ginzel}}, \bibinfo {author} {\bibfnamefont {Hans-J\"urgen}\
  \bibnamefont {Briegel}}, \bibinfo {author} {\bibfnamefont {Ullrich}\
  \bibnamefont {Martini}}, \bibinfo {author} {\bibfnamefont {Berthold-Georg}\
  \bibnamefont {Englert}}, \ and\ \bibinfo {author} {\bibfnamefont {Axel}\
  \bibnamefont {Schenzle}},\ }\bibfield  {title} {\enquote {\bibinfo {title}
  {Quantum optical master equations: The one-atom laser},}\ }\href {\doibase
  10.1103/PhysRevA.48.732} {\bibfield  {journal} {\bibinfo  {journal} {Phys.
  Rev. A}\ }\textbf {\bibinfo {volume} {48}},\ \bibinfo {pages} {732--738}
  (\bibinfo {year} {1993})}\BibitemShut {NoStop}%
\bibitem [{\citenamefont {Briegel}\ \emph {et~al.}(1994)\citenamefont
  {Briegel}, \citenamefont {Englert}, \citenamefont {Ginzel},\ and\
  \citenamefont {Schenzle}}]{Briegel1994}%
  \BibitemOpen
  \bibfield  {author} {\bibinfo {author} {\bibfnamefont {Hans-J\"urgen}\
  \bibnamefont {Briegel}}, \bibinfo {author} {\bibfnamefont {Berthold-Georg}\
  \bibnamefont {Englert}}, \bibinfo {author} {\bibfnamefont {Christian}\
  \bibnamefont {Ginzel}}, \ and\ \bibinfo {author} {\bibfnamefont {Axel}\
  \bibnamefont {Schenzle}},\ }\bibfield  {title} {\enquote {\bibinfo {title}
  {One-atom maser with a periodic and noisy pump: An application of damping
  bases},}\ }\href {\doibase 10.1103/PhysRevA.49.5019} {\bibfield  {journal}
  {\bibinfo  {journal} {Phys. Rev. A}\ }\textbf {\bibinfo {volume} {49}},\
  \bibinfo {pages} {5019--5041} (\bibinfo {year} {1994})}\BibitemShut {NoStop}%
\bibitem [{\citenamefont {Carruthers}\ and\ \citenamefont
  {Nieto}(1965)}]{Carruthers1965}%
  \BibitemOpen
  \bibfield  {author} {\bibinfo {author} {\bibfnamefont {P.}~\bibnamefont
  {Carruthers}}\ and\ \bibinfo {author} {\bibfnamefont {M.~M.}\ \bibnamefont
  {Nieto}},\ }\bibfield  {title} {\enquote {\bibinfo {title} {Coherent states
  and the number-phase uncertainty relation},}\ }\href {\doibase
  10.1103/PhysRevLett.14.387} {\bibfield  {journal} {\bibinfo  {journal} {Phys.
  Rev. Lett.}\ }\textbf {\bibinfo {volume} {14}},\ \bibinfo {pages} {387--389}
  (\bibinfo {year} {1965})}\BibitemShut {NoStop}%
\bibitem [{\citenamefont {Carruthers}\ and\ \citenamefont
  {Nieto}(1968)}]{Carruthers1968}%
  \BibitemOpen
  \bibfield  {author} {\bibinfo {author} {\bibfnamefont {P.}~\bibnamefont
  {Carruthers}}\ and\ \bibinfo {author} {\bibfnamefont {Michael~Martin}\
  \bibnamefont {Nieto}},\ }\bibfield  {title} {\enquote {\bibinfo {title}
  {Phase and angle variables in quantum mechanics},}\ }\href {\doibase
  10.1103/RevModPhys.40.411} {\bibfield  {journal} {\bibinfo  {journal} {Rev.
  Mod. Phys.}\ }\textbf {\bibinfo {volume} {40}},\ \bibinfo {pages} {411--440}
  (\bibinfo {year} {1968})}\BibitemShut {NoStop}%
\end{thebibliography}%

\end{document}